\documentclass{jfm}

\usepackage{graphicx}
\usepackage{epstopdf, epsfig}
\usepackage{amsmath,amssymb,stmaryrd,array} 
\usepackage{float}
\usepackage{color,soul,rotating,overpic}
\usepackage{multicol}
\usepackage{hyperref}
\hypersetup{
	colorlinks=true,
	linktoc=all,
	linkcolor=blue,
	citecolor=blue,
}

\definecolor{red}{rgb}{0.8500, 0.1250, 0.0480} 
\definecolor{blue}{rgb}{0, 0.4470, 0.7410}
\definecolor{green}{rgb}{0.4660, 0.6740, 0.1880}
\definecolor{gray}{rgb}{0.7, 0.7, 0.7}
\newcolumntype{M}[1]{>{\centering\arraybackslash}m{#1}}

\shorttitle{Biglobal instabilities of compressible open-cavity flows}
\shortauthor{Y. Sun, K. Taira, L. N. Cattafesta and L. S. Ukeiley}

\title{Biglobal instabilities of compressible open-cavity flows}

\author{Yiyang Sun\aff{1}
  \corresp{\email{ys12d@my.fsu.edu}},
  Kunihiko Taira\aff{1},
  Louis N. Cattafesta III\aff{1},
 \and Lawrence S. Ukeiley\aff{2}}

\affiliation{\aff{1}Department of Mechanical Engineering, Florida State University, Tallahassee, FL 32310, USA
\aff{2}Department of Mechanical and Aerospace Engineering, University of Florida, Gainesville, FL 32611, USA}

\begin{document}

\maketitle

\begin{abstract}

The stability characteristics of compressible spanwise-periodic open-cavity flows are investigated with direct numerical simulation (DNS) and biglobal stability analysis for rectangular cavities with aspect ratios of $L/D=2$ and 6. This study examines the behavior of instabilities with respect to stable and unstable steady states in the laminar regimes
for subsonic 
as well as transonic conditions 
where compressibility plays an important role. 
It is observed that an increase in Mach number destabilizes the flow in the subsonic regime and stabilizes the flow in the transonic regime.  Biglobal stability analysis for spanwise-periodic flows over rectangular cavities with large aspect ratio is closely examined in this study due to its importance in aerodynamic applications.
Moreover, biglobal stability analysis is conducted to extract 2D and 3D eigenmodes for prescribed spanwise wavelengths $\lambda/D$ about the 2D steady state. The properties of 2D eigenmodes agree well with those observed in the 2D nonlinear simulations. In the analysis of 3D eigenmodes, it is found that an increase of Mach number stabilizes dominant 3D eigenmodes. For a short cavity with $L/D=2$, the 3D eigenmodes primarily stem from centrifugal instabilities. For a long cavity with $L/D=6$, other types of eigenmodes appear whose structures extend from the aft-region to the mid-region of the cavity, in addition to the centrifugal instability mode located in the rear part of the cavity. A selected number of 3D DNS are performed at $M_\infty=0.6$ for cavities with $L/D=2$ and 6. For $L/D=2$, the properties of 3D structures present in the 3D nonlinear flow correspond closely to those obtained from linear stability analysis. However, for $L/D=6$, the 3D eigenmodes cannot be clearly observed in the 3D DNS, due to the strong nonlinearity that develops over the length of the cavity. In addition, it is noted that three-dimensionality in the flow helps alleviate 
violent oscillations for the long cavity. The analysis performed in this paper can provide valuable insights for designing effective flow control strategies to suppress undesirable aerodynamic and pressure fluctuations in compressible open-cavity flows.
\end{abstract}



\section{Introduction}\label{sec:intro}

Flow over an open rectangular cavity has been a fundamental research topic for decades because of its ubiquitous nature in many engineering settings, including landing gear wells, gaps between plates, and aircraft weapon bays \citep{Cattafesta:PAS08,Lawson:PAS11}. In open cavity flow, a shear layer emanates from the leading edge of the cavity and spans the length of the cavity. The perturbations in the shear layer are amplified through the Kelvin--Helmholtz instability, which introduce large vortical structures that impinge on the aft wall of the cavity creating intense pressure fluctuations and acoustic waves. These waves propagate upstream and induce new disturbances in the shear layer from the leading edge, which forms an acoustic feedback process that can lead to self-sustained oscillations. 

The properties of these oscillations can be affected by various parameters including cavity geometry as well as the hydrodynamic and acoustic characteristics of the incoming flow \citep{Rockwell:JFE78,Lawson:PAS11, Rowley:JFM02, Sun:AIAA14}. The early work of \cite{Rossiter:ARCRM64} predicts the 2D shear-layer oscillation frequency through a semi-empirical formula based on the free stream Mach number, $M_\infty$. \cite{Heller:AIAA75} later modified the formula to better match the experimental measurements. This modified formula is used as reference in the present work. The mode associated with the resonance is referred to as the Rossiter mode and its frequency $f_n$ can be predicted in terms of the Strouhal number as
\begin{equation}
St_L=\frac{f_n L}{u_\infty}=\frac{n-\alpha}{1/\kappa+M_\infty/\sqrt{1+(\gamma-1)M_\infty^2/2}},
\label{eqRossiter}
\end{equation}
where $L$ is the length of cavity, empirical constant $\kappa$ $(= 0.57)$ is the average convective speed of disturbance in shear layer, $\alpha$ ($= 0.25$) \citep{Rossiter:ARCRM64} corresponds to phase delay, $\gamma$ (=1.4) is specific heat ratio, and $n=1,2, \dots $ leads to the $n$th Rossiter mode. We also define a Strouhal number based on the cavity depth ($St_D=f_n D/u_\infty$) to quantify the frequencies of 3D modes of open-cavity flows. 

The oscillations associated with open-cavity flows are generally undesirable, because they may damage the cavity contents due to the unsteady aerodynamic forces and intense pressure fluctuations. During the past few decades, researchers have developed techniques to suppress the oscillations through various passive and active flow control strategies. A comprehensive review on active control of high Reynolds number cavity flow for a wide range of Mach numbers is given by \cite{Cattafesta:PAS08}. Both open and closed-loop control techniques have demonstrated the ability to significantly reduce the pressure fluctuations and noise emission \citep{Samimy:JFM07, Cattafesta:PAS08, Zhang:AIAA15, Lusk:EF12}. However, there has not been a clear control strategy that can be universally applied to open-cavity flows in the most general manner. As the baseline flow changes with different cavity geometry and operating conditions, the current control approaches for suppressing aerodynamic fluctuations have to be tailored for each operating condition and cavity configuration. Although flow control has been applied in some high-speed flows \citep{Cattafesta:PAS08, Cattafesta:ARFM11}, both effective and efficient control of transonic and supersonic cavity flows remains challenging due to a lack of actuator control authority at these flow conditions. Due to the large energy input required for high-speed cavity flows, the control authority is typically insufficient. Moreover, current actuators are characterized by a fixed gain-bandwidth product \citep{Cattafesta:ARFM11}, which  motivates the present study that can shed light on the proper choice of actuation parameters based on intrinsic flow instabilities.



As observed in several studies on cavity flows \citep{Maull:JFM63,Ahuja:NASA95,Beresh:AIAA15,Arunajatesan:AIAA14, Beresh:JFM16, Sun:AIAA16}, three dimensionality can affect the dominant oscillation characteristics. Although the three dimensionality discussed in these experiments and simulations mostly relates to the significance of spanwise end effects on the flow, such end effects can modify spanwise instabilities in the flow. This suggests a potential to control cavity flow with 3D perturbations. To understand the characteristics of the three-dimensionality in open-cavity flow, 3D simulations can capture unsteadiness and structures of steady saturated flow but cannot reveal the instability characteristics directly. However, with the development of numerical techniques, numerical instability analysis of cavity flows has attracted increasing attention during the last decade. In particular, linear stability analysis provides deep insights into the instability mechanisms, as well as physics-based guidelines for effective 3D control strategies in terms of spatial and temporal parameters needed in flow-control designs. Hence, in this paper, the influence of compressibility on characteristics of 2D and 3D global instabilities of spanwise-periodic open-cavity flows are examined thoroughly based on linear stability theory.  

To analyze the 3D instabilities of such cavity flows, we utilize biglobal stability theory to identify the properties of the 3D instabilities associated for a given 2D base state. The computational cost associated with this analysis is much less compared to the case in which the base state and perturbations are both considered three-dimensional. Hence, it is a widely used technique to specifically examine the spanwise instabilities of 2D inhomogeneous flows \citep{Theofilis:PAS03}. In table \ref{tab:sum}, we summarize past biglobal stability analyses of open-cavity flows. \cite{Theofilis:AIAA04} presented the framework of the biglobal stability analysis for compressible open-cavity flows with aspect ratio $L/D=2$ and $M_\infty=0.325$. Following their work, several researchers reported on instability studies of open-cavity flows using linear instability analysis. \cite{Bres:JFM08} characterized the onset of 2D subsonic ($0.1 \leq M_\infty \leq 0.6$) cavity flow instabilities at low Reynolds numbers. They also performed 3D linear simulations on compressible open-cavity flow with $L/D=2$ and 4 and identified the spanwise wavelength of the most-unstable/least-stable modes via examination of the most amplified disturbances with respect to the steady base flow. They found that the 3D modes have an order-of-magnitude lower frequency than those of the 2D resonant modes in the cavity, with the wavelength of the most-unstable mode being $\lambda/D\approx1$. \cite{Yamouni:JFM13} performed global stability analysis to investigate the interaction between feedback aeroacoustic mechanism and acoustic resonance in the flow over cavity with $L/D=1$. Moreover, \cite{Vicente:JFM14} conducted global stability analysis on incompressible open-cavity flows and observed that 3D instability modes can split into different branches depending on their spanwise wavelengths. They also compared their numerical results to experiments and reported that their numerical results resembled the fully saturated nonlinear flow features seen in experiments. Considering lateral wall effects on the 3D structures present in finite-span cavity flows, \cite{Liu:JFM16} performed triglobal instability analysis to unravel the transition of steady laminar flow over a three-dimensional cavity for incompressible flow. All of these studies provide insights into the characteristics of spanwise instabilities associated with open-cavity flows. Nonetheless, there is still a gap in the literature with respect to instabilities of open-cavity flow in the transonic regime. Moreover, the instabilities of flows over the cavities with large $L/D=6$, which are particularly relevant in aircraft bays, have been rarely studied. Furthermore, three-dimensional flow control strategies applied on realistic compressible open cavity flows \citep{Lusk:EF12,George:AIAA15,Zhang:AIAA15} are functions of spanwise wavelength and Mach number, for which the present work can offer insights.

\begin{table}
\begin{center}
\begin{tabular}{l c M{1in}M{1in}}
					&$L/D$	&$M_\infty$			&$\beta ~(=2\pi/\lambda)$	\\ \hline
\cite{Bres:JFM08}		& 1,~2,~4	&0.1-0.6				&3.14 - 12.56		\\ 
\cite{Yamouni:JFM13}	& 1~\&~2	&0.0-0.9				&0				\\ 
\cite{Garrido:JFM14}		& 1 - 3	&0.0			 		&0 - 22			\\ 
\cite{Vicente:JFM14}		& 2		&0.0					&0 - 22			\\
Present				&2~\&~6	&0.1-1.4				&0 \& 3.14 - 12.56
\end{tabular}
\end{center}
\caption{A summary of biglobal stability analysis of laminar open-cavity flows studied in past literature and the present work. Mach number 0 represents incompressible flow.}
\label{tab:sum}
\end{table}

One of the objectives of this paper is to perform 2D DNS of flows over rectangular cavities with $L/D=2$ and 6 to characterize the effects of free stream Mach number $M_\infty$, Reynolds number $Re_\theta$ and aspect ratio $L/D$ on the 2D flow instabilities. Furthermore, the stable/unstable steady states obtained from 2D simulations serve as base states in the biglobal stability analysis to reveal characteristics of 2D ($\lambda/D=\infty$) and 3D ($\lambda/D=0.5 -2.0$) eigenmodes associated with the flows. In the linear stability analysis component of this study, 2D and 3D global eigenmodes are identified for $M_\infty=0.3-1.4$ and $L/D=2$, 6. These global eigenmodes are also compared to the flow fields from the 2D and 3D nonlinear simulations. We will show that most of the linear stability predictions of flow properties are in a good agreement to those captured from the nonlinear flows, which could serve a foundation for parameter choice in flow control designs. 

In what follows, the computational approach and numerical validation are presented in \S{\ref{sec:approach}}. In \S{\ref{sec:results}}, the characteristics of the 2D instabilities in open-cavity flows are investigated via DNS. With the stable/unstable steady states obtained from 2D DNS, 2D eigenmodes captured via biglobal stability analysis are discussed and compared to the flow characteristics revealed in the nonlinear simulations in \S{\ref{sec:global2D}}. Furthermore, 3D eigenmodes with specified spanwise wavelengths are examined in \S{\ref{sec:global3D}}, in which the instabilities of 3D modes show significant dependence on Mach number and spanwise wavelength. A comparison of the results from the DNS and biglobal stability analysis is provided in \S{\ref{sec:global3D}}. Finally, concluding remarks are offered in \S{\ref{sec:summary}}.

\section{Computational approach}\label{sec:approach}
\subsection{Direct numerical simulation setup and validation}
Two-dimensional DNS of compressible flows over a rectangular cavity are performed using a high-fidelity compressible flow solver {\it{CharLES}} \citep{Khalighi:ASME2011, Khalighi:AIAA11, Bres:AIAAJ17} to solve the full compressible Navier--Stokes equations. A second-order finite-volume method and the third-order Runge--Kutta temporal scheme are implemented. 
The Harten-Lax-van Leer contact (HLLC) scheme is used to capture the shocks formed in supersonic flow  \citep{Toro:09}. The variables including the spatial coordinate $x_i$, time $t$, density $\rho$, velocity $u_i$, energy $e$, pressure $P$, temperature $T$, are non-dimensionalized as
\[
x_i = \frac{{x_i} ^d}{D^d},~
t = \frac{t^d a_\infty^d}{D^d},~
\rho = \frac{\rho ^d}{\rho_{\infty}^d},~
u_i = \frac{{u_i} ^d}{a_{\infty}^d},~
e = \frac{e ^d}{\rho_{\infty}^d (a_{\infty}^d)^2},~
P =  \frac{P ^d}{\gamma P_\infty^d},~
T = \frac{T ^d}{T_\infty^d},~
\]
where variables with superscript $d$ refer to the dimensional quantities and those with the subscript $\infty$ denote the free stream values. The $x$-, $y$-, and $z$-directions represent the streamwise, wall-normal, and spanwise directions, respectively. A structured mesh with non-uniform spacing in both $x$- and $y$-directions is used for the simulations. Open-cavity flows are specified by $L/D$, where $L$ and $D$ represent the length and depth of the cavity, respectively, initial boundary layer momentum thickness $\theta_0$ at the leading edge of the cavity, and free stream Mach number $M_\infty \equiv u^d_\infty/a^d_\infty$. The free stream sonic speed is denoted by $a_\infty^d$. The Reynolds number based on the initial momentum boundary layer thickness $\theta_0$ and the Prandtl number are respectively defined as
\[
Re_\theta\equiv \frac{\rho_\infty u_\infty \theta_0}{\mu_\infty}~~~~\text{and}~~~~Pr \equiv \frac{c_p \mu_\infty}{k},
\]
where $\mu_\infty$ is the dynamic viscosity, $c_p$ is the specific heat, and $k$ is the thermal conductivity.
 
\begin{figure}
\begin{center}
   \includegraphics[width=0.65\textwidth]{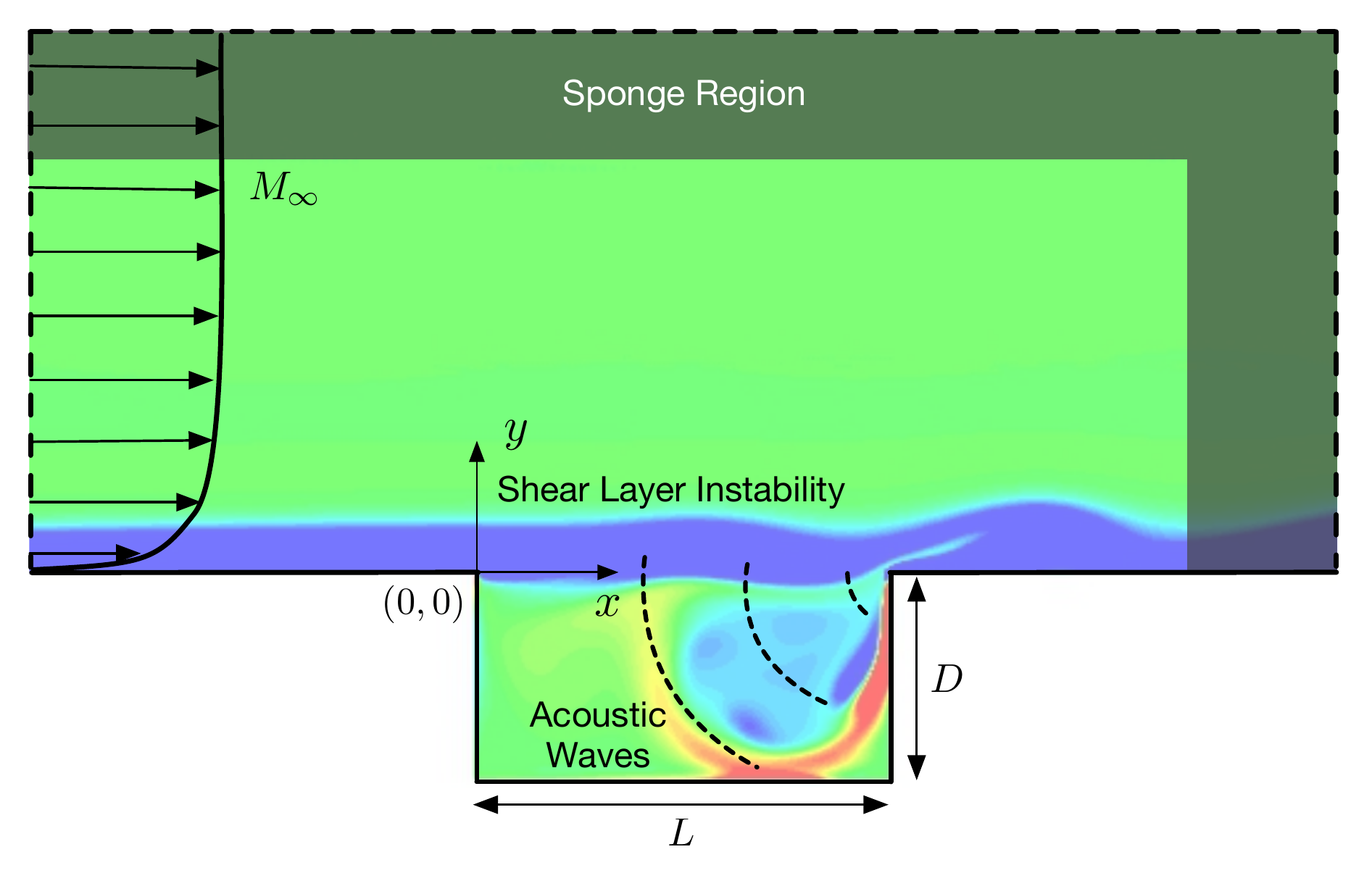}
   \caption{Computational setup for open-cavity flow (not to scale).}
   \label{fig:setup}
\vspace{-0.2in}
\end{center}
\end{figure}

In the present investigation, we consider two-dimensional cavities with $L/D = 2$ and 6. The former geometry serves as the basis for comparison with those reported in the literature, while the latter is representative of a prototypical cavity application on aircraft. As illustrated in figure \ref{fig:setup}, the origin is placed at the leading edge of the cavity. The initial momentum boundary layer thickness $\theta_0$ is prescribed at the leading edge, and the distance between the upstream wall boundary and the leading edge is adjusted accordingly for the chosen $Re_\theta$. The outflow boundary is placed $7D$ from the trailing edge of the cavity. The normal distance from the cavity surface to the top boundary of the computational domain is maintained at $9D$. The size of computational domain follows the work of \cite{Colonius:99}, in which they studied the effects of the computational domain on the flows and identified the appropriate size of the domain. No-slip and adiabatic boundary conditions are specified at the upstream and downstream floor as well as the walls of the cavity. To damp out exiting acoustic waves and wake structures, sponge zones \citep{Freund:AIAAJ97} are applied to the outlet and top boundaries spanning a length of $2D$ from computational boundaries. The computational domain for three-dimensional DNS extends the two-dimensional setup with spanwise periodicity for a width-to-depth ratio of $W/D=2$ (which is suitable for $\lambda/D\le2$) with 64 grid points spaced uniformly in the spanwise direction.

The effect of Mach number is analyzed from the subsonic regime to the transonic regime with Reynolds number $Re_\theta$ from 5 to 144 for flows over cavities with $L/D = 2$ and 6. To initialize the flow field, an incompressible Blasius boundary layer profile is imposed over the entire computational domain above the floor, while the flow inside the cavity is set to be quiescent. Consistent with the chosen Reynolds number range of this study, the incompressible Blasius boundary layer profile is utilized as the variation in boundary layer thickness for the range of Mach numbers from $0$ to $1.4$ is less than $10\%$ \citep{white91}. As the Blasius profile is characterized by the momentum boundary layer thickness, we fix the ratio of the cavity depth to the initial momentum thickness $D/\theta_0 = 26.4$ for all the cases considered in the present study.

\begin{figure}
\begin{center}
\begin{tabular}{cc}
   \includegraphics[width=0.46\textwidth]{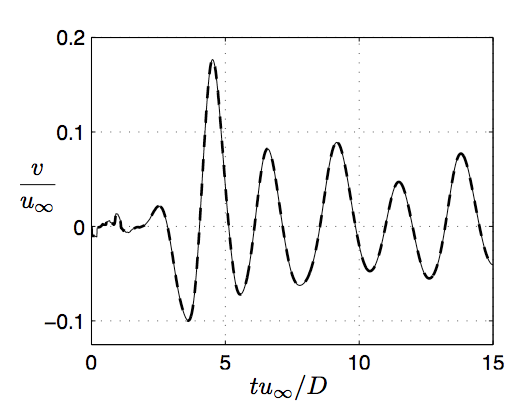}&\includegraphics[width=0.45\textwidth]{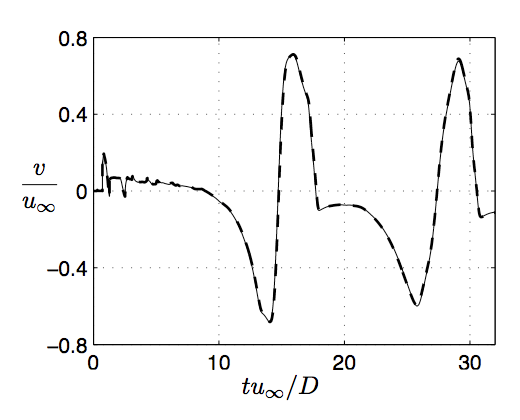}\\
   $(a)$ $L/D=2$, $M_\infty=0.1$ and $Re_\theta=144$&$(b)$ $L/D=6$, $M_\infty=0.3$ and $Re_\theta=46$
   \end{tabular}
      \caption{Comparison of the $v$-velocity at the midpoint ($x=L/2$, $y=0$) of the cavity. The solid line shows the baseline case with half a million grid points and the dashed line represents the refined case with 1.1 million grid points.}
   \label{fig:resolution}
\end{center}
\end{figure}

A grid convergence study for both rectangular-cavity geometries ($L/D=$ 2 and 6) is performed. Presented in figure \ref{fig:resolution} are two grid convergence comparisons performed with $Re_\theta=144$ and $M_\infty = 0.1$ for $L/D=2$, and $Re_\theta = 46$ and $M_\infty=0.3$ for $L/D=6$. The baseline computation is conducted on a structured mesh with approximately half a million grid points. A finer mesh with one million grid points is also performed for comparison. The $v$-velocity history at the midpoint location ($x =L/2$, $y = 0$) over the cavity is shown in figure \ref{fig:resolution}. The baseline mesh of half a million grid points is shown to be sufficient to achieve numerical convergence. Moreover, the frequencies of the oscillations in the flow for $L/D=2$, $Re_\theta=56.8$ and $M_\infty=0.6$ are compared to the prediction from Rossiter's semi-empirical formula and the work by \cite{Bres:2007} in table \ref{tableSt_validation}, exhibiting good agreement. 

\begin{table}
\begin{center}
\begin{tabular}
      {l c c}  
                      					& $St_{L1}$	& $St_{L2}$  \\ \hline
            \cite{Rossiter:ARCRM64} 	& 0.321  	& 0.750 \\  
            \cite{Bres:2007} 			& 0.404  	& 0.698   \\ 
            Present					& 0.412	& 0.715 \\
\end{tabular}
\end{center}
\caption{Comparison of Rossiter modes I and II frequencies for open-cavity flow with $L/D=2$, $Re_\theta = 56.8$ and $M_\infty=0.6$.}
\label{tableSt_validation}
\end{table} 

\begin{figure}
\begin{center}
{\scriptsize
  \begin{tabular}
      {ccc}
      \includegraphics[width=1.4in]{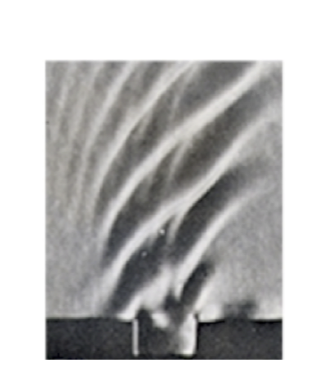}  &  \includegraphics[width=1.4in]{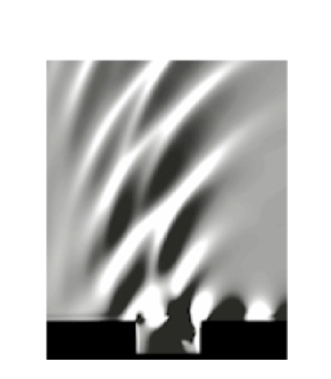} &
      \includegraphics[width=1.4in]{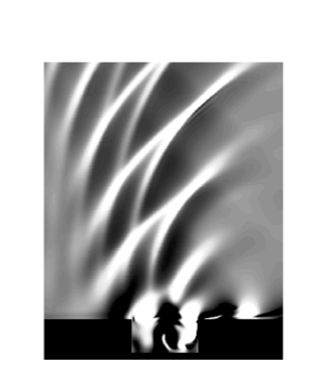} \\
       Krishnamurty, $M_\infty = 0.82$   &Rowley et al., $M_\infty = 0.8$ &Present, $M_\infty = 0.8$  \\   
                        & $Re_\theta = 56.8$  & $Re_\theta = 67$  
  \end{tabular}
  \begin{tabular}
      {ccc}
            \includegraphics[width=1.35in]{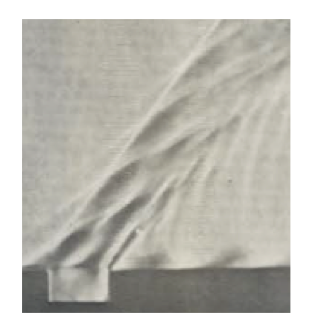}  & &
      \includegraphics[width=1.35in]{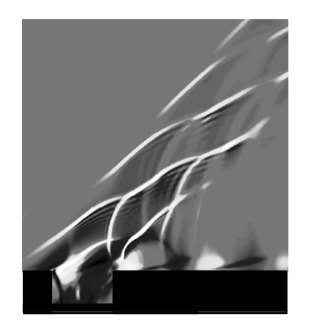} \\ 
    Krishnamurty, $M_\infty = 1.38$ & & Present, $M_\infty = 1.4$  \\
                 &   & $Re_\theta = 56.8$  
  \end{tabular}
 }
\end{center}
   \caption{Comparison of schlieren images by \cite{Krishnamurty:1956}, DNS by \cite{Rowley:JFM02} (reproduced with permission from Cambridge University Press) and the present work ($\partial \rho/\partial x$) at $M_\infty = 0.8$ and 1.4.}
   \label{fig:krish}
\end{figure}

Furthermore, instantaneous density gradient flow fields at $M_\infty=0.8$ and 1.4 are compared to the experimental schlieren images by \cite{Krishnamurty:1956} and numerical results by \cite{Rowley:JFM02} in figure \ref{fig:krish}. In the experiments conducted by \cite{Krishnamurty:1956}, the cavity width is almost $40$ times the depth ($W/D \approx 40$). Hence, the results from this experimental study can be regarded as approximately two-dimensional. As shown in figure \ref{fig:krish}, the present results exhibit good agreement with the experiments and previous simulation results. For $M_\infty = 0.8$, the acoustic waves generated at the trailing edge of cavity propagate upstream, and for the $M_\infty = 1.4$ case, the acoustic wave structures are aligned in an oblique manner. Further analysis of the open cavity flow features are provided in \S{\ref{sec:results}}. 

\subsection{Biglobal stability analysis setup and validation}

Biglobal stability analysis is performed with respect to the base flow obtained from the 2D nonlinear simulation to uncover the compressible spanwise-periodic global instabilities. The state vector $\boldsymbol q = [\rho,  \rho u, \rho v, \rho w, e]^T$ is decomposed into base state $\boldsymbol {\bar q}(x,y)$ and perturbations $\boldsymbol {q}'(x,y,z,t)$ as
\begin{equation}
\label{bi}
\boldsymbol q(x,y,z,t) = \boldsymbol {\bar q}(x,y) + \boldsymbol {q}'(x,y,z,t).
\end{equation}
The steady solution (base state) $\boldsymbol {\bar q}$ satisfies the Navier--Stokes equations and $\boldsymbol q'$ is small in its magnitude compared to the base state ($|\boldsymbol q'| \ll |\boldsymbol {\bar q}|$). In the present work, the base state $\boldsymbol {\bar q}$, obtained from the 2D DNS, is either a time-invariant stable state or an unstable steady state calculated by the selective frequency damping method \citep{Akervik:PF06}. By substituting equation (\ref{bi}) into the Navier--Stokes equations and the equation of state, the governing equations are linearized by retaining only the linear terms of $\boldsymbol q'$ which yields the following set of linear equations for $\boldsymbol {q}'$
\begin{equation}
\label{lin}
\begin{split}
& \frac{\partial \rho '}{\partial t} +\frac{\partial }{\partial x_j} (\rho'\bar u_j + \bar \rho u_j' ) = 0, \\
& \frac{\partial }{\partial t} (\rho'\bar u_i + \bar \rho u_i' )+\frac{\partial}{\partial x_j} (\bar{\rho}\bar u_i  u_j' +\bar{\rho} u_i'     \bar u_j + \rho' \bar u_i \bar u_j + P'\delta_{ij}) \\
& \qquad = \frac{1}{Re}\frac{\partial}{\partial x_j}\left(\frac{\partial u_i'}{\partial x_j}+ \frac{\partial u_j'}{\partial x_i}- \frac{2}{3}\frac{\partial u_k'}{\partial x_k}\delta_{ij}\right),\\
& \frac{\partial e'}{\partial t}+ \frac{\partial}{\partial x_j}((\bar e + \bar P) u_j' + (e' + P')\bar u_j) \\
& \qquad =\frac{1}{Re}\frac{\partial}{\partial x_j}\left[\bar u_i \left(\frac{\partial u_i'}{\partial x_j}+\frac{\partial u_j'}{\partial x_i}-\frac{2}{3}\frac{\partial u_k'}{\partial x_k}\delta_{ij}\right)  +u_i'\left(\frac{\partial \bar u_i}{\partial x_j}+ \frac{\partial \bar u_j}{\partial x_i}-\frac{2}{3}\frac{\partial \bar u_k }{\partial x_k}\delta_{ij}\right)\right]\\
& \qquad \qquad +\frac{1}{Re}\frac{1}{Pr}\frac{\partial ^2 T'}{\partial x_k\partial x_k}
\end{split}
\end{equation}
along with the linearized equation of state 
\begin{equation}
P'=R(\rho' \bar T+\bar \rho T'),
\end{equation}
where $R$ is the gas constant.

The above linear governing equations permit modal perturbations of the form
 \begin{equation}
 \begin{split}
\boldsymbol q'(x,y,z,t) =\boldsymbol { \hat q} (x,y)e^{i(\beta z - \omega t)}+\text{complex conjugate}.
 \end{split}
 \label{modal}
\end{equation}
Upon substitution of this modal expression into the linearized Navier--Stokes equations (\ref{lin}), we can transform the instability analysis from solving an initial value problem to an eigenvalue problem of
\begin{equation}
\mathcal{A( \boldsymbol {\bar q};\beta)}\boldsymbol{\hat q}=\omega \boldsymbol{\hat q}.
\label{modal2}
\end{equation}
Temporal instability is examined by inserting a real wavenumber $\beta$ with its corresponding wavelength $\lambda/D=2\pi/\beta$. The corresponding eigenmodes consist of eigenvectors $\boldsymbol {\hat q}(x,y)=\boldsymbol {\hat q}_\text{r}(x,y)+i\boldsymbol{\hat q}_\text{i}(x,y)$ and their complex eigenvalues $\omega=\omega_r + i\omega_i$, where $\boldsymbol{\hat q}_\text{r}$ and $\boldsymbol{\hat q}_\text{i}$ represent real and imaginary components of eigenvectors; $\omega_r$ and $\omega_i$ are the modal frequency and growth ($\omega_i>0$) or decay ($\omega_i<0$) rate, respectively. While solving for the eigenmodes, the linear operator $\mathcal{A}\in \mathbb{C}^{5n \times 5n}$ and eigenvector $\boldsymbol{\hat q}\in \mathbb{C}^{5n}$ for $n={O}(10^6)$ grid points, can become extremely large. For this large-scale eigenvalue problem, the ARPACK library \citep{Arpack:96}, with an implicitly restarted Arnoldi method, is used in the current study to solve for the eigenmodes. The present approach is based on a matrix-free computation. Since the linear governing equations for the perturbation variables are explicit in their forms, we use the regular mode (not shift-and-invert transformation) in ARPACK and only provide the matrix vector products $A(\bar q; \beta)\hat q$ repeatedly to the solver, which avoids requesting large memory space to store matrix entries while solving the eigenvalue problem. All the eigenmodes reported in this paper are converged with $||-i\omega \hat q-A\hat q||\le O(10^{-10})$. Along the cavity walls, velocity perturbations and the wall-normal gradient of pressure perturbation are set to zero. According to the governing equations, the boundary condition for density perturbation is not required because the momentum perturbation flux is zero due to zero velocity along the wall. For the inlet, density and velocity perturbations, as well as the pressure gradient are prescribed to be zero. For the outflow and far field boundaries, gradients of density, velocity and pressure are prescribed as zero. Moreover, an adiabatic condition is assumed for all boundaries. The base state is interpolated on a coarse mesh for the eigenvalue problem \citep{Bergamo:AST15}. A grid resolution study was performed to ensure accurate stability results. Additional details on the computational procedures can be found in \cite{Sun:TCFD16}.

\begin{figure}
\begin{center}
  \includegraphics[width=0.9\textwidth]{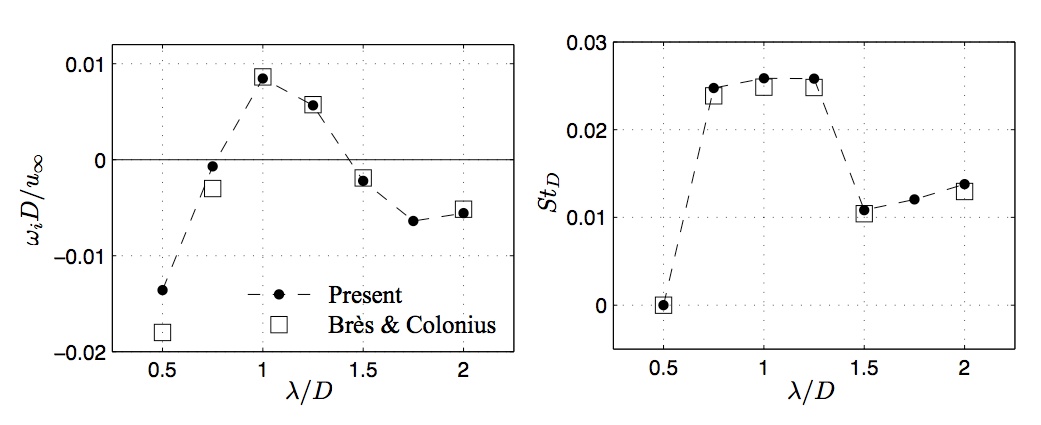}
    \includegraphics[width=1.0\textwidth]{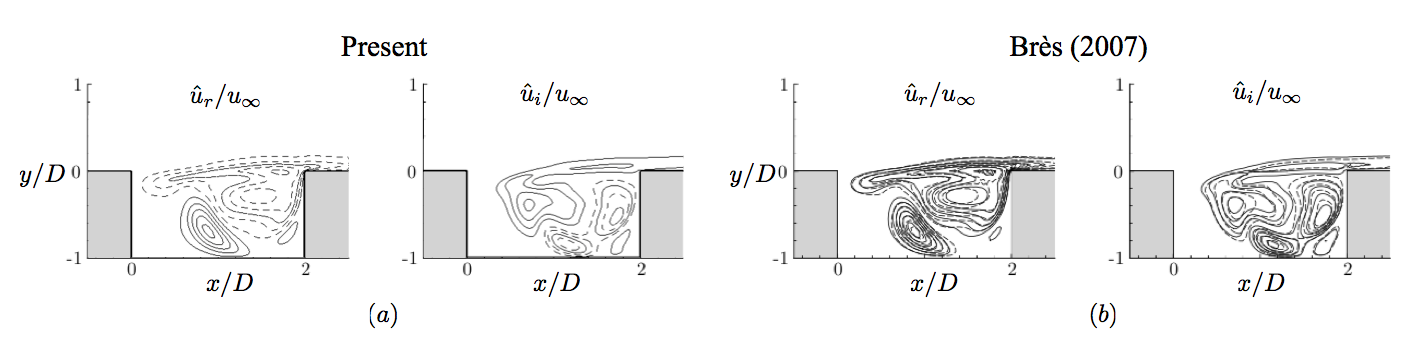}
   \caption{Comparison of dominant 3D modes properties (growth/decay rate $\omega_i D/u_\infty$ and frequency $St_D$) between the present biglobal analysis and linear simulation by \cite{Bres:2007} at $Re_\theta=56.8$. Eigenvectors of the most-unstable mode with $\lambda/D=1.0$ are in good agreement with those in \cite{Bres:2007}, contour levels of real and imaginary components of streamwise velocity $\hat u_r/u_\infty$ and $\hat u_i/u_\infty \in[-0.058,0.058]$ are used. $(a)$ Present: $M_\infty=0.3$; $(b)$ \cite{Bres:2007}: $M_\infty=0.325$ (reproduced with permission from Cambridge University Press).}
   \label{fig:BiG_Bres}
\end{center}
\end{figure}

The validation of global instability analysis is performed on the flow at $M_\infty=0.3$ and $Re_\theta=56.8$ with $L/D=2$. As shown in figure \ref{fig:BiG_Bres}, the dominant 3D modes from this study are compared to the results from \cite{Bres:JFM08}, in which they resolved the dominant 3D modes by solving an initial value problem based on the linearized Navier--Stokes equations. The growth/decay rate, frequencies and eigenvector of the dominant eigenmode obtained from present instability study agree well with those from \cite{Bres:JFM08}.

\section{2D direct numerical simulation and analysis}\label{sec:results}

We focus on examining the effects of $L/D$, Mach number, and Reynolds number on 2D flow oscillation mechanisms in this section. These parameters are known to significantly influence cavity flows \citep{Krishnamurty:1956, Rowley:JFM02, Bres:JFM08, Lawson:PAS11}. In general, flows over open cavities have been broadly classified into stable, shear-layer and wake-mode dominated flows. For a stable case, the flow reaches a time-invariant steady state. With the shear-layer (Rossiter) mode, disturbances convecting downstream are amplified in the shear layer. The wake mode is rarely observed in experiments but is captured in simulations at low Reynolds number when the cavity length is relatively large compared to the momentum boundary layer thickness at the leading edge \citep{Rowley:JFM02, Sun:TCFD16}. It was also observed by \cite{Gharib:JFM87} and \cite{Zhang:EF11} for incompressible axisymmetric cavity flow in their experiments. The primary feature of the wake mode is the shedding of large vortices leading to the interaction between the vortex and the trailing edge, causing violent fluctuations in the cavity. 

Here, we discuss the flow characteristics, instabilities, and behavior of the Rossiter mode by performing 2D DNS. The simulations were conducted over a sufficiently long convective time (with a minimum of 150 convective units $D/u_\infty$) for the flow to reach a steady state. Such flow in this study is categorized as stable (asymptotically stable) if the flow is devoid of any oscillation and otherwise unstable. A number of cases that span the range of Mach numbers from $M_\infty = 0.1$ to 1.4 and Reynolds number ($Re_\theta$) up to 144 with $L/D=2$ and 6 are analyzed in detail. For $L/D=2$, the parameters of $M_\infty$ and $Re_\theta$ are chosen to greatly expand upon the subsonic stability analysis performed by \cite{Bres:JFM08}. We extend their analysis to the transonic Mach number regime and determine unstable steady states of oscillatory cavity flows in preparation for the biglobal stability analysis in \S{\ref{sec:biglobal}}. Flow over a long cavity with $L/D=6$ is also examined in detail as this configuration is representative of long cavities used in aircraft.

\subsection{Flow field characteristics}
\label{sec:foc2}

\begin{figure}
\begin{center}
{\scriptsize
  \begin{tabular}
      {>{\centering}m{0.2in}>{\centering}m{1.5in}>{\centering}m{1.3in}m{1.3in}} \hline 

      $M_\infty$ & $\partial \rho/\partial x$ &$\omega_zD/u_\infty$  & Time-averaged streamlines\\ \hline
      
            \vspace{-0.1in} 0.6 &  \includegraphics[width=1.5in]{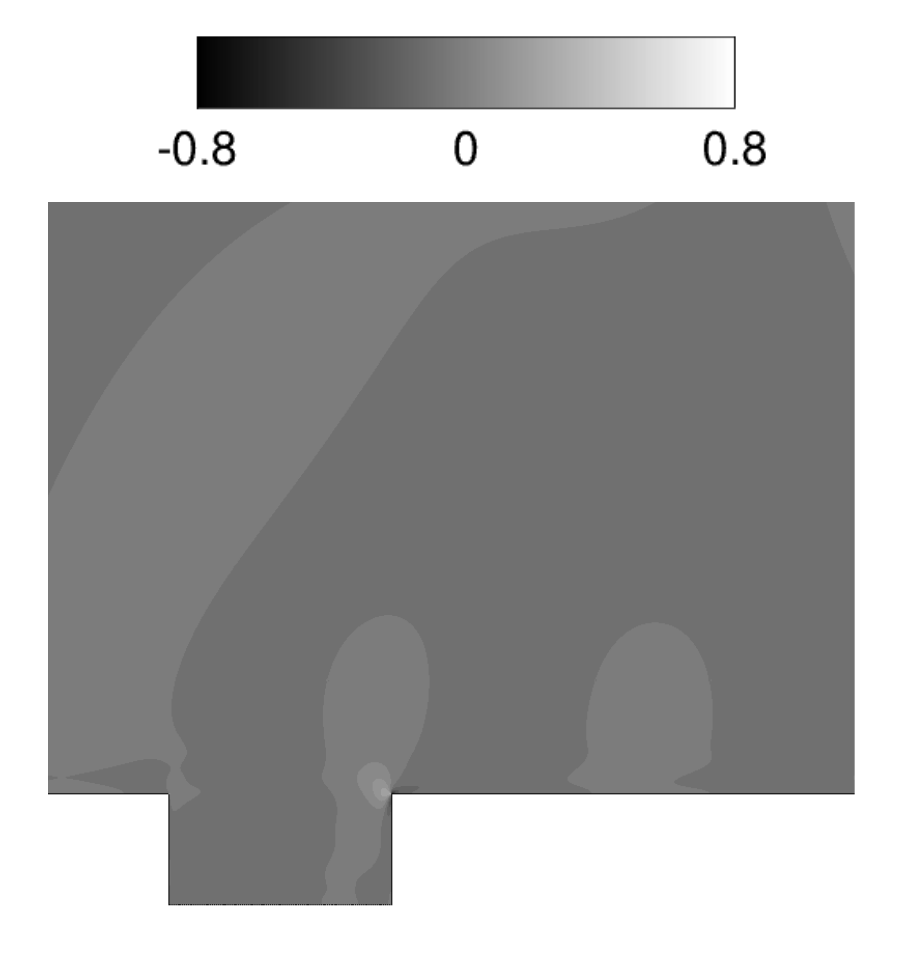}  & \vspace{-0.31in}\includegraphics[width=1.2in]{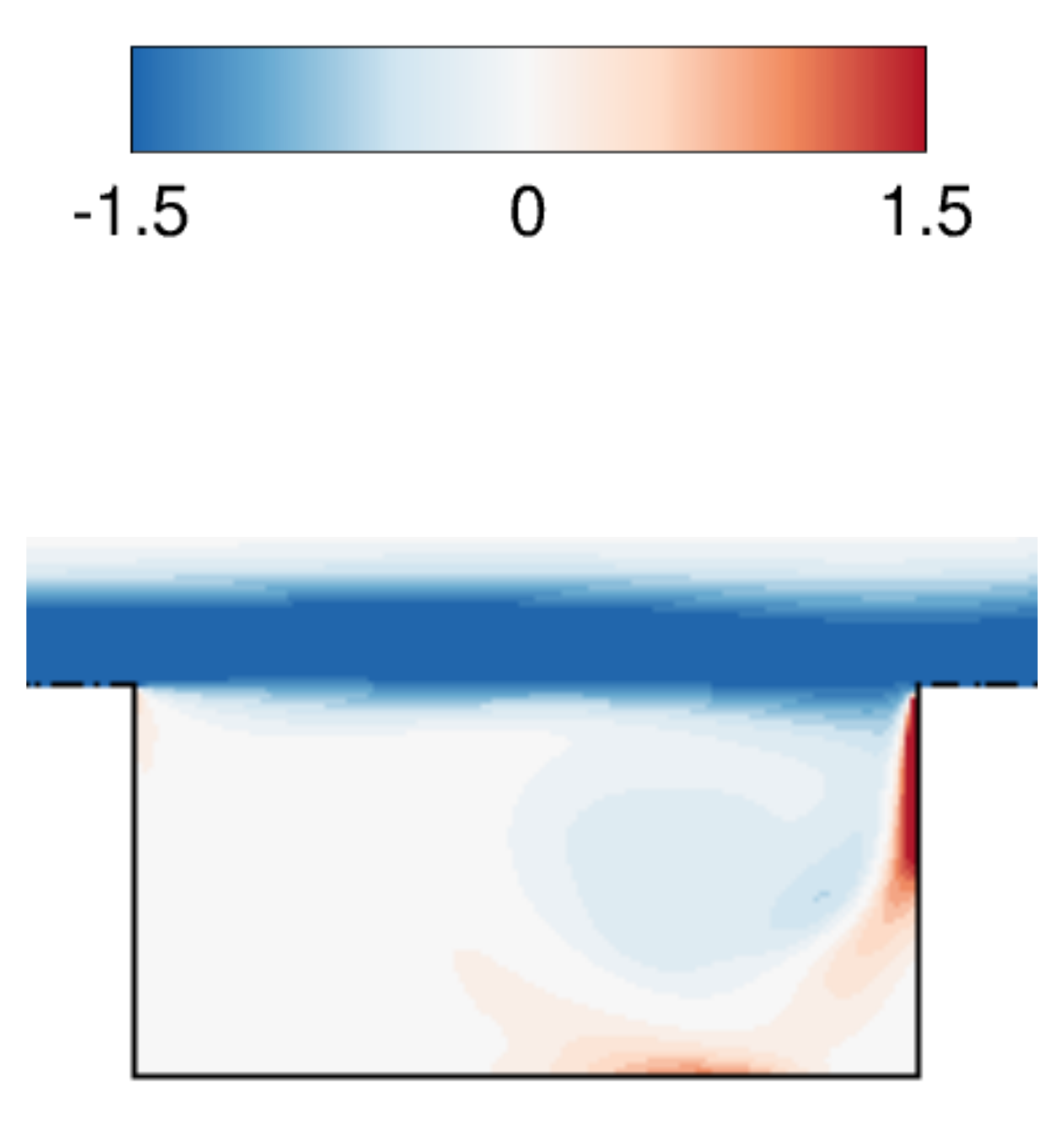} &\vspace{+0.265in}\includegraphics[width=1.25in]{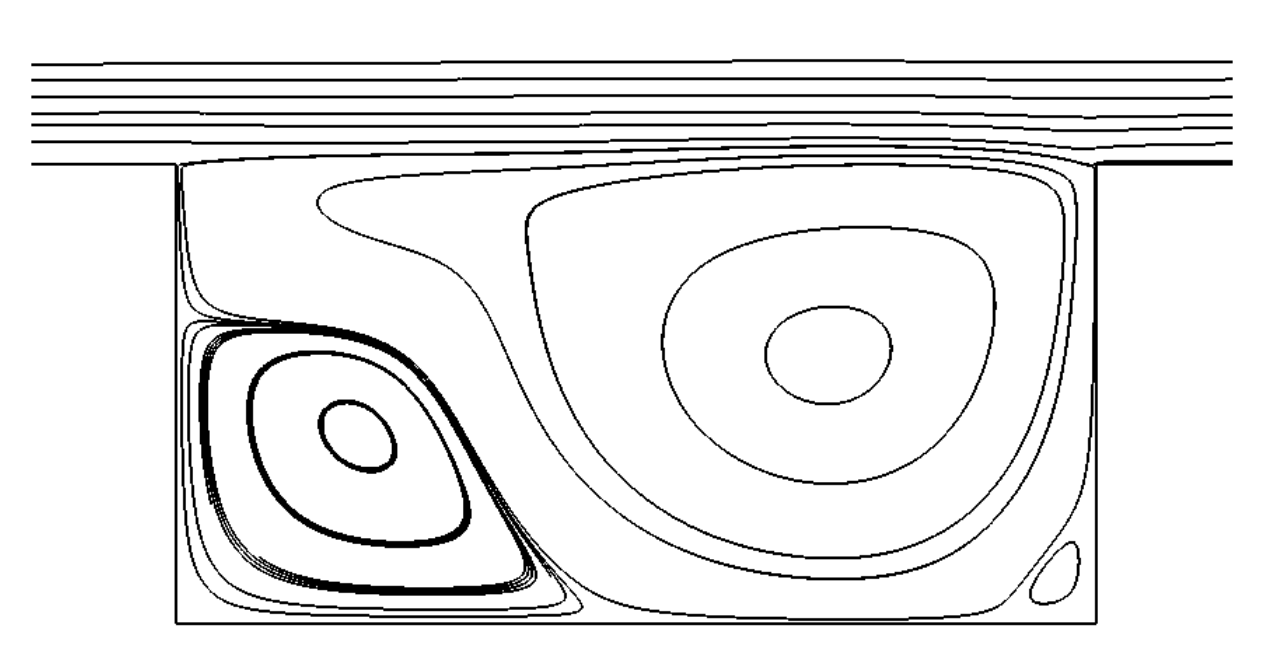}   \\
             \vspace{-0.15in}0.8 &   \includegraphics[width=1.5in]{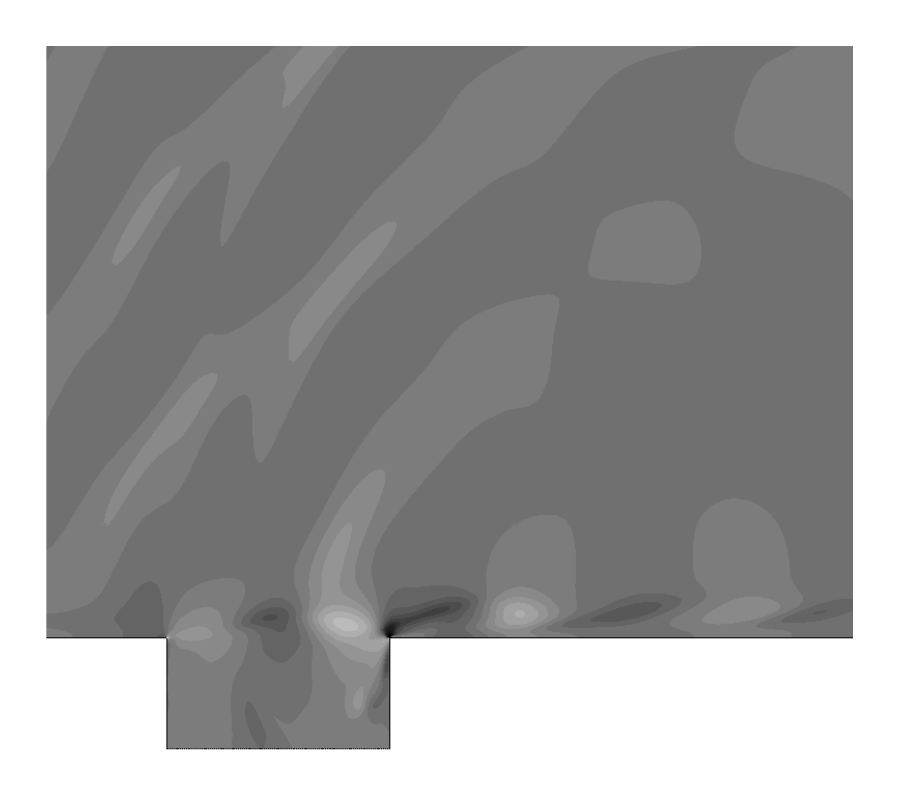}  &\includegraphics[width=1.2in]{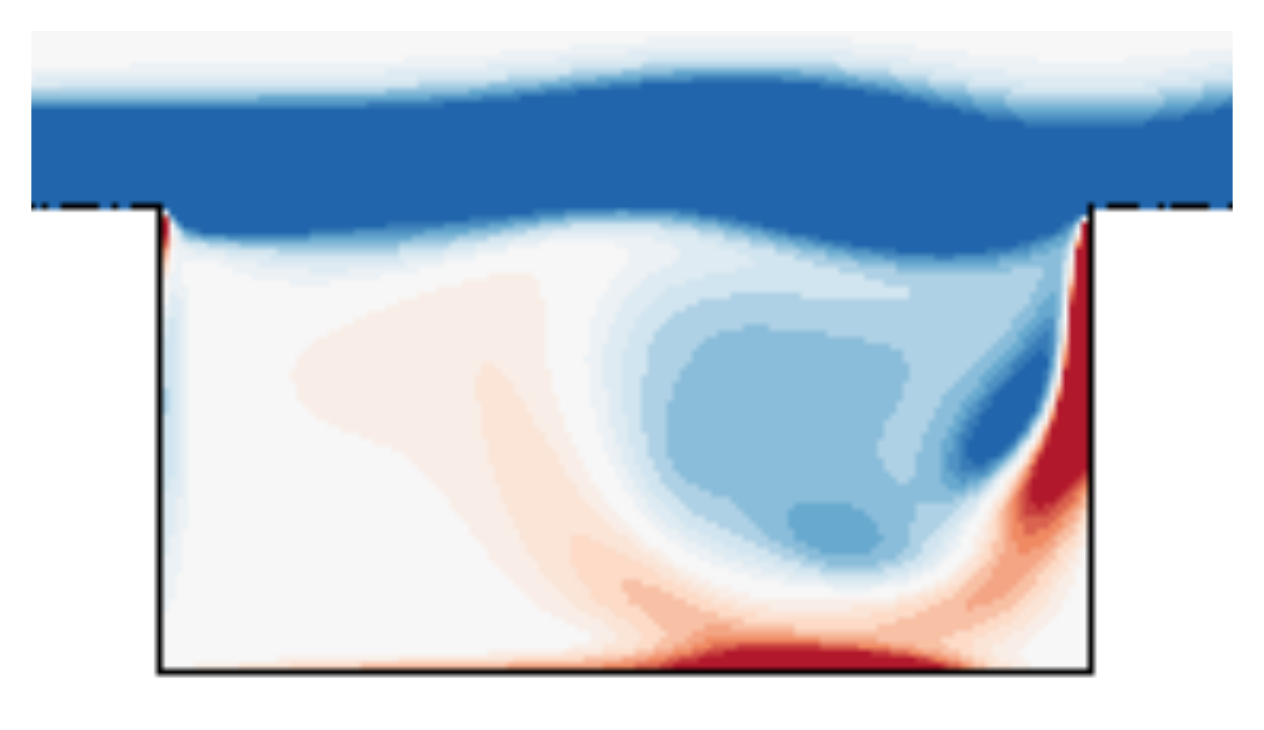}  & \includegraphics[width=1.25in]{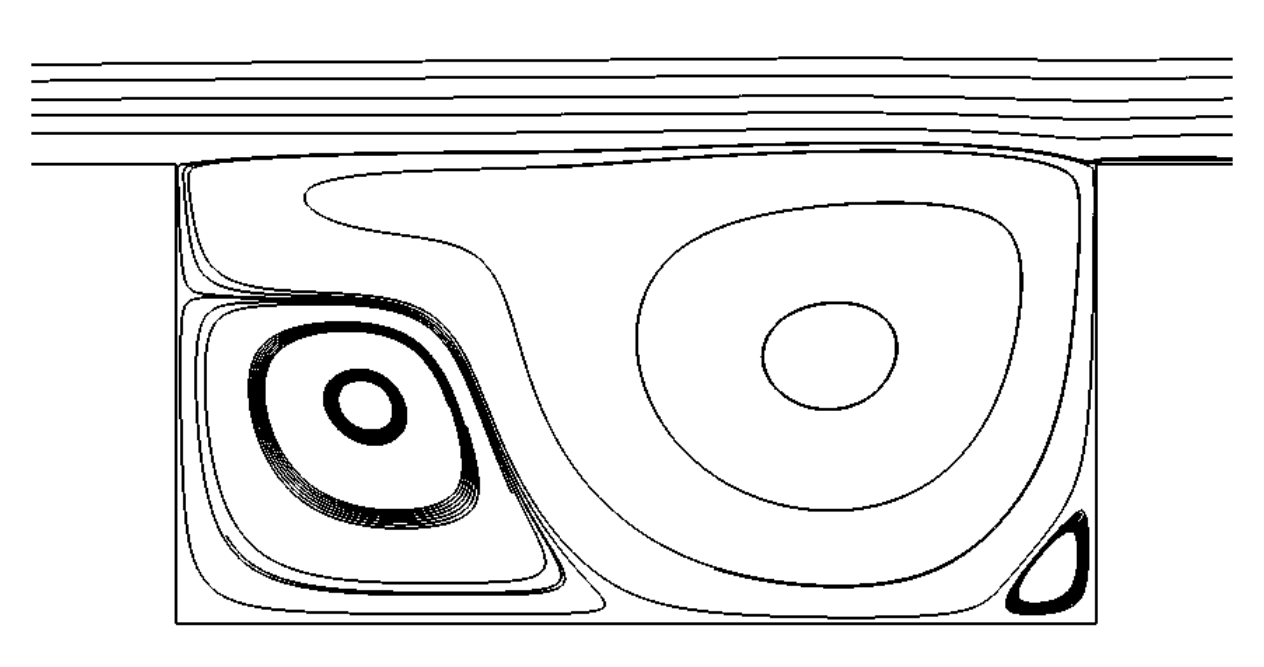}    \\ 
             \vspace{-0.15in}1.0 &   \includegraphics[width=1.5in]{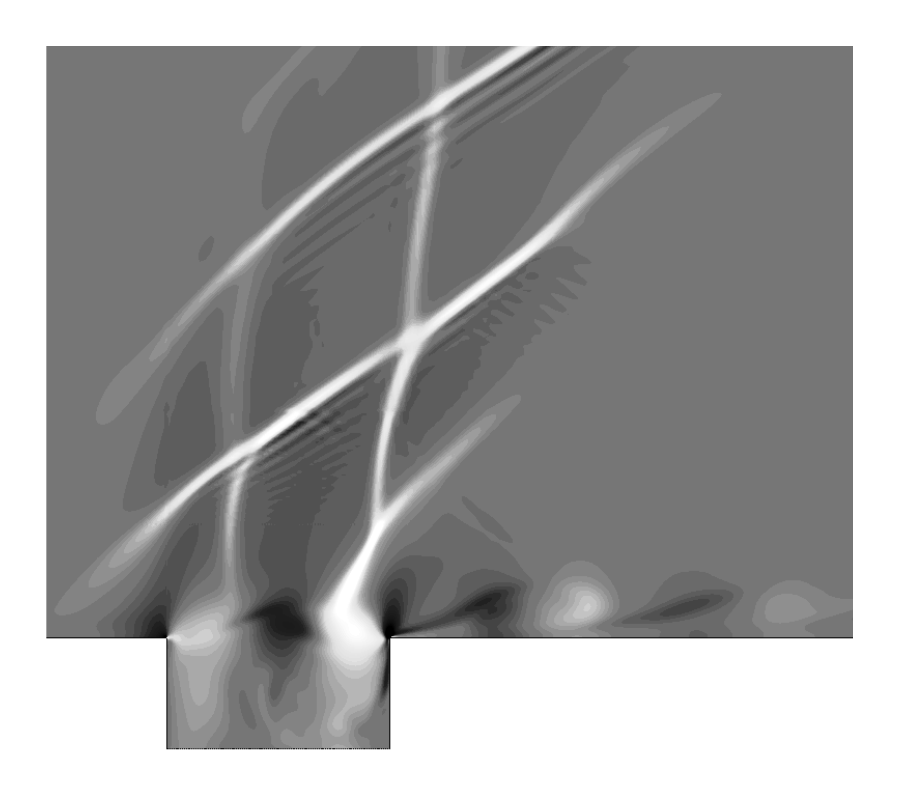}  & \includegraphics[width=1.2in]{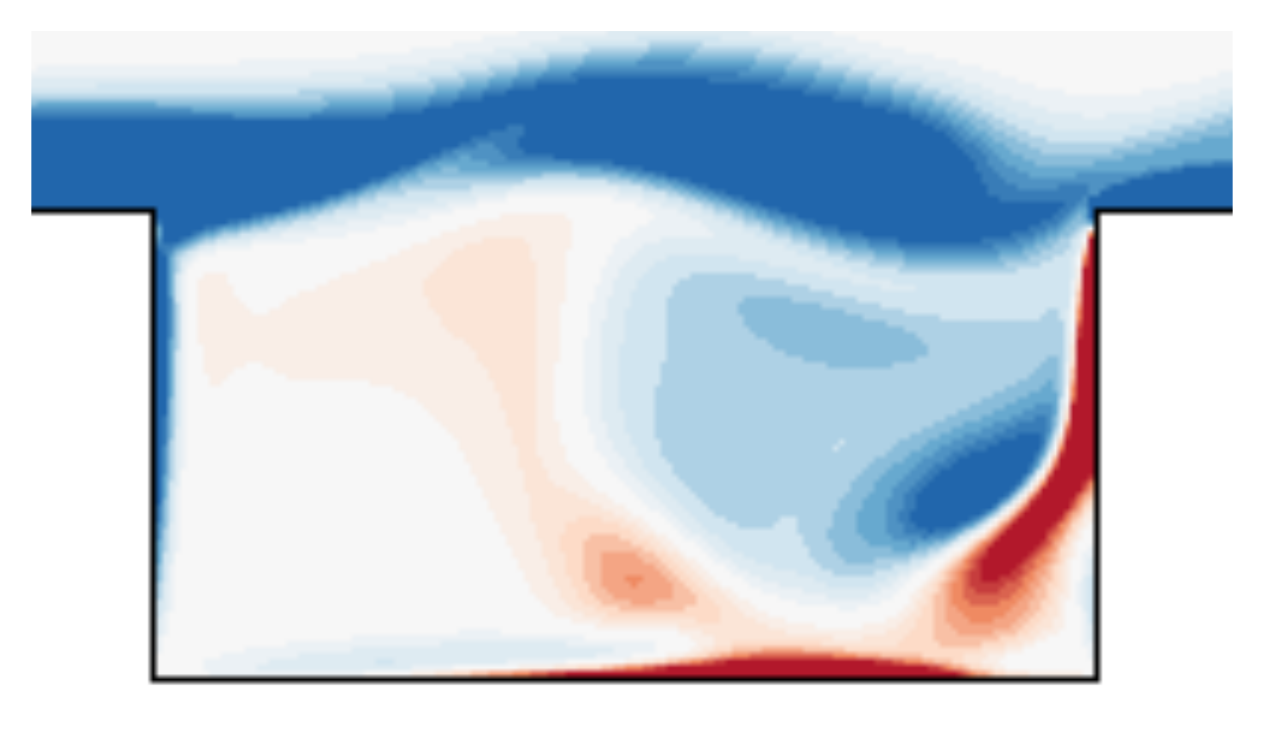} & \includegraphics[width=1.25in]{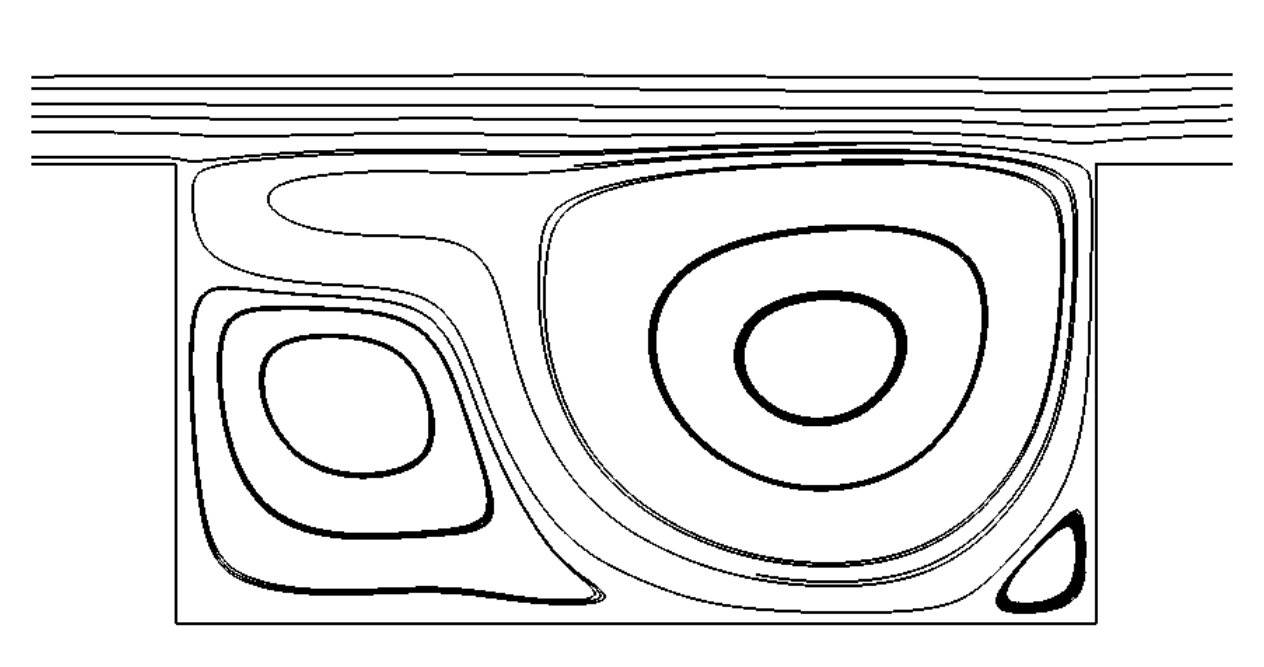}     \\
             \vspace{-0.15in}1.2 & \includegraphics[width=1.5in]{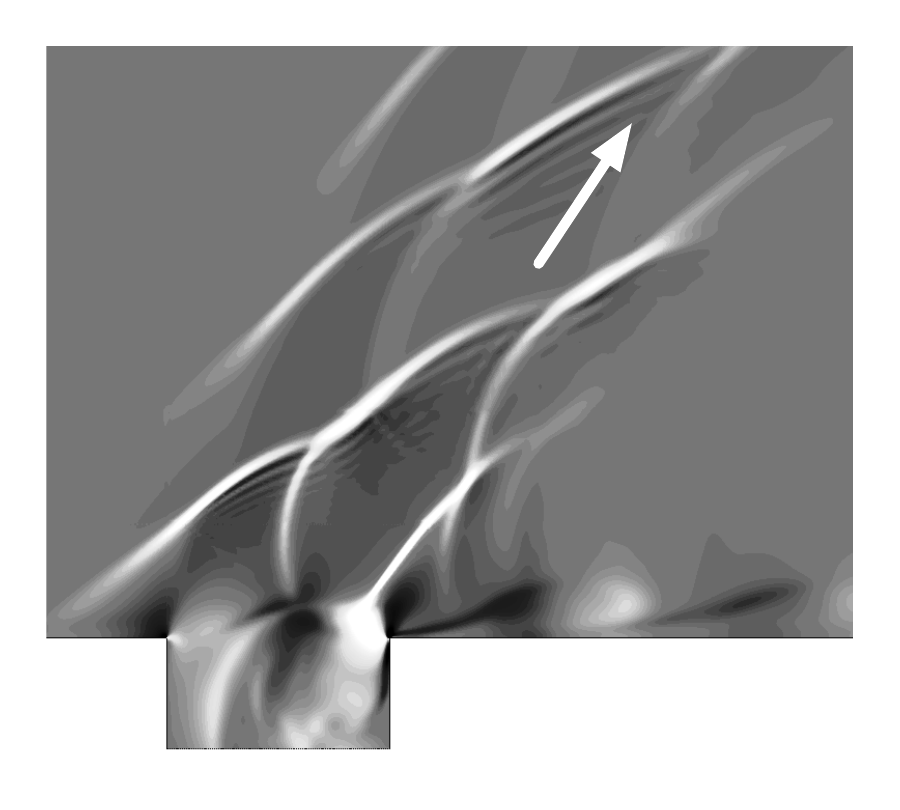}  & \includegraphics[width=1.2in]{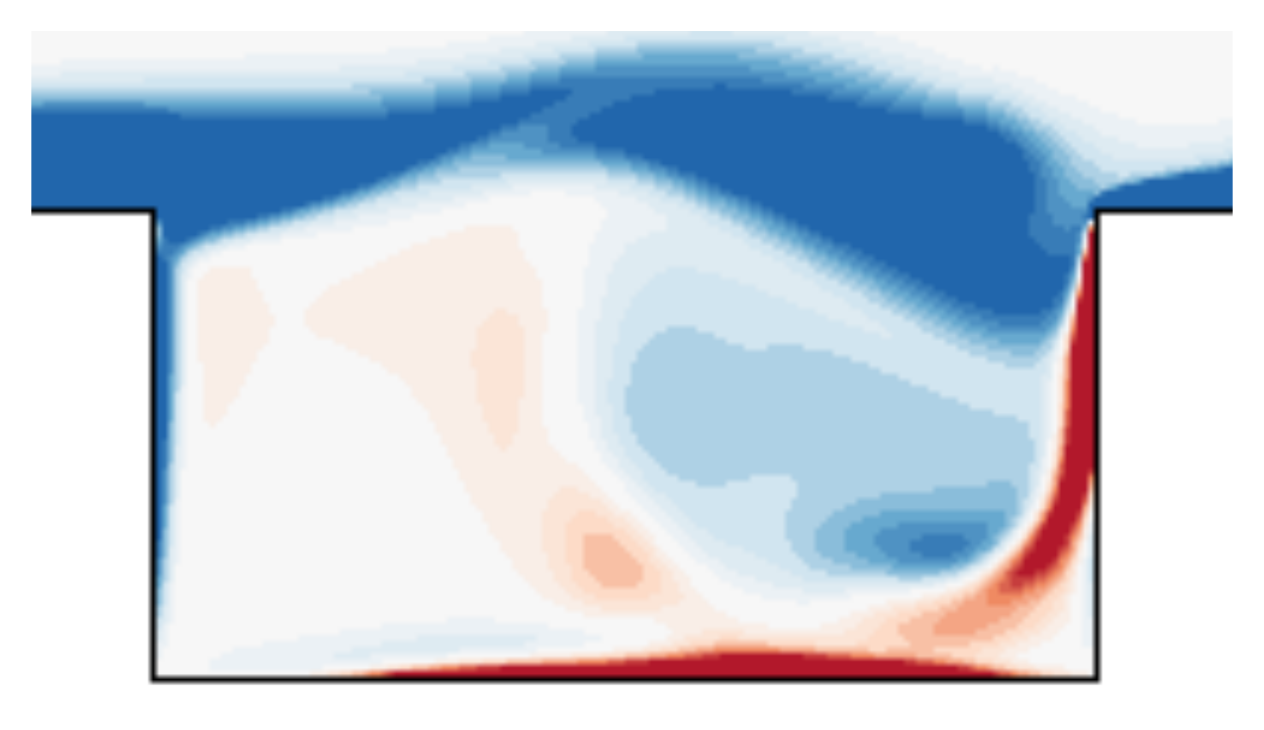} &  \includegraphics[width=1.25in]{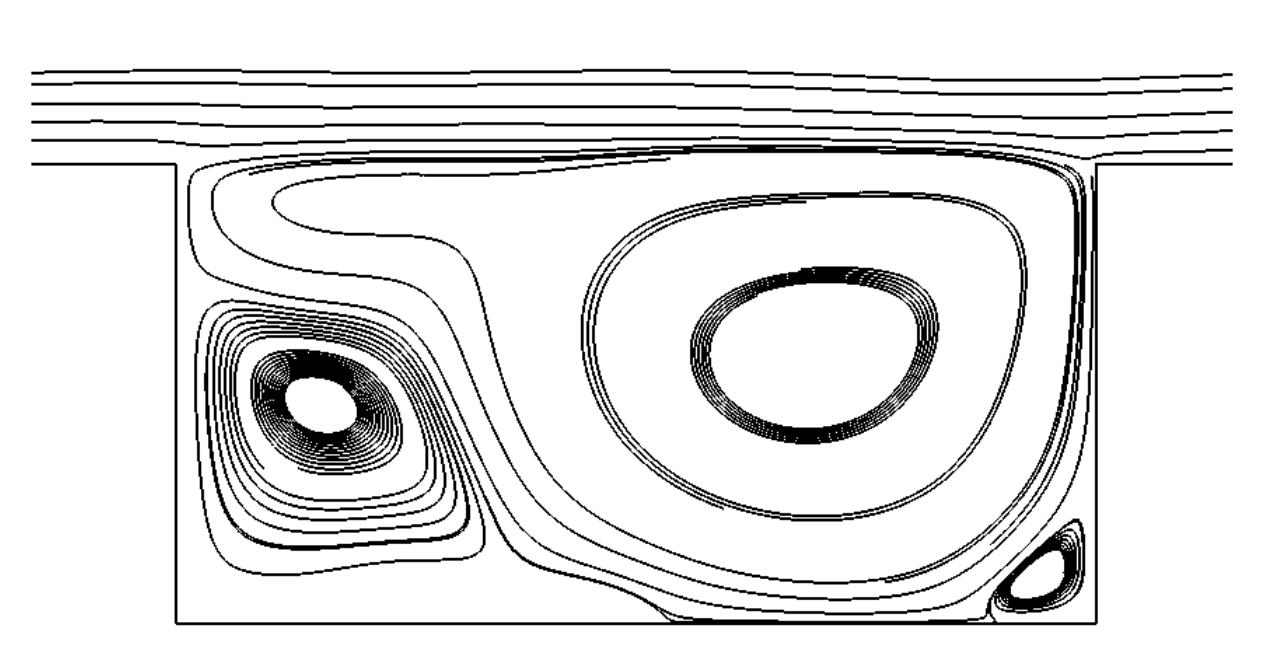}    \\ 
             \vspace{-0.15in}1.4 &  \includegraphics[width=1.5in]{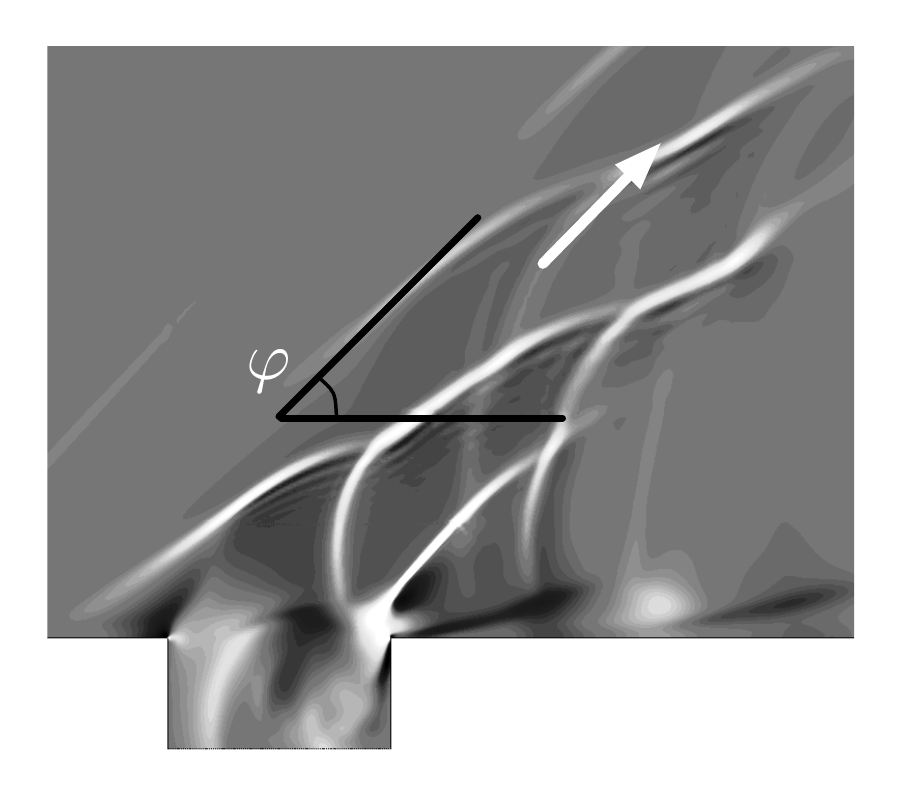}  & \includegraphics[width=1.2in]{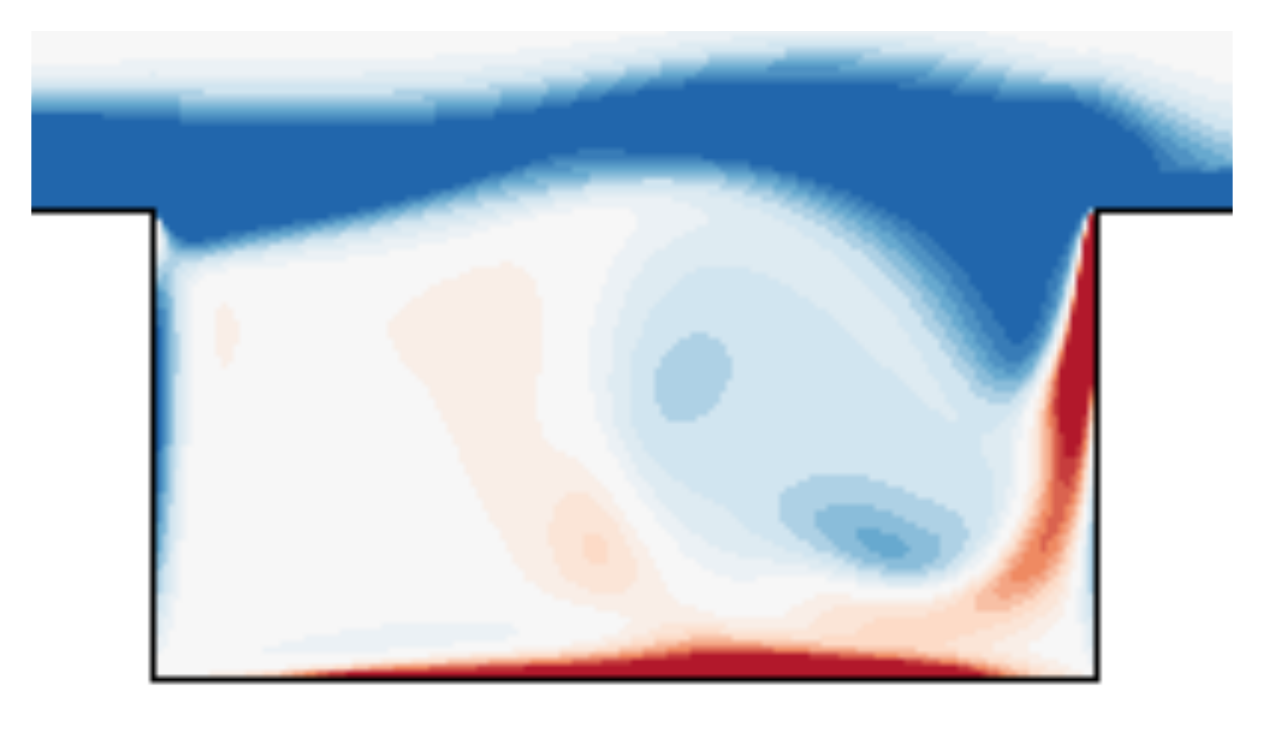} &  \includegraphics[width=1.25in]{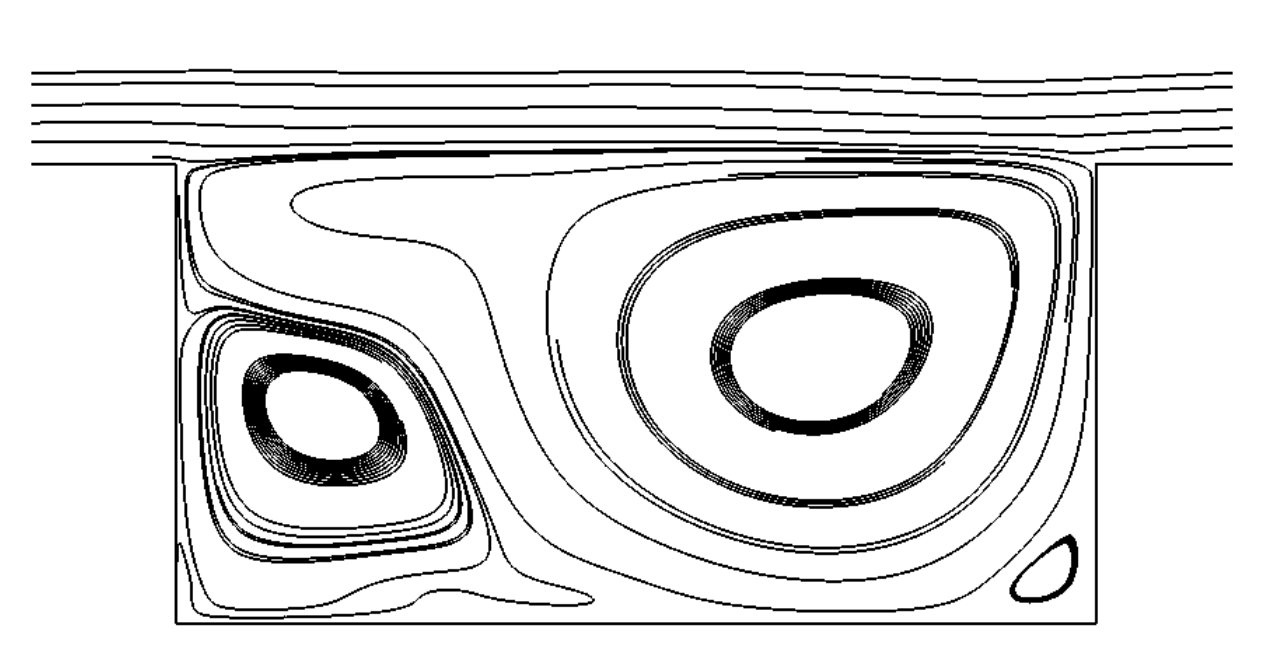}      \\   \hline
  \end{tabular}
 }
\end{center}
\caption{Instantaneous numerical schlieren $\partial \rho/\partial x \in [-0.8,0.8]$, instantaneous vorticity $\omega_zD/u_\infty \in [-1.5,1.5]$ contours and time-averaged streamlines are shown for $M_\infty= 0.6$ -- 1.4, $L/D=2$, and $Re_\theta = 46$. The arrows in numerical schlieren images indicate the propagation directions of compression waves.}
\label{tableLD2}
\end{figure}

\begin{figure}
\begin{center}
{\scriptsize
  \begin{tabular}
      {>{\centering}m{0.2in}>{\centering}m{1.5in}>{\centering}m{1.5in}m{1.5in}} \hline
      $M_\infty$ & $\partial \rho/\partial x$& $\omega_z D/u_\infty$ &~~~Time-averaged streamlines  \\ \hline
             0.3 & \includegraphics[width=1.5in]{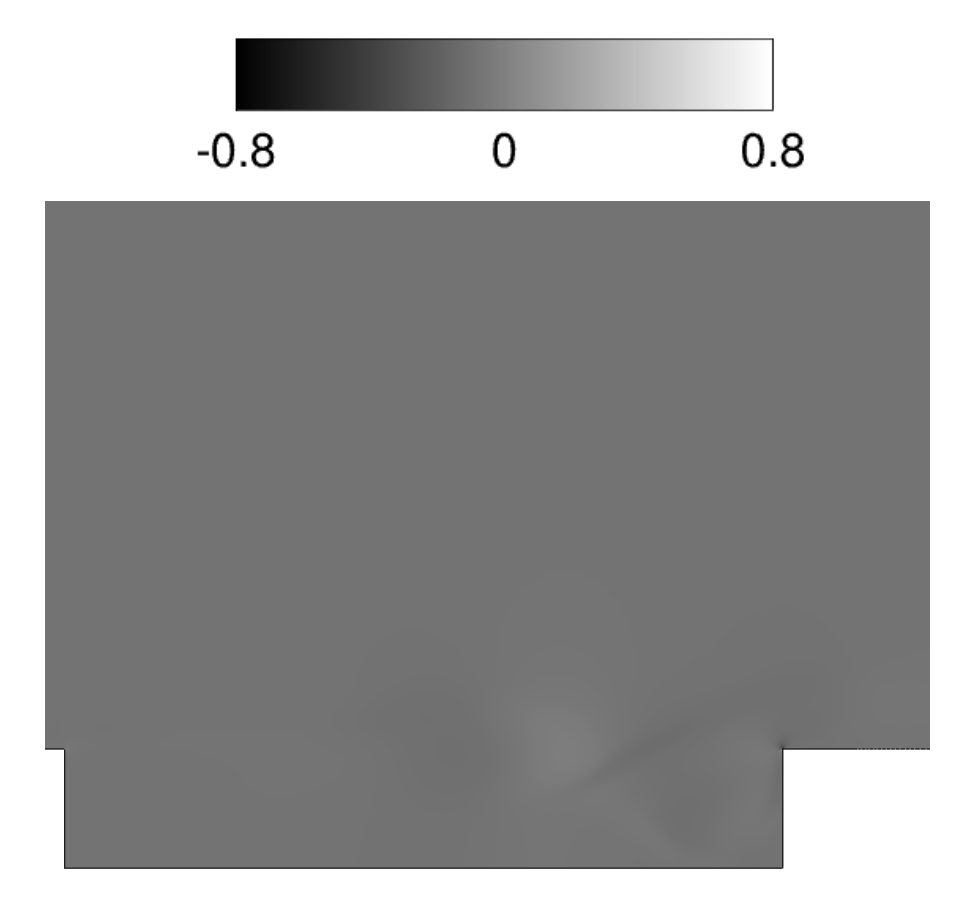} &\vspace{-0.38in}\includegraphics[width=1.5in]{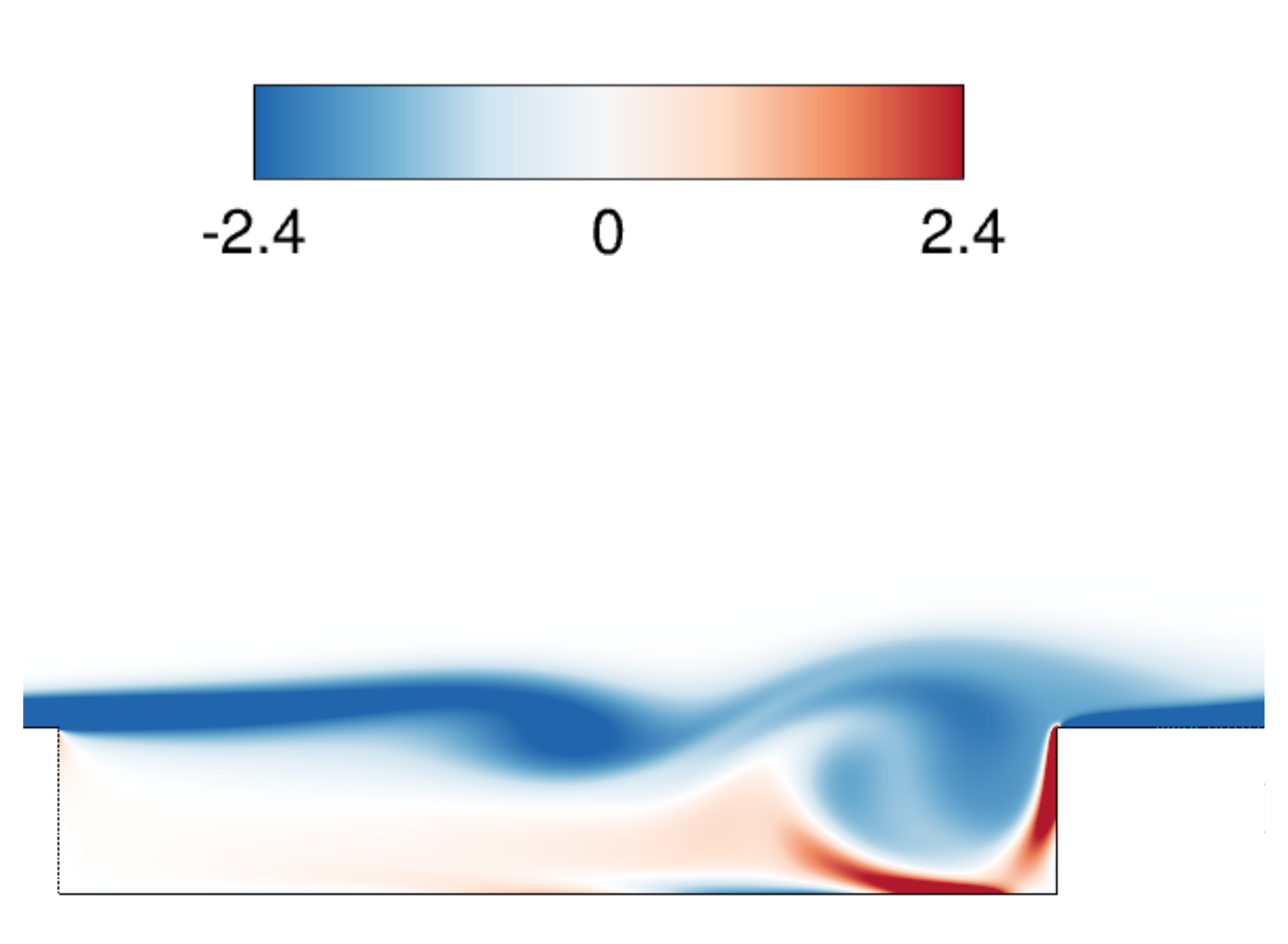}&   \includegraphics[width=1.5in]{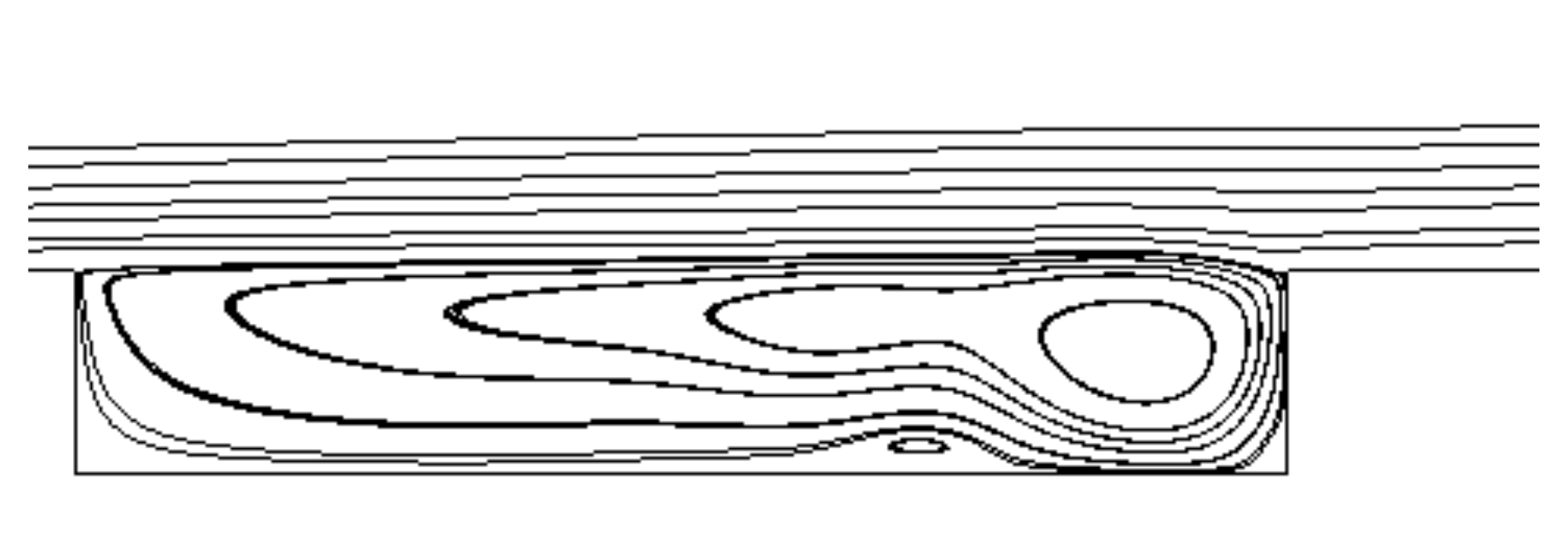}    \\ 
            0.6 &  \includegraphics[width=1.5in]{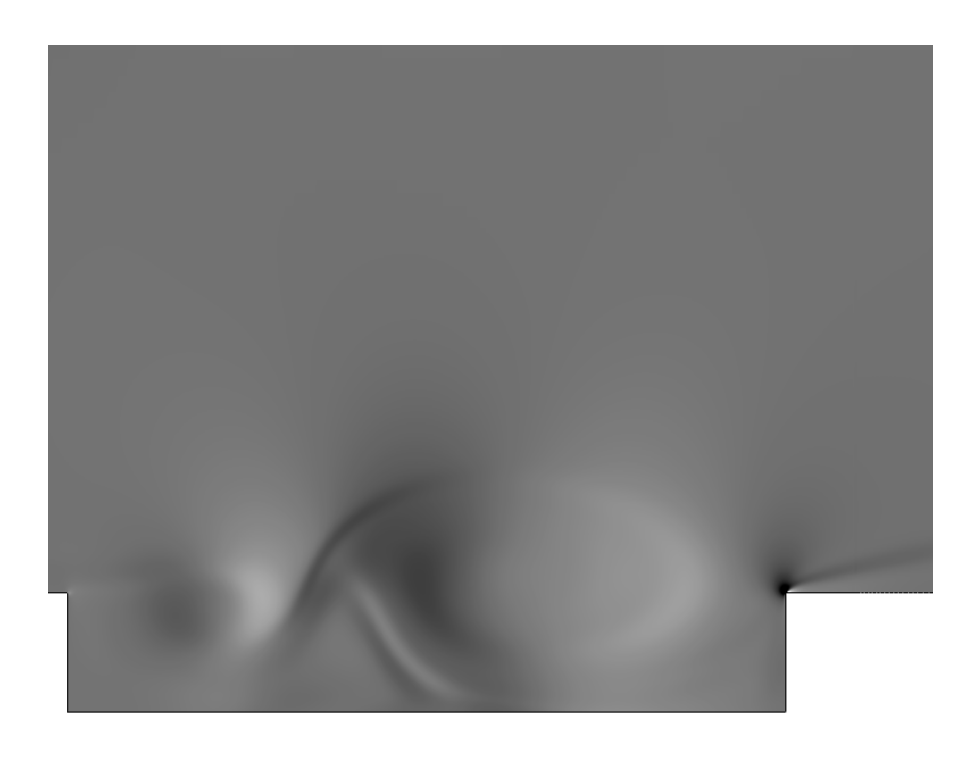} &\includegraphics[width=1.5in]{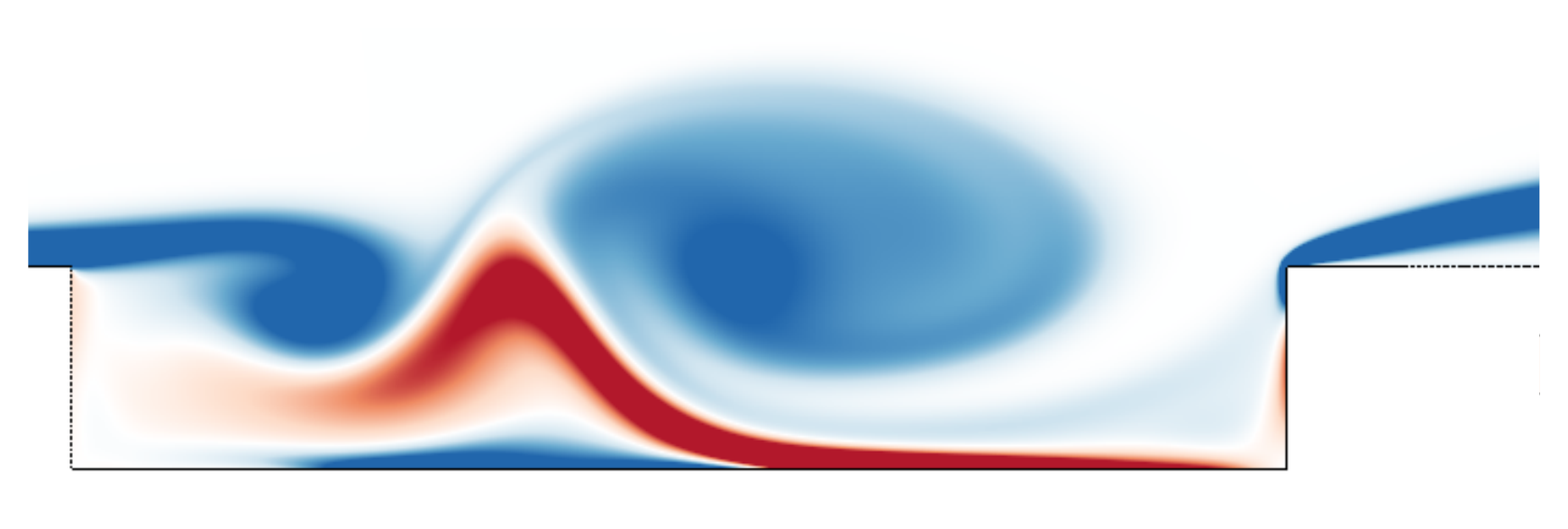}  &   \includegraphics[width=1.5in]{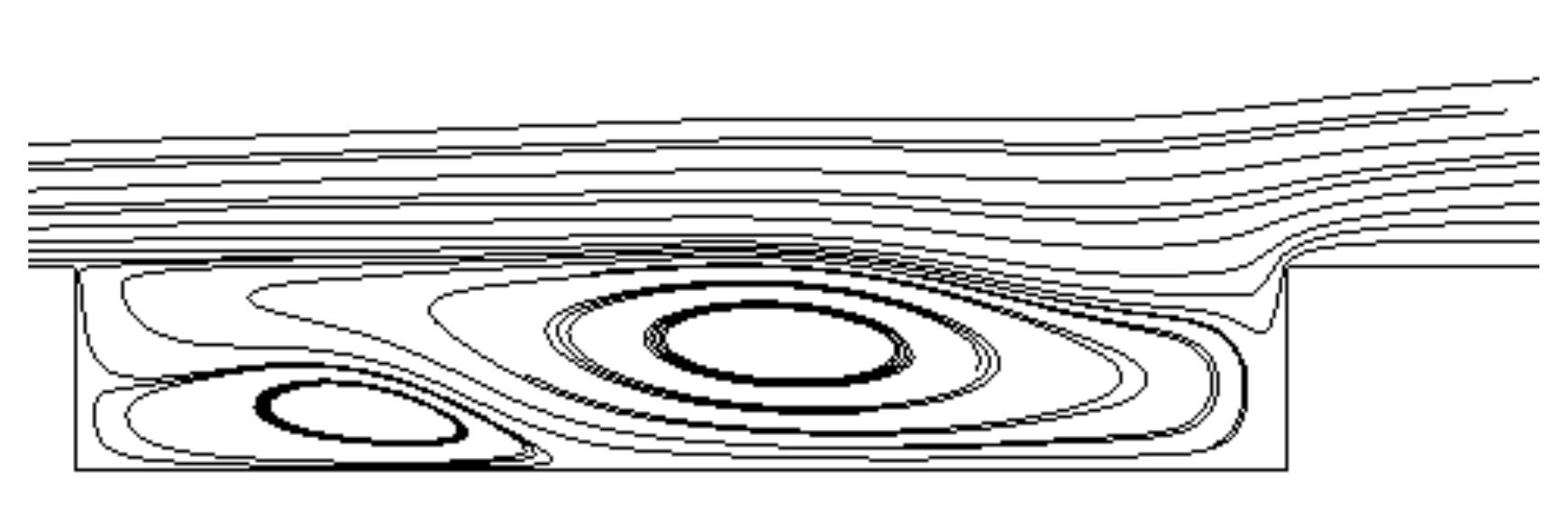}   \\ 
            0.9 &  \includegraphics[width=1.5in]{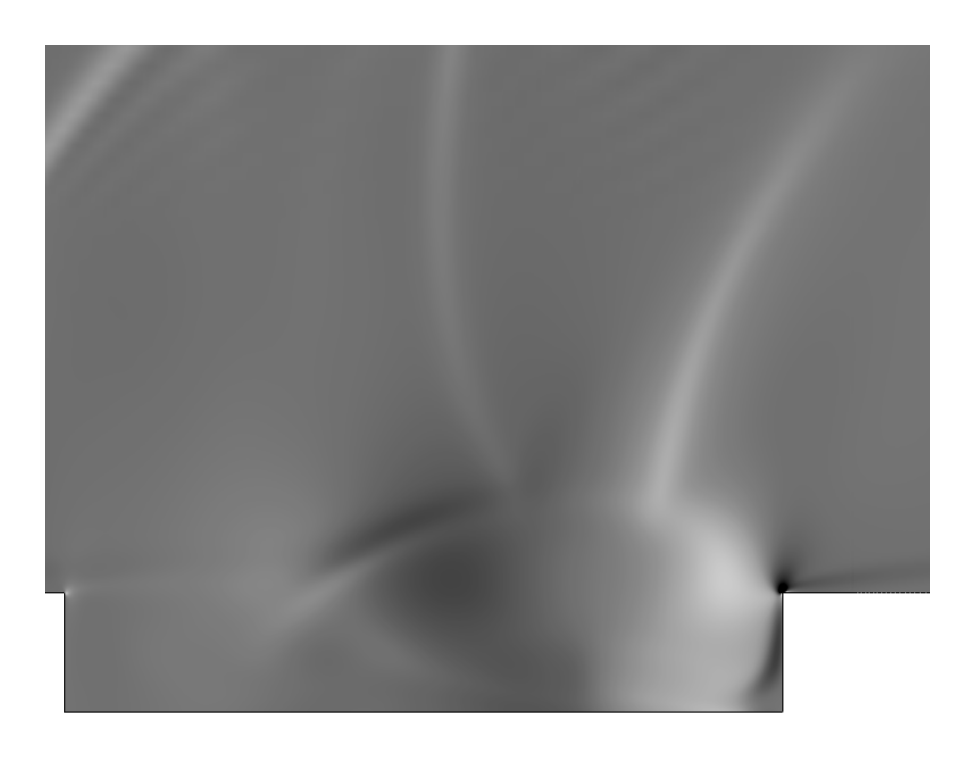} &\includegraphics[width=1.5in]{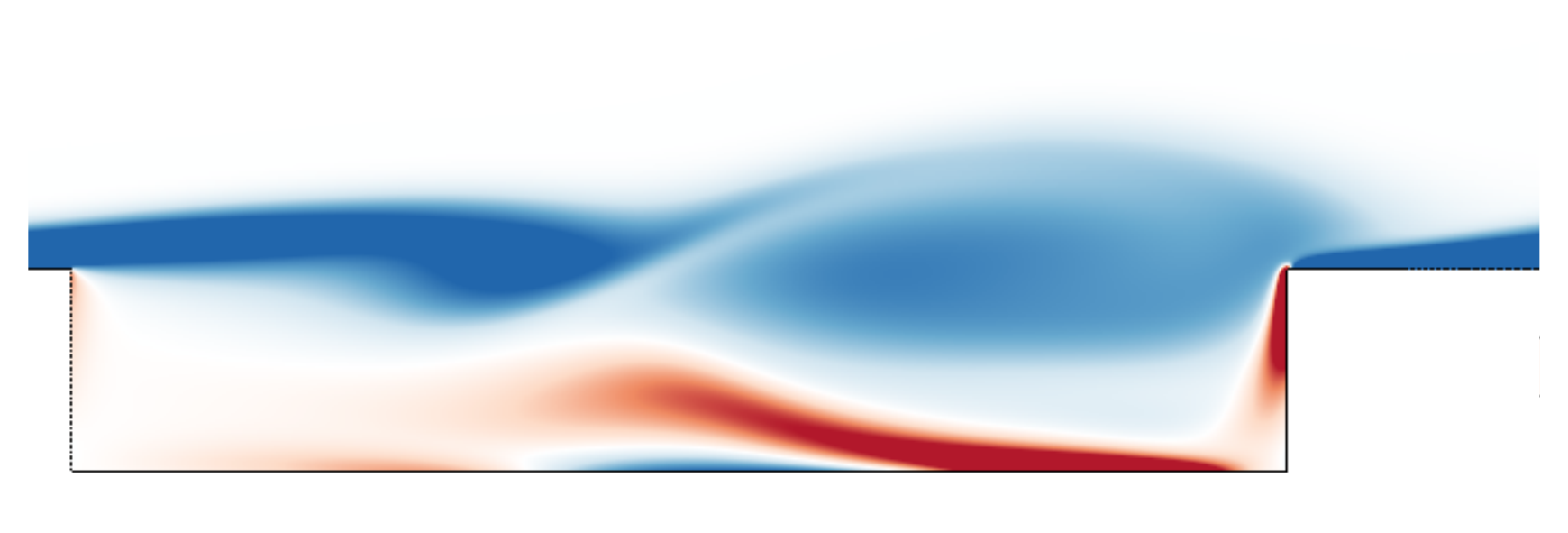}  &   \includegraphics[width=1.5in]{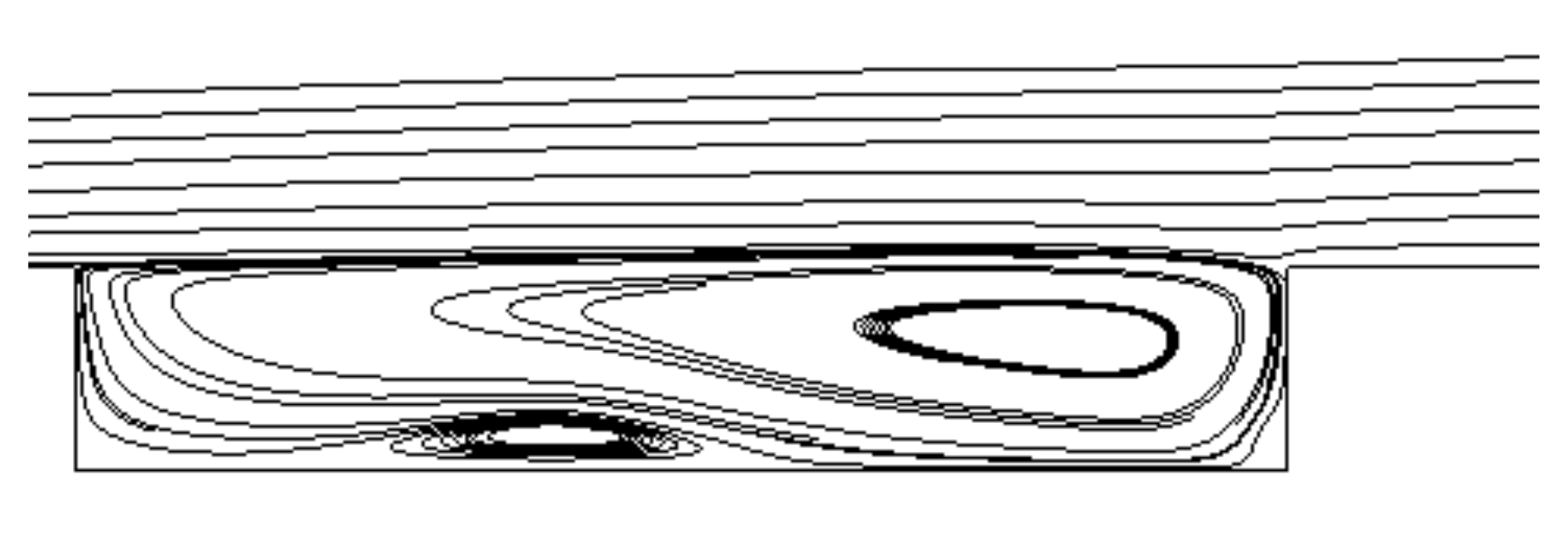}  \\       
            1.2 &  \includegraphics[width=1.5in]{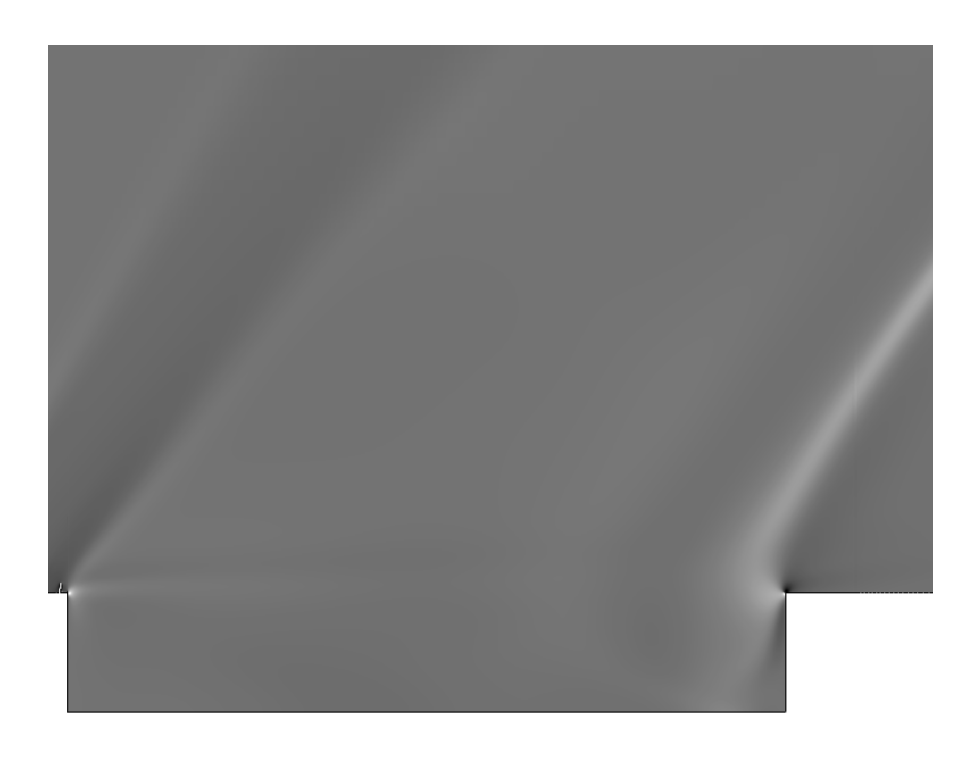} &\includegraphics[width=1.5in]{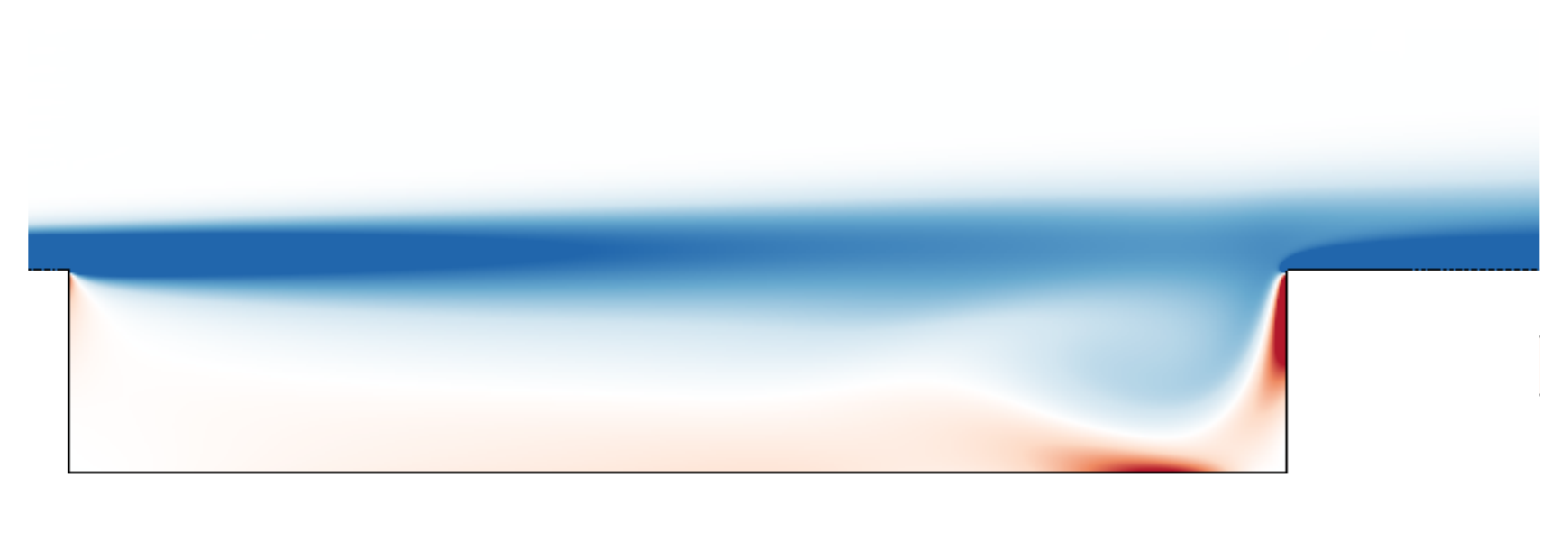}  &   \includegraphics[width=1.5in]{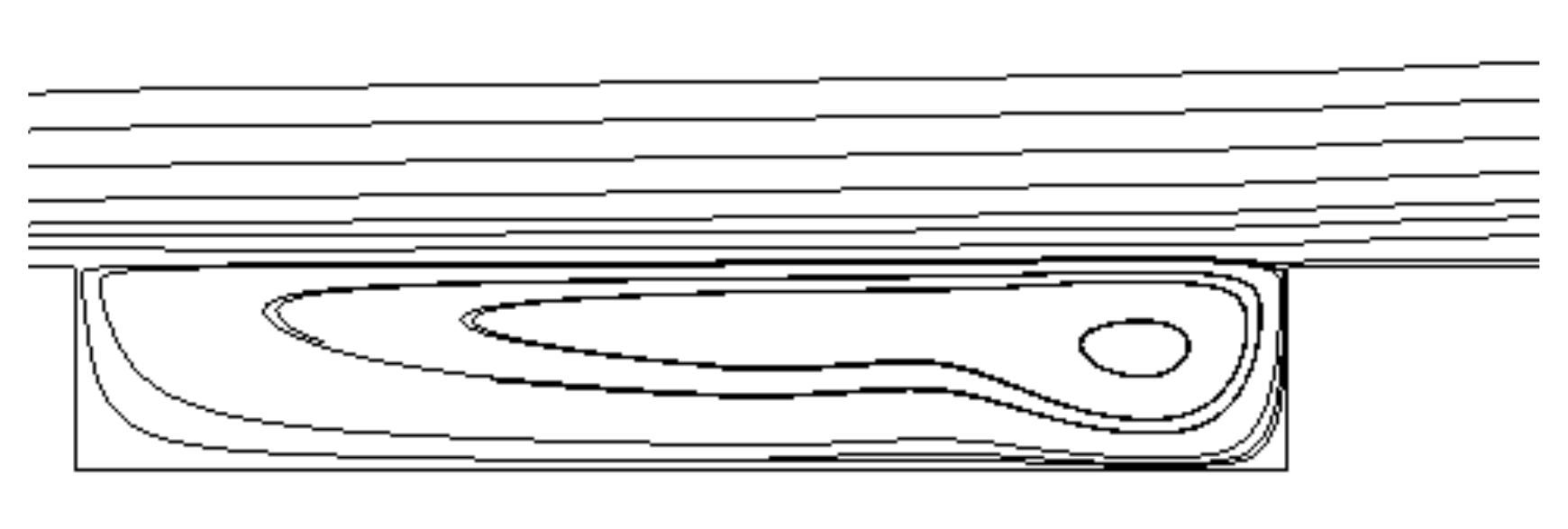}   \\ \hline
  \end{tabular}
 }
\end{center}
\caption{Instantaneous numerical schlieren $\partial \rho/\partial x \in [-0.8,0.8]$, instantaneous vorticity $\omega_zD/u_\infty\in[-2.4,2.4]$ contours and time-averaged streamlines are shown for $M_\infty= 0.3$ -- 1.2, $L/D=6$, and $Re_\theta = 19$.}
\label{tableLD6}
\end{figure}

To examine how compressibility affects flow features, the instantaneous density gradient field, the instantaneous vorticity field, and the time-averaged streamlines are shown in figures \ref{tableLD2} and \ref{tableLD6} for cavities with $L/D=2$ and 6, respectively.

For a cavity with $L/D=2$, the instantaneous density gradient fields for Mach numbers from 0.6 to 1.4 are presented in figure \ref{tableLD2} at $Re_\theta = 46$. At $M_\infty=0.6$, the compression waves in the flow field are not prominent compared to those at higher Mach numbers. When $M_\infty$ increases to 0.8, the acoustic radiation emitted from the trailing edge becomes noticeable and its wavelength becomes smaller, which has also been discussed by \cite{Rowley:JFM02}. For $M_\infty \ge 1.0$, the acoustic waves are more prominent and its structure over the cavity becomes directional. This phenomenon is observed in experiments by \cite{Krishnamurty:1956} as well. Part of the compression waves are generated at the rear corner of cavity. At the trailing edge, for the oncoming flow with a large deflection angle, a shock is formed due to the impingement of the shear layer on the trailing edge. The remaining waves are generated from the periodic expansion and compression of shear-layer oscillations in the supersonic flow. Inside the cavity, animations (not shown) reveal that waves travel back and forth due to reflections. Outside of the cavity, compression waves propagate upstream in the subsonic cases ($M_\infty < 1.0$). On the other hand, the compression waves are swept downstream in the supersonic cases ($M_\infty \ge 1.0$). This leads to the formation of oblique compression waves propagating downstream. A beam structure is formed which consists of the compression waves described above. The angle $\varphi$ of the beam composed of these compression waves above the cavity is measured in the density gradient flow fields. This angle $\varphi$ approximates the corresponding Mach wave angle $\sin^{-1}(1/M_\infty)$ for $M_\infty >1$. For $M_\infty=1.2$ and 1.4, we find $\varphi=60.0^\circ$ and $46.0^{\circ}$ with $\sin^{-1} (1/M_\infty)=56.4^\circ$ and $45.6^\circ$, respectively.

As shown in figure \ref{tableLD6}, for a cavity with $L/D=6$ at $Re_\theta=19$, large density gradients are caused by expansion and compression of the shear layer for the subsonic regime ($M_\infty=0.3$ and 0.6). However, for $M_\infty=0.9$, the acoustic waves become observable with their wavelength corresponding to the length of the cavity. At $M_\infty=1.2$, the flow becomes stable with weak Mach waves emanating from the leading and trailing edges.

For a cavity with $L/D$ = 2, we only observe the shear-layer mode for unstable flows, with a large clockwise-rotating vortex present in the rear part of the cavity as shown in figure \ref{tableLD2}. In all cases considered for the short cavity, the streamlines reveal two major recirculation zones inside the cavity. The lack of variation in the time-averaged streamlines versus Mach number suggests that compressibility does not affect the mean flow inside the cavity significantly. 

However, the flow features in the $L/D=6$ cases vary greatly with Mach number. At $M_\infty = 0.3$, instantaneous vorticity flow fields shown in figure \ref{tableLD6}, reveal a shear-layer mode with a clockwise rotating vortex sitting near the trailing edge. An increase in $M_\infty$ from 0.3 to 0.6 causes the flow oscillate more violently with a transition from the shear-layer mode to the wake mode. The wake mode (at $M_\infty=0.6$) dominates the flow with a large-scale vortex rolling up with an opposite sign vortex sheet being engulfed between the vortices. 
Details on the appearance of the wake mode are reported in \cite{Sun:TCFD16}.
For $M_\infty = 0.9$, the opposite sign vorticity patch pulled between the shedding vortices decreases, and the fluctuations over the cavity weaken. In the transonic regime, increasing the Mach number from 0.9 to 1.2 leads to a steady flow. Compressibility apparently affects the flow instabilities by first destabilizing at subsonic speeds and then eventually stabilizing at transonic speeds.

A large difference between flow states versus Mach number is revealed in the time-averaged streamlines. When the free stream Mach number increases from 0.3 to 0.6, the center of the main recirculation region moves towards the leading edge, then moves back towards to the trailing edge for $M_\infty=0.9$. The features of flows inside the longer cavity are somewhat more complex than those of the shorter length. 
 
\begin{figure}
\begin{tabular}{cc}
   \includegraphics[width=0.49\textwidth]{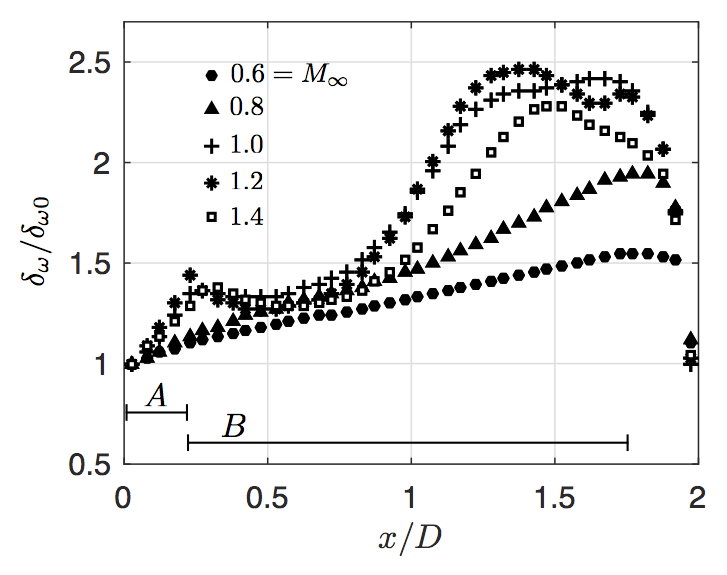}&   \includegraphics[width=0.48\textwidth]{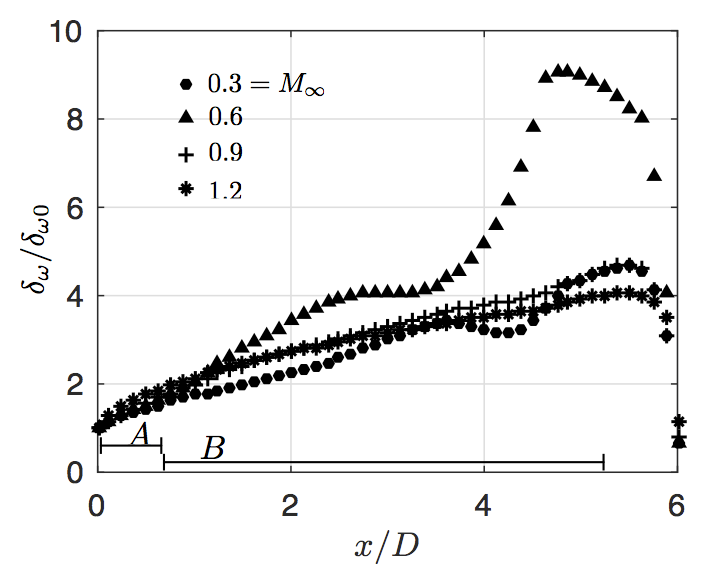}\\
   $(a)$  $L/D=2$, $Re_\theta = 46$ & $(b)$  $L/D=6$, $Re_\theta = 19$
   \end{tabular}
\caption{Vorticity thickness normalized by its value $\delta_{\omega 0}$ at the leading edge is shown as a function of streamwise location. Different regions are indicated by $A$ ($x/L\in[0,0.12]$) and $B$ ($x/L\in[0.12,0.86]$).}
\label{fig:vtLD} 
\end{figure}

The complexity of open-cavity flow results from the feedback process associated with the shear-layer development. Let us further investigate the characteristics of the shear layer over the cavity. Several researchers have noted the similarities between the shear layer spanning the cavity and a free shear layer, such as their linear spreading rate \citep{Sarohia:1975,Cattafesta:AIAA97,Rowley:JFM02}. We evaluate the development of the vorticity thickness
\begin{equation}
\delta_\omega = M_\infty\left[\left(\frac{\partial \bar u}{\partial y}\right)_\text{max}\right]^{-1}.
\end{equation}
In figure \ref{fig:vtLD}, the vorticity thickness is normalized by its value at the leading edge $\delta_{\omega 0}$. We examine the spreading rate dependence on different regions in the shear layer and only report ${\rm d}\delta_\omega/{\rm d}x$ for the cases with linear spreading rate, which are summarized in table \ref{vt_compare}.

\begin{table}
\begin{center}
\begin{tabular}
      {m{0.3in} m{0.4in}cm{0.4in}m{0.3in}m{0.4in}c}
      	\multicolumn{2}{l}{$L/D=2$}	&				&	&\multicolumn{2}{l}{$L/D=6$} 		&			\\ 
	$M_\infty$	& \centering$A$	&\centering$B$		&	&$M_\infty$  	&\centering$A$		&$B$	\\
	0.6 		& \centering0.105	& \centering0.062	&      	& 0.3 		&\centering0.134	&0.097	\\
	0.8 		& \centering0.118	& \centering0.094	&	& 0.6 		&\centering0.180	&-		\\
	1.0 		& \centering0.341	& \centering-		&	& 0.9 		&\centering0.186	& 0.104	\\
	1.2 		& \centering0.446	& \centering-		&	& 1.2 		&\centering0.170	& 0.061	\\
	1.4 		& \centering0.237	& \centering-		&	&			&		&		\\
\end{tabular}
\end{center}
\caption{Spreading rate of vorticity thickness ${\rm d} \delta_\omega/{\rm d} x$ of flows over $L/D=2$ cavity for $M_\infty=0.6$ to 1.4 at $Re_\theta=46$ and  $L/D=6$ cavity for $M_\infty=0.3$ to 1.2 at $Re_\theta=19$ as illustrated in figure \ref{fig:vtLD}. Different regions are indicated by $A$ ($x/L\in[0,0.12]$) and $B$ ($x/L\in[0.12,0.86]$). }
\label{vt_compare}
\end{table}

As shown in figure \ref{fig:vtLD} for cavity with $L/D=2$, for the cases ($M_\infty=0.6$ and 0.8) without intense acoustic compression waves, the overall spreading rate is still nearly linear. \cite{Rowley:JFM02} found ${\rm d}\delta_\omega/{\rm d}x=0.05$ for $D/\theta_0=26.5$ at $Re_\theta=56.8$ and $M_\infty=0.6$, and the value is close to the spreading rate  ${\rm d}\delta_\omega/{\rm d}x=0.062$ (region $B$) for $D/\theta_0=26.4$ at a lower $Re_\theta=46$ and $M_\infty=0.6$ in the present work. In the cases with $M_\infty \ge 1.0$, a double-hump distribution of vorticity thickness is observed in figure \ref{fig:vtLD} (a). As the numerical schlieren reveals in figure \ref{tableLD2}, strong compression waves are formed in supersonic flows. When these waves propagate upstream, they interact with shear layer over the cavity, which distorts the mean profile of the shear layer and results in the double-hump feature. However, for subsonic flows, the compression waves are not strong enough to affect the mean profile, resulting in a linear spreading rate. Near the leading edge of cavity (region $A$), the spreading rate is increased as Mach number increases until $M_\infty=1.2$. Further increasing Mach number to $M_\infty=1.4$ reduces the spreading rate. The stabilizing effect of compressibility in the transonic regime will be further addressed in a later discussion in \S \ref{sec:global2D}. 

For $L/D=6$, except for the wake-mode case at $M_\infty=0.6$, all the other cases (shear-layer modes) reveal approximately linear spreading rates. We note that the growth rates here (region $B$) are in agreement with the values reported by \cite{Gharib:JFM87}, which are almost constant ${\rm d} \delta_\omega/{\rm d} x=0.124$ when $L/\theta_0 > 103$. The flow at $M_\infty=1.2$ is stable, in which there is no roll-up of the shear layer bridging the cavity; thus the spreading rate in region $B$ is reduced compared to the other shear-layer cases. 

\subsection{Stability diagram} 

\begin{figure}
\centering
   \includegraphics[width=0.7\textwidth]{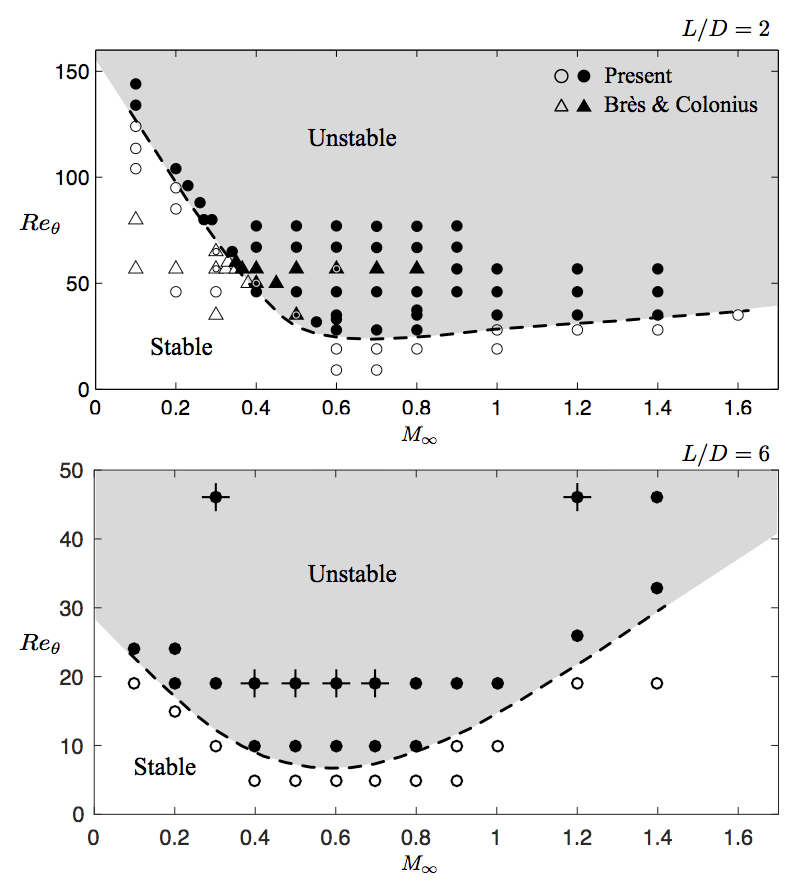}
   \caption{Stability diagram of 2D open-cavity flow for $L/D = 2$ and 6 obtained from DNS. Stable cases: {\scriptsize $\triangle$}, {\large $\circ$}; unstable cases: {\footnotesize $\blacktriangle$}, {\large $\bullet$}. The triangles are from \cite{Bres:JFM08} and the circles are from the present study. The dashed lines represent the approximate neutral stability curves. For the cavity with $L/D=6$, the wake-mode dominated cases are indicated by $+$.}
   \label{fig:map} 
\end{figure}

In the present work, the parameters of $M_\infty$ and $Re_\theta$ are selected to expand upon the subsonic stability analysis performed by \cite{Bres:JFM08}. Through an extensive parametric study, the influence of $M_\infty$ and $Re_\theta$ on the stability of 2D open-cavity flows are revealed, as shown in figure \ref{fig:map}. We find the approximate neutral stability curve via nonlinear flow simulations to separate the stable and unstable zones in terms of $M_\infty$ and $Re_\theta$. For both $L/D=2$ and 6, when the Mach number decreases towards the incompressible limit, the flow becomes more stable for a wider range of Reynolds numbers as shown in figure \ref{fig:map}. In the subsonic regime below $M_\infty=0.6$, an increase in Mach number destabilizes the flow, which was also documented by \cite{Bres:JFM08}. However, in the present study, as the Mach number increases above $M_\infty=0.6$ for both $L/D=2$ and 6, the slope of the neutral stability curve increases, which corresponds in the simulations to a reduction in the amplitude of the observed oscillations. This destabilization ($M_\infty\le0.6$) and subsequent stabilization ($M_\infty>0.6$) effects of Mach number are also reported by \cite{Yamouni:JFM13} for short cavities with $L/D=1$ and 2. Compared to $L/D=2$, the neutral stability curve for $L/D=6$ shifts downwards. The flows over cavities with larger aspect ratio are thus more unstable because of the increased spatial extent for the shear layer to develop and amplify disturbances.

\subsection{Rossiter modes} 
\label{RM}
\begin{figure}
\begin{center}

  \begin{tabular}{cc}
    \vspace{-0.1in} \includegraphics[width=2.6in]{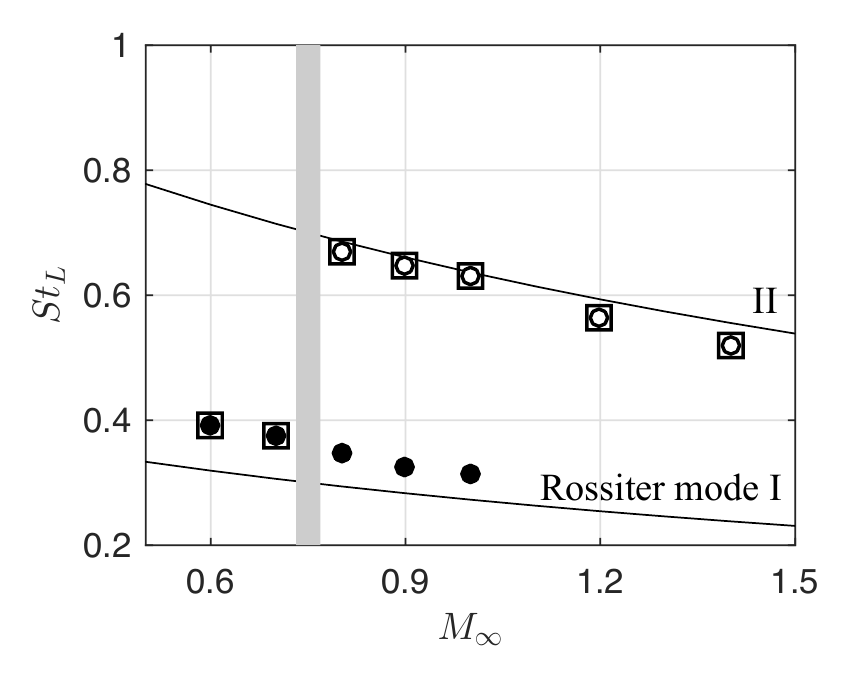}  & \includegraphics[width=2.6in]{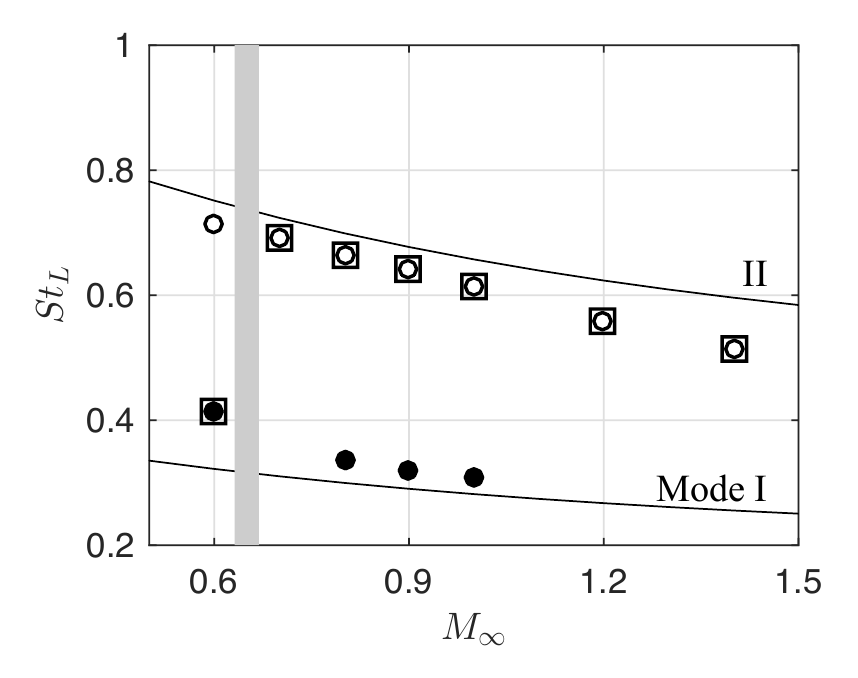} \\
     $(a)$ $Re_\theta=46$&$(b)$ $Re_\theta=56.8$\\
    \vspace{-0.1in}  \includegraphics[width=2.5in]{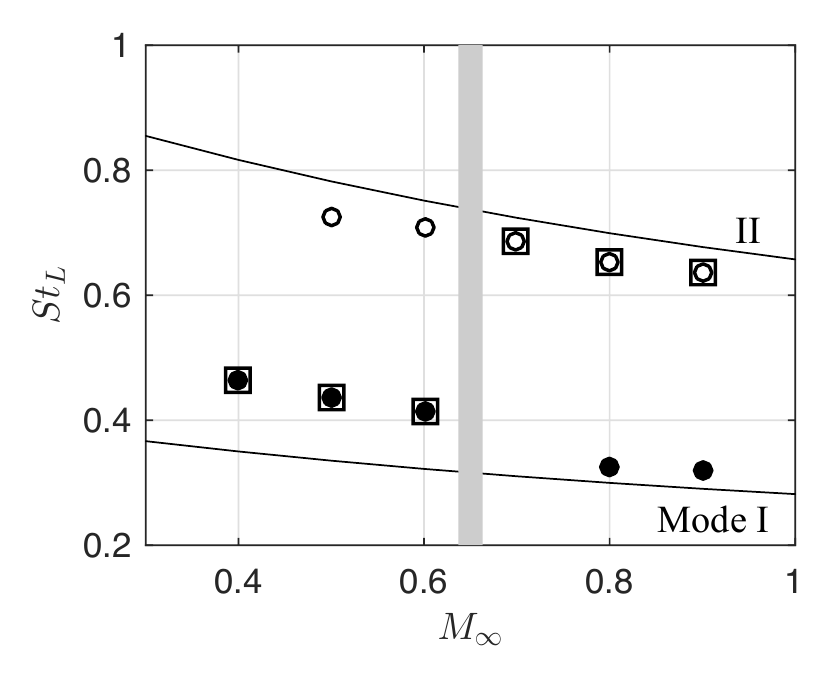}  &  \includegraphics[width=2.5in]{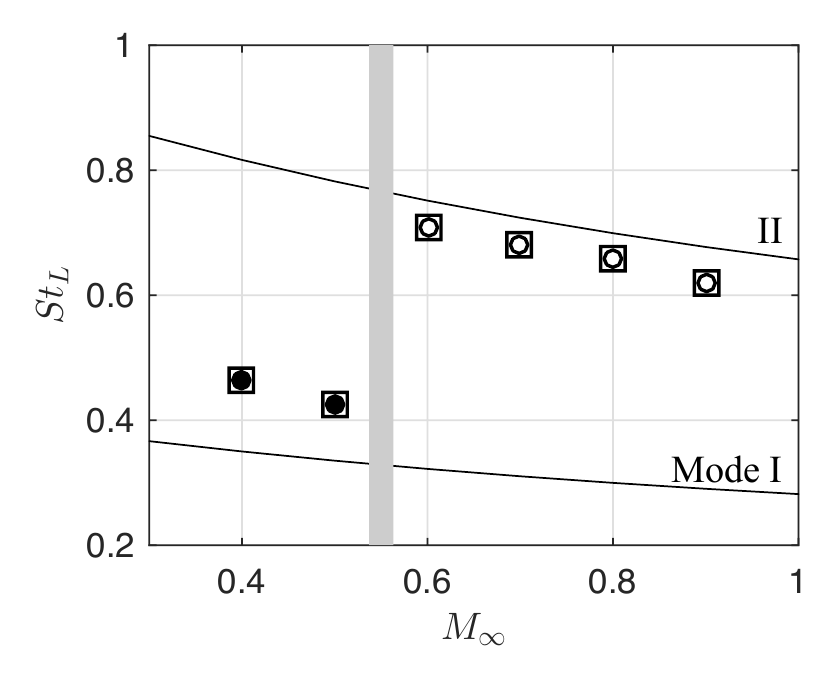}   \\
      $(c)$ $Re_\theta=67$ &$(d)$ $Re_\theta=77$\\                 
  \end{tabular}

\end{center}
   \caption{Comparison of $St_L (=St_D \cdot L/D)$ from the classic Rossiter semi-empirical formula, Eq. (\ref{eqRossiter}) (solid line) and the present work for $L/D=2$. Rossiter mode I: $\bullet$; mode II: $\circ$. The dominant modes are indicated by $\square$. The shaded grey line shows the estimate of the Mach number where the dominant mode shifts. }
   \label{fig:RossiterLD2} 
\end{figure}

The primary oscillation mechanism in open-cavity flow is associated with the Rossiter modes. For all cases considered herein, the time history of the vertical velocity at the mid-point of the cavity shear layer ($x=L/2,$ $y=0$) is recorded. A discrete Fourier transform is performed on the probe data collected after a minimum of $50$ convective units to eliminate the initial transients. For the unstable cases, the extracted oscillation frequencies  are compared with Rossiter's prediction based on the modified formula \citep{Heller:AIAA75}, Eq. (\ref{eqRossiter}).
The dominant and subdominant Rossiter modes as a function of Mach number are presented in figure \ref{fig:RossiterLD2}. The frequencies (Strouhal numbers) of the Rossiter modes show a decreasing trend with increasing $M_\infty$, which follows the prediction of the semi-empirical formula.

\begin{figure}
\begin{center}
  \begin{tabular}{cc}
   \vspace{-0.1in}\includegraphics[width=2.6in]{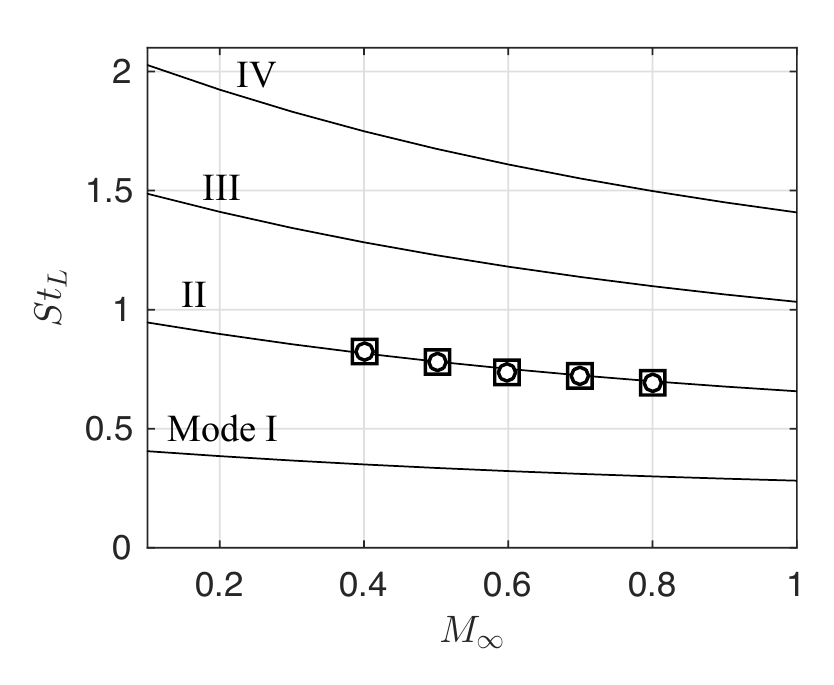} & \includegraphics[width=2.6in]{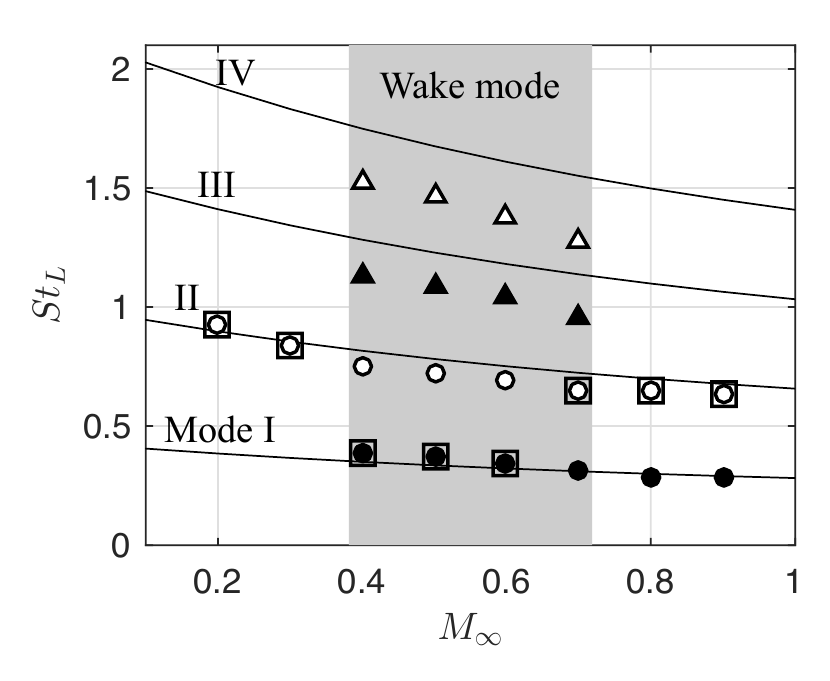}\\
   ($a$) $Re_\theta=10$ & ($b$) $Re_\theta = 19$
     \end{tabular}
   \caption{Comparison of $St_L (=St_D \cdot L/D)$ from the classic Rossiter semi-empirical formula, Eq. (\ref{eqRossiter}) (solid line) and present work for $L/D=6$. The wake-mode dominated cases are indicated in the shaded region with $M_\infty \in[0.4,0.7]$, while other cases represent shear-layer modes. In the wake-mode dominated flow, symbols $\bullet$, $\circ$, $\blacktriangle$ and $\vartriangle$ represent wake mode and its harmonics. In the shear-layer mode, these symbols denote Rossiter mode I to IV.  The dominant modes are indicated by $\square$. }
   \label{fig:RossiterLD6} 
\end{center}
\end{figure}

From the previous work by \cite{Bres:2007} on $L/D=2$, it was inferred that Rossiter mode I dominates the flow oscillation in the subsonic regime ($0.3\le M_\infty \le0.6$). The present investigation reveals that the dominant Rossiter mode shifts to mode II with an increase in Mach number. This transition indicated by the shaded grey line in figure \ref{fig:RossiterLD2} occurs at progressively lower Mach numbers for higher Reynolds numbers. The dominant mode switching was also observed in the experimental results from \cite{Kegerise:04PF} for cavity with $L/D=2$ at a much higher Reynolds number. As shown above in figure \ref{tableLD2} at $Re_\theta=46$, the acoustic radiation becomes stronger at $M_\infty=0.8$, where the dominant Rossiter mode shifts from mode I to II. Thus, it appears that the strong acoustic wave emission can be correlated with Rossiter mode II rather than mode I for the short cavity.

For flow over a cavity with $L/D = 6$, the frequencies (Strouhal numbers) of Rossiter mode for $Re_\theta=10$ and 19 are shown in figure \ref{fig:RossiterLD6}. At $Re_\theta = 10$, Rossiter mode II is the dominant mode and it is the sole mode present. As $Re_\theta$ increases to 19, the flow is unstable for $M_\infty=0.2$ and 0.3. Moreover, the wake mode appears in the flows for a range of $M_\infty\in[0.4,0.7]$, in which the dominant wake-mode frequency and its harmonics are observed. Although the wake-mode frequencies are close to the value of Rossiter mode frequencies, the features of the flow fields are significantly different from the shear-layer modes. Relating back to the stability diagram in figure \ref{fig:map} of $L/D=6$ at $Re_\theta=19$, we also find that the flow exhibits a shear-layer mode near the neutral stability boundary.  For longer cavity flows with $L/D=6$, there is no dominant Rossiter mode shifting observed in the shear-layer mode cases; and the wake mode dominates the flow at $M_\infty \in[0.4,0.7]$.

The analysis performed above reveals the 2D open-cavity flow physics and the tendency of the flow towards specific Rossiter modes. Noting that the results from this section are obtained from 2D numerical simulations. In \S \ref{sec:biglobal}, biglobal stability analysis is employed to gain deeper insights into the flow instabilities for a range of spanwise wavenumbers, $\beta$. Two-dimensional global stability can be uncovered with $\beta=0$, while spanwise-periodic 3D global eigenmodes can be revealed by specifying $\beta>0$. Because 2D base states are required for performing the biglobal stability analysis, the results from the above nonlinear simulations serve as the foundation for the subsequent stability analysis, in addition to the unstable steady states discussed below.

 
\subsection{Unstable steady states}
A steady-state flow is a time-invariant solution of the Navier--Stokes equations, which can be used as base state in linear stability analysis. This base state might be stable or unstable depending on the flow conditions. Time-averaged flow is obviously time-invariant but is not necessarily a solution of the Navier--Stokes equation. Numerical techniques can be used to determine the unstable steady state even if the flow is naturally unstable. 

\begin{figure}
\begin{center}
   \includegraphics[width=1.0\textwidth]{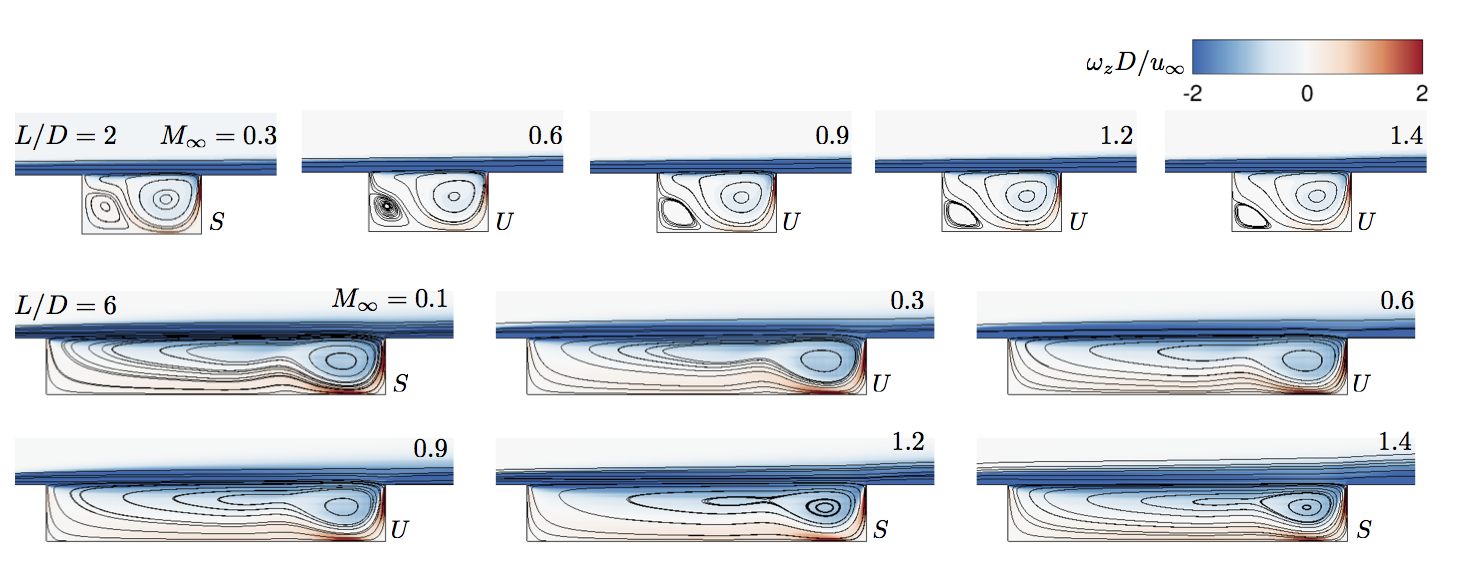}
   \caption{Streamlines of stable (S)/unstable (U) steady state of flows at various Mach number. $L/D=2$: $Re_\theta=56.8$; $L/D=6$: $Re_\theta=19$. Contours represent vorticity $\omega_z D/u_\infty\in[-2,2]$.}
   \label{fig:uss} 
\end{center}
\end{figure}

In the present work, the selective frequency damping method \citep{Akervik:PF06} is used to find the unstable steady state for unstable flows with the feedback gain chosen to be $\chi=0.1$. It has been verified that the numerically found time-invariant solutions from the selective frequency damping method are indeed the unstable steady states by substituting them into the Navier--Stokes equations and ensuring that $|\dot q|<10^{-7}$. As shown in figure \ref{fig:uss}, we find that the unstable steady states remain similar regardless of the Mach number variation in the short cavity ($L/D=2$), exhibiting the same features of the time-averaged streamlines as shown in figure \ref{tableLD2}. For the cavity with $L/D=6$, all the stable and unstable steady states share similar features regardless of the Mach number, while the time-averaged flow fields vary significantly depending on the Mach number as shown in figure \ref{tableLD6}.  In the present paper, we use these steady states (in figure \ref{fig:uss}) as the base states to perform global stability analysis. 

As we discuss in the next section, we find that the use of the stable/unstable steady states shown in figure \ref{fig:uss} as the base states in the biglobal stability analysis yields excellent agreement in the oscillatory features in most of the cases except for a very limited number of cases where the wake mode dominates the flow. For the wake-mode dominated flow (at $Re_\theta=19$), the mean flow can also be considered for its use as the base flow, as discussed in our companion work \citep{Sun:TCFD16}.  

Using the stable and unstable steady states, we perform extensive 2D and 3D linear global stability analysis.  The 2D and 3D results are provided below in \S \ref{sec:global2D} and \S \ref{sec:global3D}, respectively.
 
\section{Biglobal stability analysis}
\label{sec:biglobal}
\subsection{2D eigenmodes $(\beta=0)$}
\label{sec:global2D}
In the biglobal stability analysis, 2D global eigenmodes can be found by specifying $\beta=0$ in Eqs. (\ref{modal}) and (\ref{modal2}), which theoretically translates to the perturbation being associated with infinite wavelength in the spanwise direction. In this section, the stability properties are found for the base states. The eigenvalues $\omega=\omega_r+i\omega_i$ obtained are reported in the non-dimensional form of the growth/decay rate $\omega_i D/u_\infty$ and frequency $St_L=\omega_r L/2\pi u_\infty$ for 2D eigenmodes while $St_D=\omega_r D/2\pi u_\infty$ for spanwise periodic 3D eigenmodes.

\begin{figure}
\begin{center}
 \includegraphics[width=0.98\textwidth]{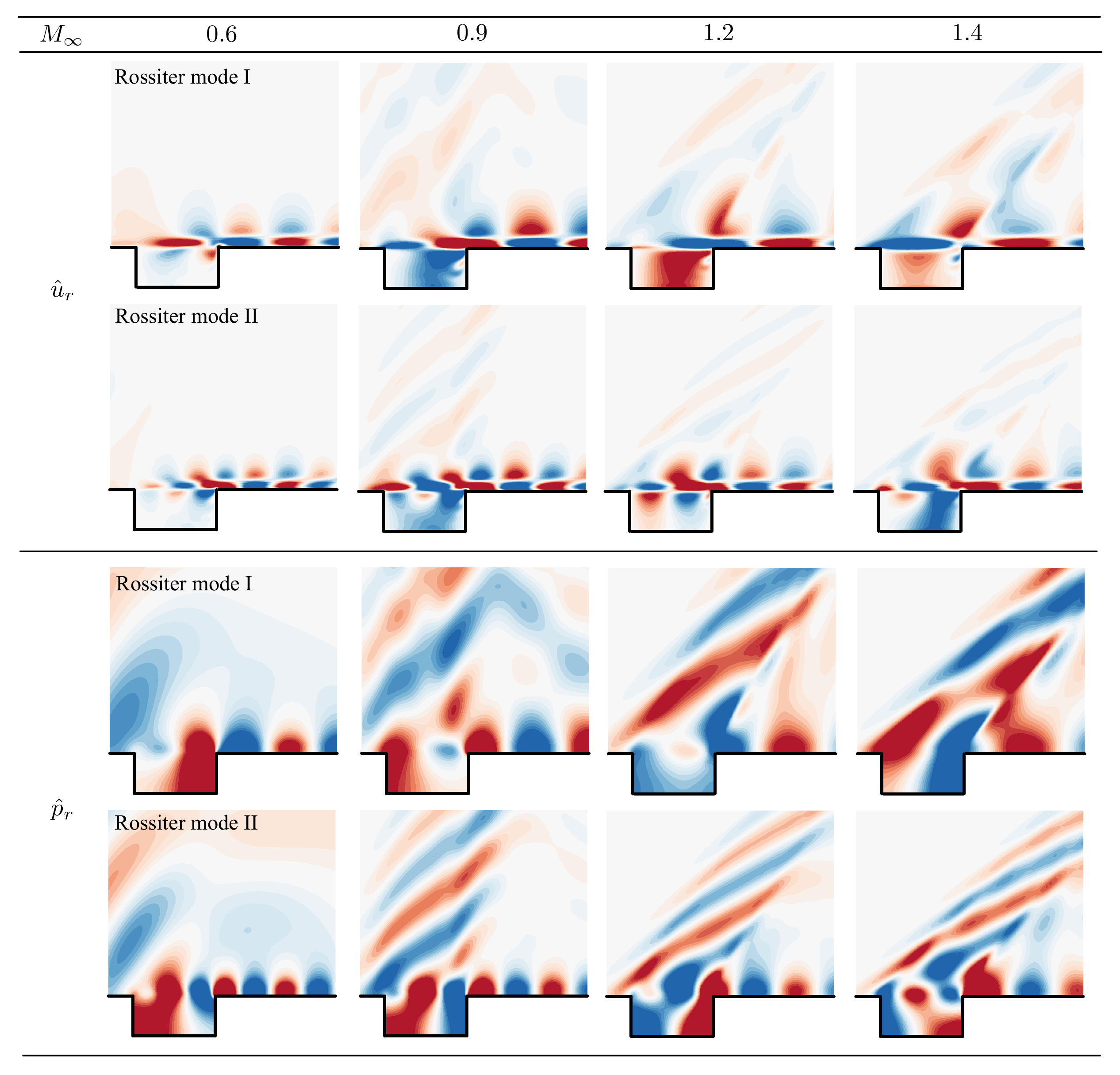}
   \caption{Spatial structures of the real component of the eigenmodes $\hat u$ and $\hat p$ for cavity of $L/D=2$ at $Re_\theta=56.8$.} 
      \label{fig:beta0_eigenmode_LD2} 
\end{center}
\end{figure}

\begin{figure}
\begin{center}
 \includegraphics[width=1.0\textwidth]{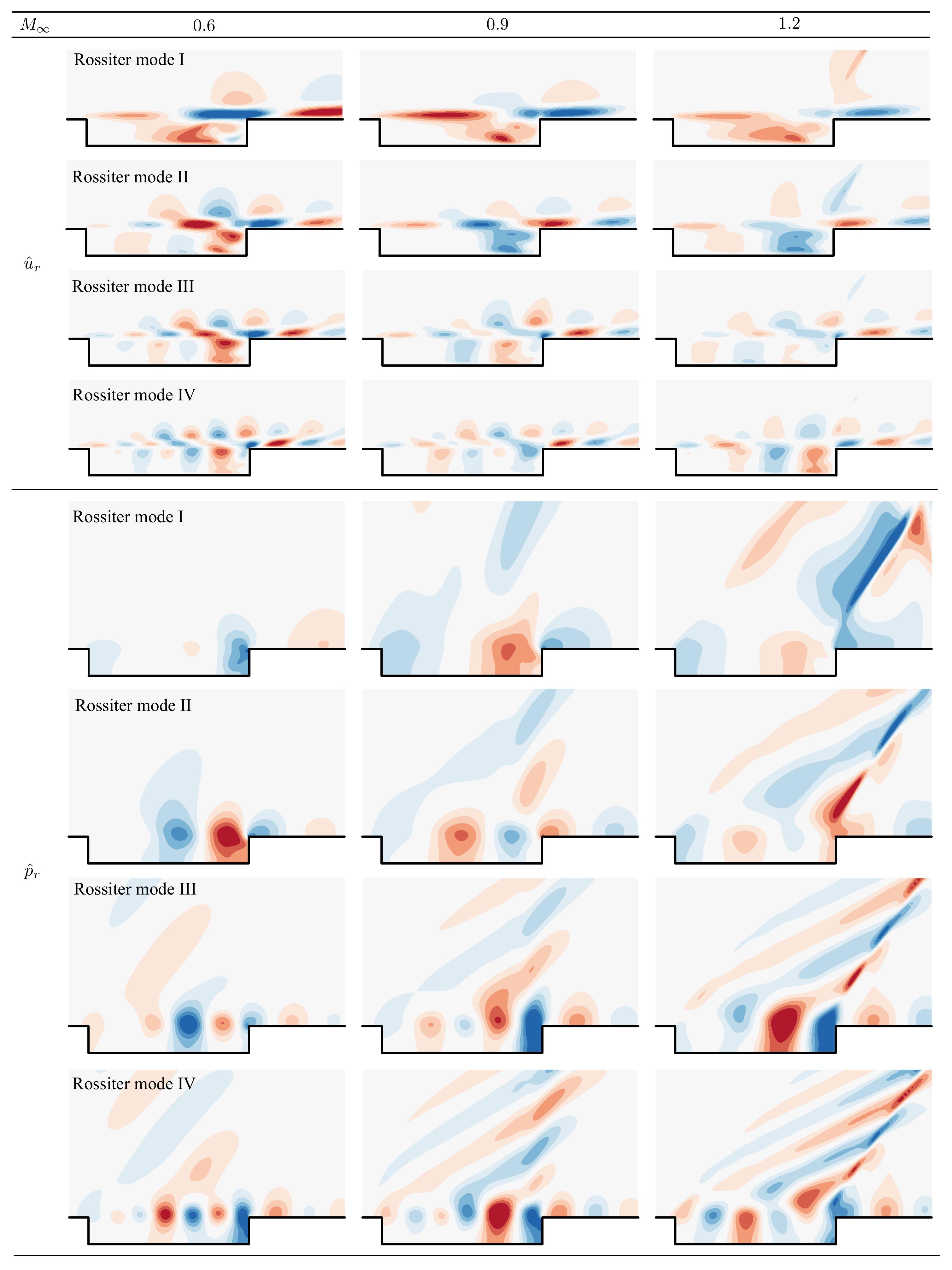}
 \caption{Spatial structures of the real component of the eigenmodes $\hat u$ and $\hat p$ for cavity of $L/D=6$ at $Re_\theta=19$.} 
       \label{fig:beta0_eigenmode_LD6} 
\end{center}
\end{figure}

The leading eigenmodes are shear-layer modes whose spatial structures are mainly located in the shear layer region. Realizing that not all shear-layer modes can be called Rossiter modes, we further check that the frequencies of these shear-layer modes match the Rossiter mode frequencies predicted by the semi-empirical formula. Hence, all the leading eigenmodes herein are associated with Rossiter modes and correspond to shear-layer modes (i.e., Kelvin--Helmholtz instabilities) reported in the works by \cite{Garrido:JFM14} and \cite{Yamouni:JFM13}. The eigenvalues of Rossiter modes are listed in table \ref{RM_table} and their eigenvectors are shown in figures \ref{fig:beta0_eigenmode_LD2} and \ref{fig:beta0_eigenmode_LD6} for $L/D=2$ and 6, respectively. As predicted by the semi-empirical formula, an increase in Mach number reduces the flow oscillation frequency. The spatial structure size of Rossiter modes becomes larger as Mach number increases. The spatial structures of higher-order modes exhibits finer spatial scales than those of lower-order modes. The major spatial structures of the velocity eigenmodes $\hat u_r$ have the largest magnitude in the shear layer, which indicates that Rossiter modes are driven by the shear layer of the flow. In the pressure eigenvector contours, there are beam-shaped structures that develop above the cavity. The geometric features of these beams parallel the compression waves previously discussed in figure \ref{tableLD2}. Based on the results from both 2D nonlinear simulations and biglobal stability analysis with $\beta=0$, it indicates that Rossiter modes are indeed two-dimensional, shear-layer driven instabilities. 

\begin{table}
\begin{center}
\begin{tabular}{lM{0.1in}M{0.9in}M{0.9in}M{0.9in}M{0.9in}}
      	\multicolumn{2}{l}{$L/D=2$}		\\ 
	$M_\infty$&&I			&II					\\
	0.3 		&&$0.534-0.008i$	&$0.795-0.016i$	\\
	0.6 		&&$0.426+0.055i$	&$0.726+0.024i$	\\
	0.9 		&&$0.377+0.019i$	&$0.687+0.071i$	\\
	1.2 		&&$0.336-0.012i$	&$0.613+0.068i$	\\
	1.4 		&&$0.310-0.029i$	&$0.566+0.037i$	\\ \\
	\multicolumn{2}{l}{$L/D=6$} 		\\
	$M_\infty$	&&I			& II			&III			&IV			\\
	0.3 		&&$0.432-0.057i$	&$0.878+0.092i$	&$1.410+0.027i$	&$1.815-0.042i$		\\
	0.6 		&&$0.381-0.000i$	&$0.784+0.081i$	&$1.175+0.069i$	&$1.574-0.005i$		\\
	0.9 		&&$0.320-0.006i$	&$0.708+0.030i$	&$1.085+0.020i$	&$1.474-0.028i$		\\
	1.2 		&&$0.282-0.032i$	&$0.638-0.008i$	&$0.994-0.014i$	&$1.344-0.046i$		\\
	1.4 		&&$0.260-0.048i$	&$0.599-0.035i$	&$0.947-0.045i$	&$1.293-0.088i$			\\
\end{tabular}
\end{center}
\caption{Eigenvalues $\omega=\omega_r+i\omega_i$ associated with Rossiter mode I to IV for $\beta=0$ are reported in a form of $St_D\cdot(L/D)+i(\omega_i D/u_\infty)$ at $Re_\theta=56.8$. Frequency $\omega_r$ is normalized as Strouhal number $St_L=\omega_r L/2\pi u_\infty$, and growth/decay rate $\omega_i$ is normalized as $\omega_i D/u_\infty$.}
\label{RM_table}
\end{table}

%

\begin{figure}
\begin{center}
\includegraphics[width=0.9\textwidth]{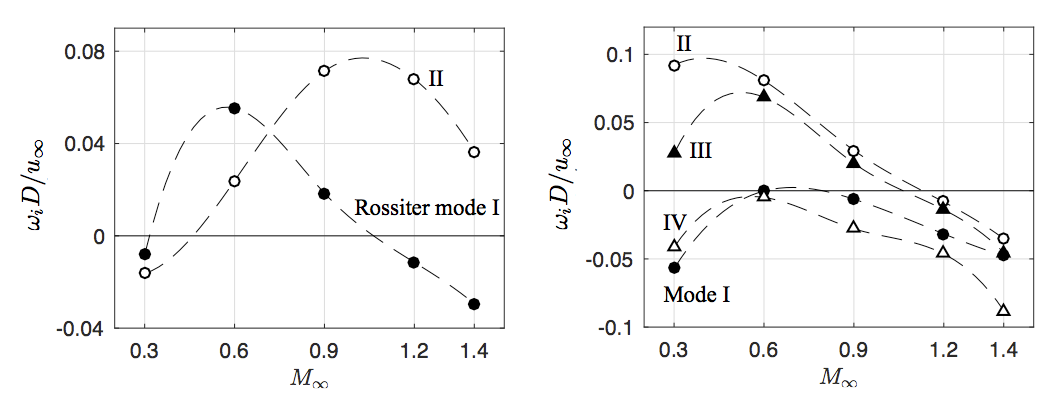}
   \caption{Growth/decay rates of Rossiter modes as a function of free stream Mach number determined from biglobal stability analysis with $\beta=0$. Left: $L/D=2$, $Re_\theta=56.8$; right: $L/D=6$, $Re_\theta=19$. Rossiter mode I: $\bullet$; mode II: $\circ$; mode III: $\blacktriangle$ and mode IV: $\vartriangle$.} 
      \label{fig:beta0_growth} 
\end{center}
\end{figure}

From the discussion in \S \ref{RM} (and figures \ref{fig:RossiterLD2} and \ref{fig:RossiterLD6}), the flow instability and dominant Rossiter mode shift with free stream Mach number. In the biglobal stability analysis, the growth/decay rate $\omega_i D/u_\infty$ over $M_\infty$ is presented in figure \ref{fig:beta0_growth}. For cavity with $L/D=2$, the growth/decay rates of Rossiter modes I and II increase as Mach number changes from 0.3 to 0.6. However, when $M_\infty$ reaches the transonic regime, the values of growth rate show a decreasing trend. The growth rate for Rossiter mode I peaks around $M_\infty=0.6$ and is surpassed by the growth rate of Rossiter mode II around $M_\infty=0.7$, which reaches its peak near $M_\infty=1$. This crossover point corresponds to the shift from Rossiter mode I to mode II seen in figure \ref{fig:RossiterLD2} (b). This peaking trends of growth/decay rates of Rossiter modes are also observed for the flows over cavity with $L/D=6$. The growth rate of the eigenmode negatively correlating to the increase in free stream Mach number can be related to the behavior of the neutral stability curve in the nonlinear stability diagrams shown in figure \ref{fig:map}. Both stability diagrams indicate that an increase of Mach number in the subsonic regime ($M_\infty<0.6$) destabilizes the flow, but over the transonic regime, an increase of Mach number ($M_\infty>0.6$) stabilizes the flow. Furthermore, in the cases of $L/D=2$ and $M_\infty=0.3$, as shown in figure \ref{fig:beta0_growth}, the values of $\omega_i D/u_\infty$ are both negative. In other words, disturbances related to Rossiter modes I and II decay in the base flow, which matches the result from the stability diagram in figure \ref{fig:map} that the flow is stable. This agreement is also noticed in the cases of cavity with $L/D=6$ at $M_\infty=1.2$ and 1.4.     

\begin{figure}
\begin{center}
  \includegraphics[width=0.9\textwidth]{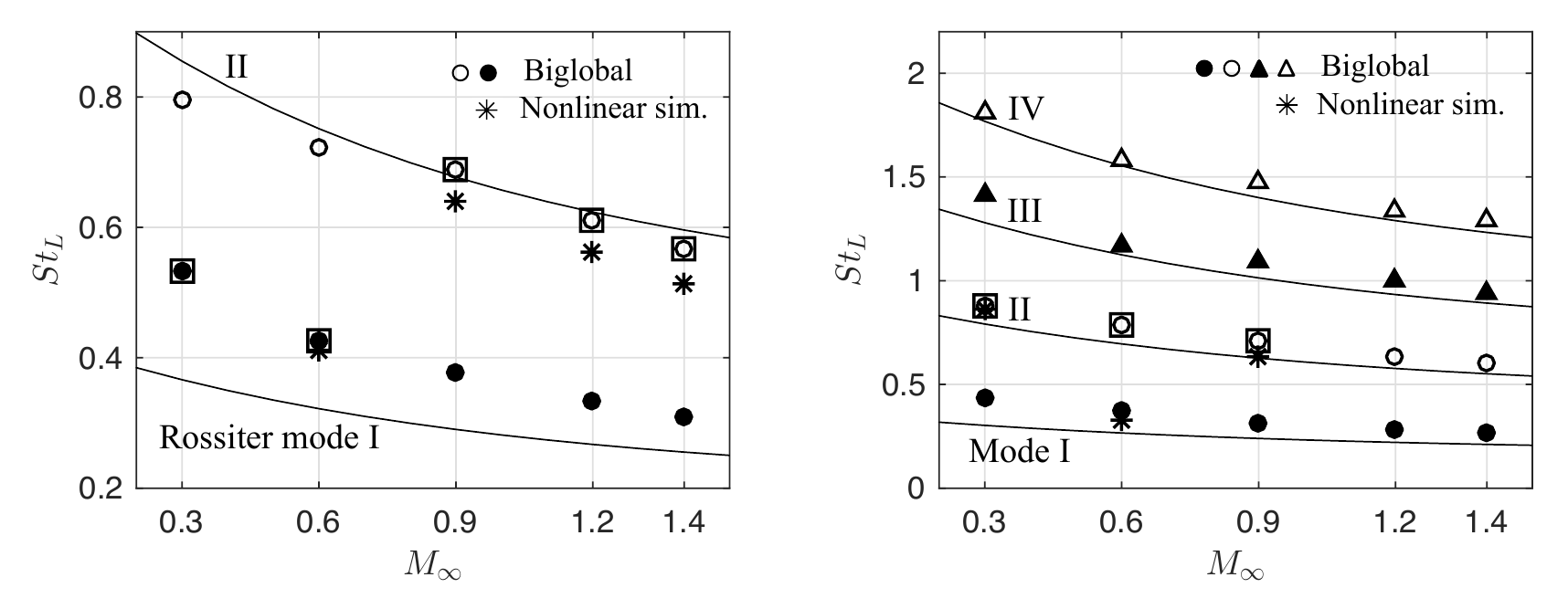}
   \caption{Dominant Rossiter modes captured from the 2D nonlinear simulations and the biglobal stability analysis. Left: $L/D=2$, $Re_\theta=56.8$; right: $L/D=6$, $Re_\theta=19$. Rossiter mode I: $\bullet$; mode II: $\circ$; mode III: $\blacktriangle$ and mode IV: $\vartriangle$. The dominant Rossiter modes from the biglobal stability analysis and the nonlinear simulations are indicated by $\square$ and $*$, respectively. } 
      \label{fig:beta0_stl} 
\end{center}
\end{figure}
From the growth/decay rate trend shown in figure \ref{fig:beta0_growth}, we compare the Rossiter mode with the largest growth/decay rate at each Mach number to the dominant Rossiter mode observed from the 2D nonlinear simulations, which can explain the mode shifting presented in figures \ref{fig:RossiterLD2} and \ref{fig:RossiterLD6}. In figure \ref{fig:beta0_stl}, the dominant Rossiter mode captured from linear stability analysis and nonlinear DNS discussed in \S \ref{sec:results} are compared. In almost all cases, there is excellent agreement in the dominant Rossiter mode with respect to Mach number. At $M_\infty=0.6$ and $L/D=6$, the dominant Rossiter mode from biglobal stability analysis deviates from that in the DNS in which the wake mode is dominant. Moreover, the eigenvector shown in figure \ref{fig:beta0_eigenmode_LD6} at $M_\infty=0.6$ resembles more closely the Rossiter modes instead of the wake mode revealed from the 2D DNS. For this particular case, we further investigate the wake mode in our companion paper \citep{Sun:TCFD16}. By the observation that the mean (time-averaged) flow and unstable steady state are significantly different in this wake-mode dominated flow, biglobal stability analysis was performed using both states (mean flow and unstable steady state) as base states. When the mean flow is prescribed as the base state, the wake-mode eigenmode is captured by the linear stability analysis, which can also predict its frequency with only 4\% difference between that and the wake-mode frequency determined from the 2D DNS. It should be mentioned that in the linear stability theory, the nonlinear interactions among modes are neglected. There is however a strong nonlinear dynamical process in the wake-mode dominated flow, making the mean greatly deviate from the unstable steady state. Hence, the use of the mean flow as the base flow is necessary for uncovering the wake mode. 

\subsection{3D eigenmodes $(\beta \ne 0)$}\label{sec:global3D}

In this section, we discuss characteristics of the 3D global instability modes with finite wavelength $\lambda/D=2\pi/\beta$ in the spanwise direction prescribed in Eq. (\ref{modal}). The analyses performed for the 3D instabilities are analogous to those for the 2D cases shown in the previous section. In what follows, the spanwise wavelength $\lambda/D$ is set in a range of 0.5 -- 2.0 following \cite{Bres:JFM08}. Note that the 3D instabilities discussed in this section represent spanwise-periodic 3D instabilities with selected wavenumbers $\beta$. 

\begin{figure}
\begin{center}
   \includegraphics[width=1.0\textwidth]{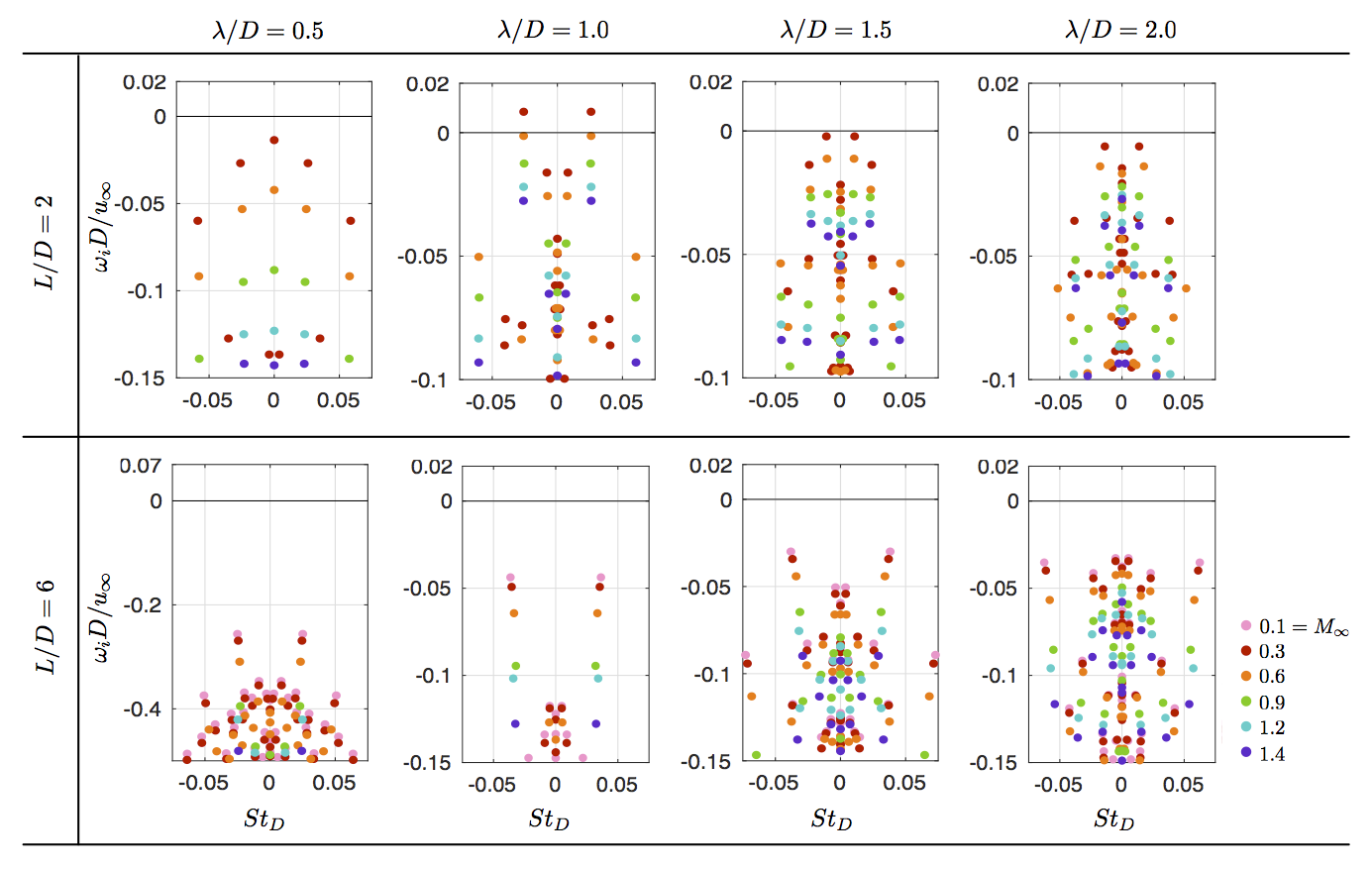}
   \caption{Eigenspectra of the 3D eigenmodes for cavity with $L/D=2$ ($Re_\theta=56.8$) and 6 ($Re_\theta=19$) with $M_\infty \in[0.1,1.4]$ and $ \lambda/D \in [0.5,2.0]$.} 
      \label{fig:3DspectraLD} 
\end{center}
\end{figure}

The eigenspectra of the 3D eigenmodes  for $L/D=2$ and 6 are presented in figure \ref{fig:3DspectraLD}. It should be noted that the eigenspectra are only shown in the vicinity of the origin. For the $L/D=2$ cavity flows at $Re_\theta=56.8$, unstable 3D modes are only observed in the subsonic cases of $M_\infty=0.3$ with $\lambda/D =1.0$, while all other 3D eigenmodes with higher Mach number ($M_\infty \ge 0.6$) are stable.  For $L/D=6$ cases at $Re_\theta=19$, all the 3D eigenmodes captured are stable. Below, the effects of cavity geometry, Mach number and spanwise wavelength $\lambda/D$ on the most-unstable/least-stable (dominant) 3D modes are further examined. 

\begin{figure}
\begin{center}
\includegraphics[width=0.98\textwidth]{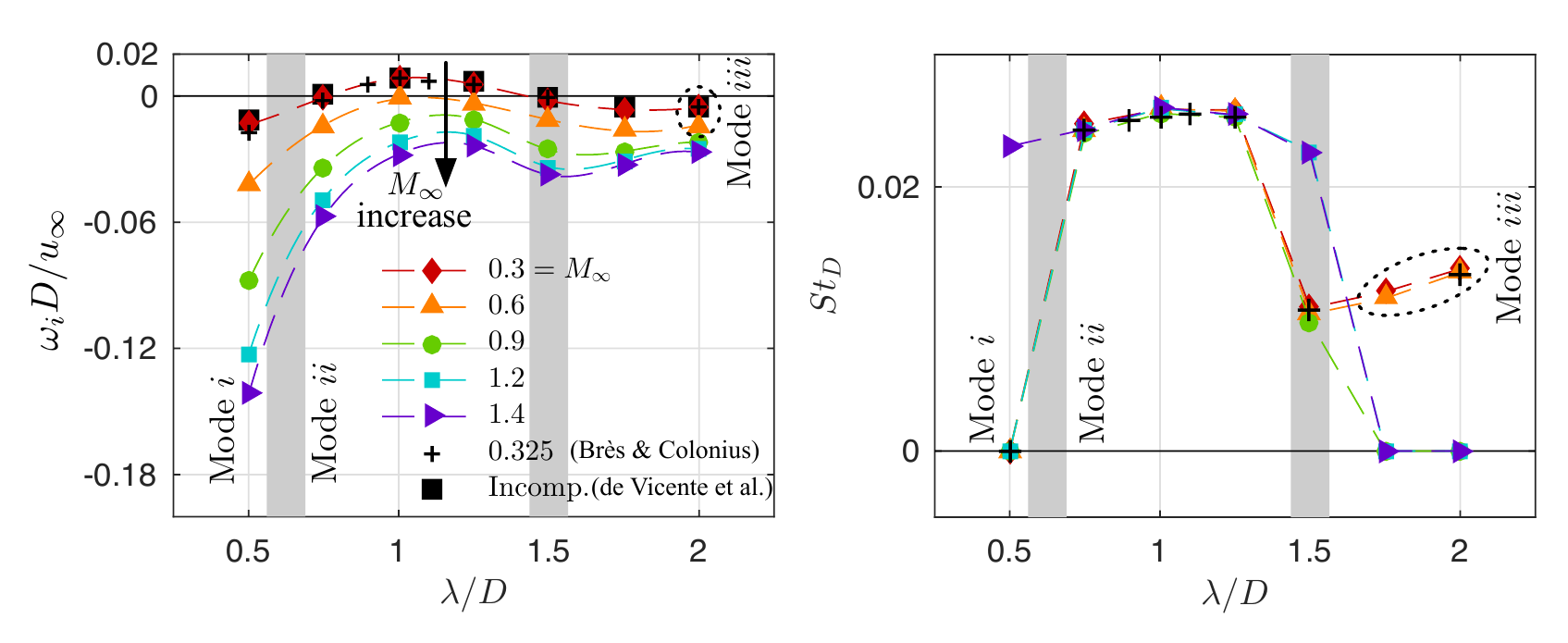}
   \caption{Eigenvalues of the dominant 3D modes as a function of $M_\infty$ and spanwise wavelength $\lambda/D$ for cavity with $L/D=2$ at $Re_\theta=56.8$. Left: growth/decay rate $\omega_i D/u_\infty$; right: frequency $St_D$. Results from \cite{Bres:JFM08} and \cite{Vicente:JFM14} are also compared to the present results. The gray regions represent the transitions of dominant modes.} 
      \label{fig:dominant_LD2} 
\end{center}
\end{figure}

\begin{figure}
\begin{center}
   \includegraphics[width=0.98\textwidth]{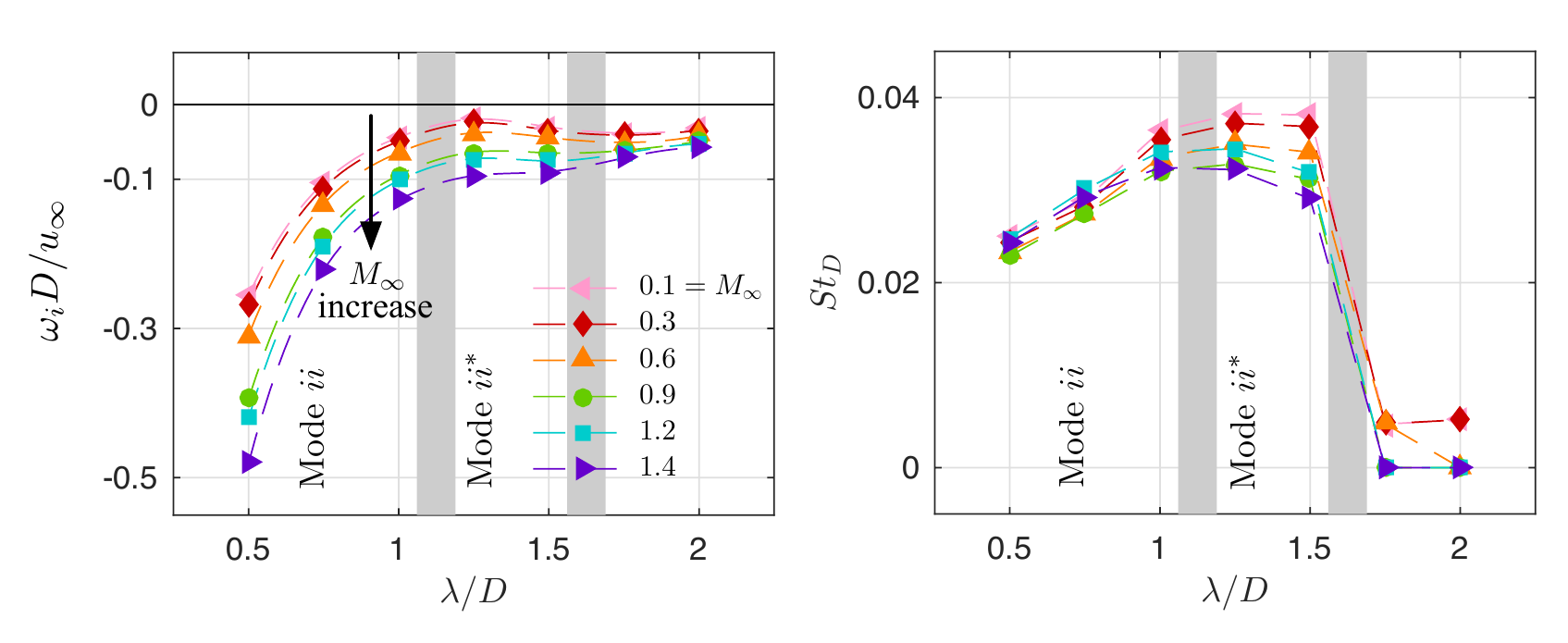}
   \caption{Eigenvalues of the dominant 3D modes as a function of $M_\infty$ and spanwise wavelength $\lambda/D$ for cavity with $L/D=6$ at $Re_\theta=19$. Left: growth/decay rate $\omega_i D/u_\infty$; right: frequency $St_D$. The gray regions represent the transitions of dominant modes.} 
      \label{fig:dominant_LD6} 
\end{center}
\end{figure}

Based on eigenspectra shown in figure \ref{fig:3DspectraLD}, the eigenvalue of the most-unstable/least-stable eigenmode for each Mach number is extracted and plotted as a function of spanwise wavelength $\lambda/D$ in figures \ref{fig:dominant_LD2} and \ref{fig:dominant_LD6}. For both cavity geometries, the overall trends of growth/decay rates are similar regardless of $M_\infty$. An increase in Mach number stabilizes all the dominant 3D eigenmodes. For each Mach number, the growth/decay rate $\omega_iD/u_\infty$ of the dominant 3D eigenmode is a function of $\lambda/D$. To identify the 3D eigenmode properties, their frequencies and spatial structures can distinguish the modes based on types of instabilities. 

We follow the nomenclature used by \cite{Bres:JFM08} for the leading modes in figures \ref{fig:dominant_LD2} and \ref{fig:dominant_LD6}. As shown in figure \ref{fig:dominant_LD2} for cavity with $L/D=2$, the leading eigenmode with $\lambda/D=0.5$ is mode $i$ with $St_D=0$ (referred to as mode $i$ by \cite{Bres:JFM08}) except at $M_\infty=1.4$. The dominant 3D mode at $M_\infty=1.4$ is mode $ii$, a traveling mode discussed next. However, its decay rate is close to that of a stationary mode as shown in figure \ref{fig:3DspectraLD}. The eigenmodes with $\lambda/D\in[0.75,1.25]$ are traveling modes with $St_D\approx0.025$ (referred to as mode $ii$ by \cite{Bres:JFM08}). For each of $\lambda/D\le1.25$, the frequency, $St_D$, of all the dominant 3D modes are independent of Mach number, but these 3D modes can be stabilized by increasing $M_\infty$ as mentioned above. However, the frequencies of the dominant 3D modes exhibit a sudden decrease near $\lambda/D=1.5$, and the modes exhibit different branches for larger $\lambda/D$ ($\ge1.75$) depending on Mach numbers. As shown in figure \ref{fig:dominant_LD2}, for the flows at $0.3\le M_\infty\le0.6$, the dominant modes are traveling modes with $St_D\approx0.015$ (referred to as mode $iii$ by \cite{Bres:JFM08}), while for the transonic flows at $0.9\le M_\infty \le1.4$, the dominant modes are stationary. Moreover, we indicate the transitions of dominant modes over the spanwise wavelength in figure \ref{fig:dominant_LD2} by grey regions. As the wavelength of the 3D eigenmode is increased from 0.5 to 2, in general, the dominant 3D mode shifts from stationary to traveling mode and back to stationary mode. This phenomenon was also reported by \cite{Vicente:JFM14} on incompressible cavity flows that the dominant 3D mode can arise from different instabilities in terms of its spanwise wavelength. 

\begin{figure}
\begin{center}
   \includegraphics[width=1.0\textwidth]{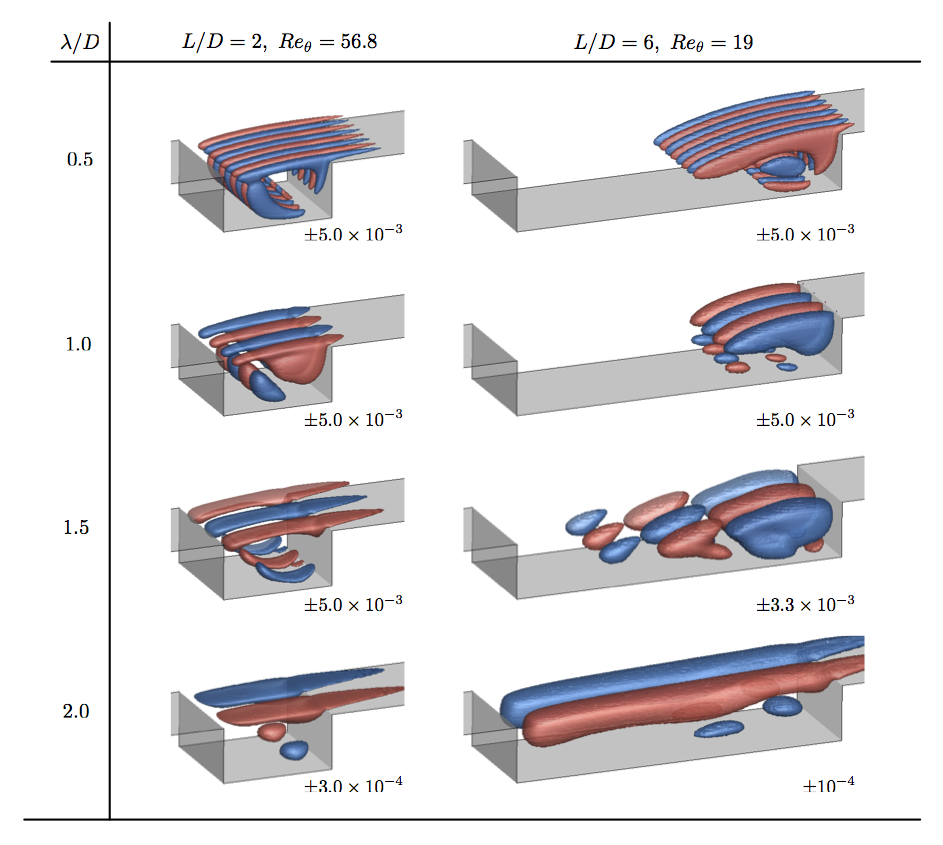}
\end{center}
\caption{Iso-surface of eigenvectors of the least-stable 3D eigenmodes at $M_\infty=0.9$ and spanwise wavelength $\lambda/D\in[0.5,2.0]$. The values selected to show the iso-surface of $\hat u_r/u_\infty$ are indicated below each subplot. Red and blue colors represent the indicated positive and negative values, respectively.}
\label{fig:3Dviews}
\end{figure}

For the eigenvalues in the case of a cavity with $L/D=6$ presented in figure \ref{fig:dominant_LD6}, the dominant 3D mode with $\lambda/D=0.5$ is the traveling mode or mode $ii$ ($St_D\approx0.025$) as mentioned above for the case of short cavity. We further note that the dominant mode has characteristics of mode $ii$ type, as we later discuss in the spatial structures of the eigenmodes (in figure \ref{fig:3Dviews}). The dominant 3D modes with $\lambda/D=0.75$ and 1.0 have higher frequencies compared to that of smaller wavelenghths, but their eigenvectors still share similarities as shown later. There is another mode having frequency $St_D\in[0.03,0.04]$ with $\lambda/D=1.25$ and 1.5, which is denoted as mode $ii^*$ due to the variation in both frequency and eigenvector compared to mode $ii$. As $\lambda/D$ increases from 1.5 to 1.75, the dominant mode changes from a traveling mode to a stationary mode. With $\lambda/D=1.75$ and 2.0, there is similarity to the cases of short cavity, in which the dominant 3D modes are nearly stationary. For the subsonic flows with $0.1\le M_\infty \le0.6$, these 3D modes possess nonzero frequencies, but with relatively small $St_D$ compared to the other traveling modes. A decreasing trend in the growth/decay rates at $\lambda/D=1.5$ is observed as shown in figure \ref{fig:dominant_LD6}, which is similar to those captured for the short cavity.  However, the possible mode transitions in the region of $0.5\le \lambda/D \le1.5$ is not distinctly clear based on the properties of the eigenvalues, which is further discussed below.  

The spatial structures of the dominant 3D eigenmodes are illustrated in figure \ref{fig:3Dviews} for different values of $\lambda/D$. For the short cavity with $L/D=2$, the eigenvectors obtained for all Mach numbers considered are similar in terms of the spatial structures. For all values of $\lambda/D$ considered, the eigenvectors show variations in the aft part of the cavity where the major recirculation zone is located as shown in the unstable steady states shown in figure \ref{fig:uss}, while their structures are also present in the shear-layer region of the flows. \cite{Bres:JFM08} examined the 3D instabilities as well at $M_\infty=0.3$ and argued that these 3D modes result from the centrifugal instability of the recirculation in the base flows. 

For the cavity with $L/D=6$, the spatial structures of the dominant 3D eigenmodes also appear independent of the Mach number, as in the case of $L/D=2$. As shown in figure \ref{fig:3Dviews}, in the cases of $\lambda/D=0.5$ and 1.0, the spatial structures of mode $ii$ are located in the rear part of the cavity, which exhibit similar features to those presented in the shorter cavity, but with the eigenmodes extending upstream along the shear layer. The spatial structures of mode $ii^*$ ($\lambda/D=1.5$) align along the floor from the rear towards the mid-region of the cavity, which is significantly different from the mode $ii$ that stems from centrifugal instabilities. The streamwise-stretched recirculation pattern in the unstable steady state shown in figure \ref{fig:uss} appears due to the large aspect ratio of the long cavity. This elongated recirculating flow is likely the reason for the formation of the mode $ii^*$, which explains the absence of mode $ii^*$ in the short cavity flows. For the stationary mode for $\lambda/D=2.0$, its spatial structure covers the entire length of the cavity. Although the frequency of the 3D mode with $\lambda/D=2.0$ in the subsonic regime ($M_\infty=0.1-0.6$) is not exactly zero, the eigenvectors still share similar features to the stationary mode. The spatial structures of 3D instability modes appear to be strongly influenced by their spanwise wavelengths in the long cavity cases, and the leading eigenmode can be a traveling or stationary mode depending on $\lambda/D$. 

In the work by \cite{Liu:JFM16} that sidewall effects are considered with triglobal stability analysis for incompressible cavity flow with $L/D=6$, they concluded that shear-layer instability is dominant in finite-span cavity flow. This is also observed in the present study on spanwise homogeneous cavity flow at $M_\infty=0.3$ that the most-unstable mode is found to be the shear-layer mode (Rossiter mode), while centrifugal instability modes are stable. The major difference appears in the spatial structure of shear-layer mode due to cavity geometries; a narrow finite-span cavity in their work and a spanwise homogeneous cavity in the present work.


\subsection{Comparison with DNS}\label{CompDNS}

\begin{figure}
\begin{center}
   \includegraphics[width=0.85\textwidth]{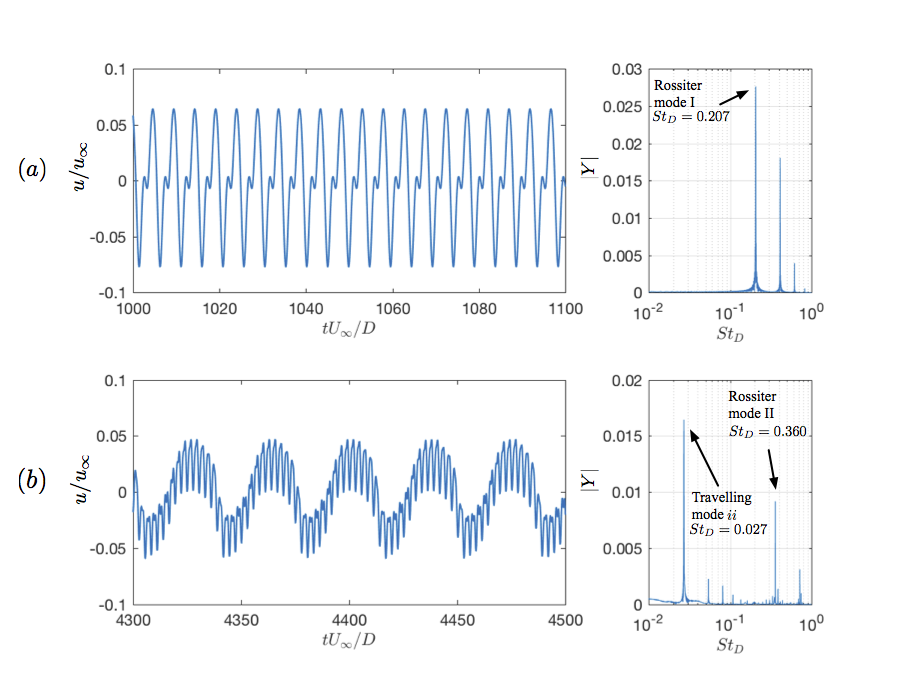}
\end{center}
\caption{Time histories of streamwise velocity $u/u_\infty$ collected by a probe located at $x/D = 1$, $y/D = 0$, $z/D = 1$ are shown on the left. The corresponding Fourier spectra are presented on the right. (a) 2D DNS and (b) 3D DNS ($W/D=2$) for $L/D=2$, $Re_\theta=56.8$ and $M_\infty=0.6$.}
\label{fig:3DFFT}
\end{figure}

\begin{figure}
\begin{center}
\begin{tabular}{cccc} \hline
\multicolumn{3}{l}{3D nonlinear simulation ($W/D=2$)}\\ 
 \includegraphics[width=0.23\textwidth]{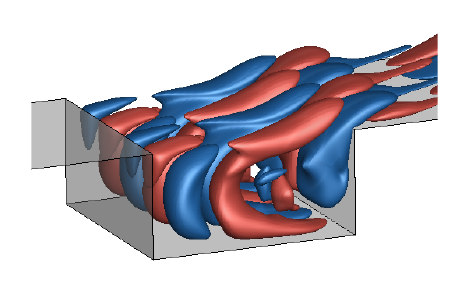}& \includegraphics[width=0.23\textwidth]{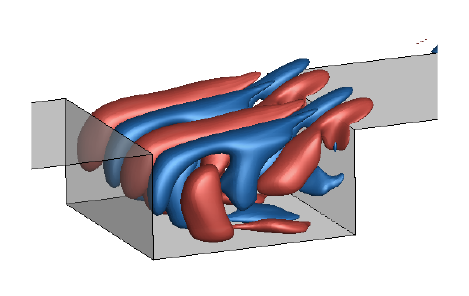}& \includegraphics[width=0.23\textwidth]{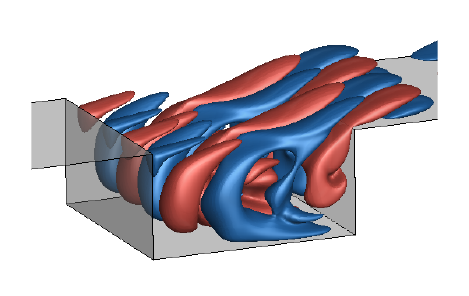}& \includegraphics[width=0.23\textwidth]{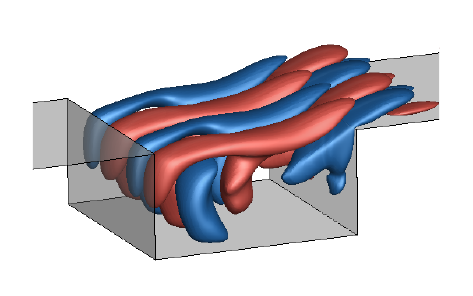}\\
 $(t-t_0)/T=0$&$1/4$&$2/4$&$3/4$\\ \hline
\multicolumn{3}{l}{3D mode $ii$ ($\lambda/D=1.0$)}\\
 \includegraphics[width=0.23\textwidth]{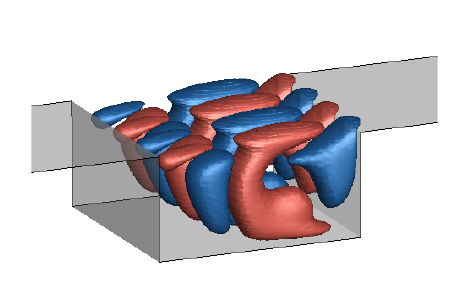}& \includegraphics[width=0.23\textwidth]{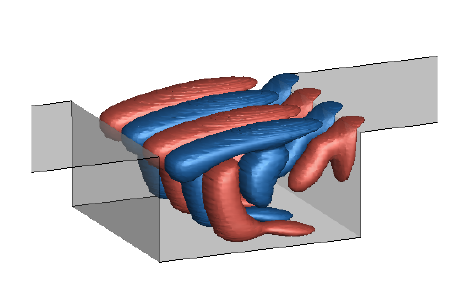}& \includegraphics[width=0.23\textwidth]{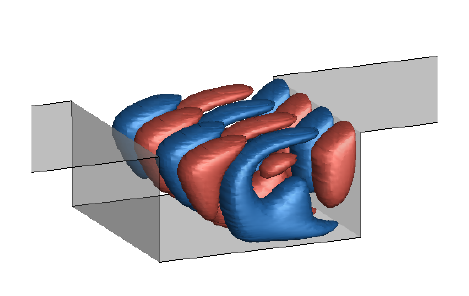}& \includegraphics[width=0.23\textwidth]{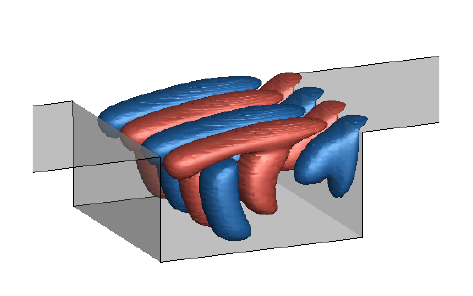}\\ 
  $(\phi-\phi_0)/2\pi=0$&$1/4$&$2/4$&$3/4$\\ \hline 
\end{tabular}
\caption{Comparison of the iso-surface of the spanwise velocity $w/u_\infty=\pm0.013$ from the 3D nonlinear simulation and $\hat w/u_\infty=\pm0.03$ from the dominant 3D eigenmode with $\lambda/D=1.0$ for the cavity with $L/D=2$ at $M_\infty=0.6$ and $Re_\theta=56.8$. Represented by $T$ is the time period of the 3D mode captured by the nonlinear simulation, $t_0$ and $\phi_0$ are reference time and phase, respectively. Red and blue colors represent positive and negative values, respectively.}
\label{fig:3Dnon_LD2}
\end{center}
\end{figure}

Next, let us compare our findings from linear stability analysis with those from DNS. Three-dimensional DNS with $W/D=2$ are performed at $M_\infty=0.6$ for both short and long cavities. For the cavity with $L/D=2$ ($Re_\theta=56.8$), Fourier spectra of velocity are shown in figure \ref{fig:3DFFT}. The comparison of dominant frequencies, $St_D$, obtained from DNS and biglobal stability analysis are listed in table \ref{tab:comp3D}. In the 3D DNS of $L/D=2$, the dominant 2D mode is Rossiter mode II. However, in 2D DNS and biglobal stability analysis with $\lambda/D=\infty$, the Rossiter mode I is the dominant mode. By performing the simulation over a sufficiently long time, the traveling mode with $St_D=0.027$ at $\lambda/D=1.0$ is observed in the 3D DNS. Its properties have excellent agreement to those of the leading 3D mode shown in figure \ref{fig:dominant_LD2}, in which mode $ii$ with $\lambda/D=1.0$ has largest $\omega_iD/u_\infty=-0.0013$ among all the $\lambda/D$ considered. Although this mode has negative growth rate, in the 3D nonlinear simulation, unstable 2D Rossiter mode and 3D modes could interact via nonlinearities and result in the existence of the 3D mode in the nonlinear flows, especially when the eigenvalue is close to the neutral stability line ($\omega_i D/u_\infty=0$). The comparison of the 3D spatial structures of nonlinear flow and traveling mode $ii$ are displayed in figure \ref{fig:3Dnon_LD2}, where instantaneous flow fields and eigenmode with a quarter period interval are presented. Mode $ii$ shown in figure 20 has qualitatively similar spatial structures to the eigenfunction of Mode I found by \cite{Garrido:JFM14} at a similar incompressible flow condition. In their work, this mode is unstable in incompressible conditions. In contrast, mode $ii$ is slightly stable in the present work at $M_\infty=0.6$, which agrees with the discussion in section 4.3 that an increase in Mach number can stabilize 3D eigenmodes. The iso-surface of spanwise velocity $w/u_\infty$ from nonlinear simulation exhibits comparable structures to those of mode $ii$. In particular, the 2D Rossiter mode also appears in the nonlinear DNS, in which streamwise distortions of the spatial structures are observable in the shear-layer region.

\begin{table}
\begin{center}
\begin{tabular}{lcl c }
				&$St_D$	&				&$\lambda/D$	\\ \hline
2D DNS			&0.207	&Rossiter mode I$^*$	&$\infty$		\\ \\
Biglobal stability	&0.213	&Rossiter mode I$^*$	&$\infty$		\\
				&0.026	&Traveling mode $ii$	&1.0			\\ \\
3D DNS			&0.360	&Rossiter mode II$^*$	&$\infty$		\\
				&0.027	&Traveling mode $ii$	&1.0			\\
\end{tabular}
\end{center}
\caption{Comparison of the frequencies of the dominant modes obtained from DNS and biglobal stability analysis for $L/D=2$, $Re_\theta=56.8$ and $M_\infty=0.6$. Dominant mode is denoted with $^*$.}
\label{tab:comp3D}
\end{table}

\begin{figure}
\centering
\begin{tabular}{m{2.2in}m{2.2in}} 
\includegraphics[width=0.4\textwidth]{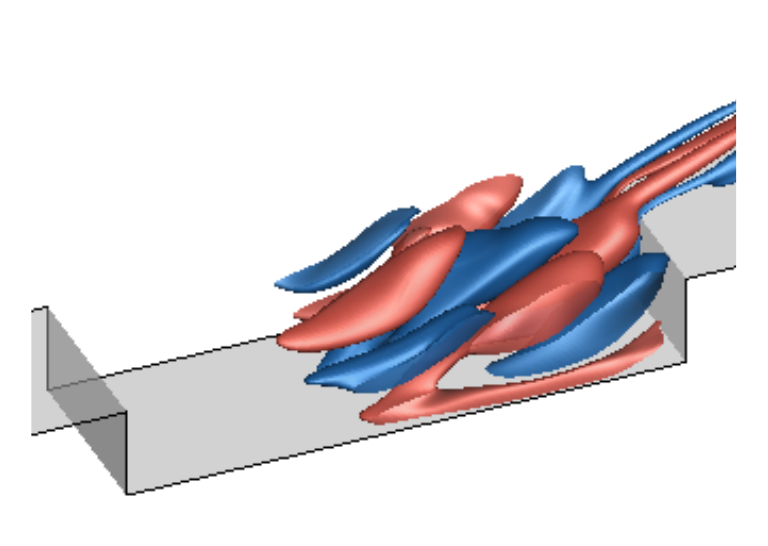}	&	\includegraphics[width=0.4\textwidth]{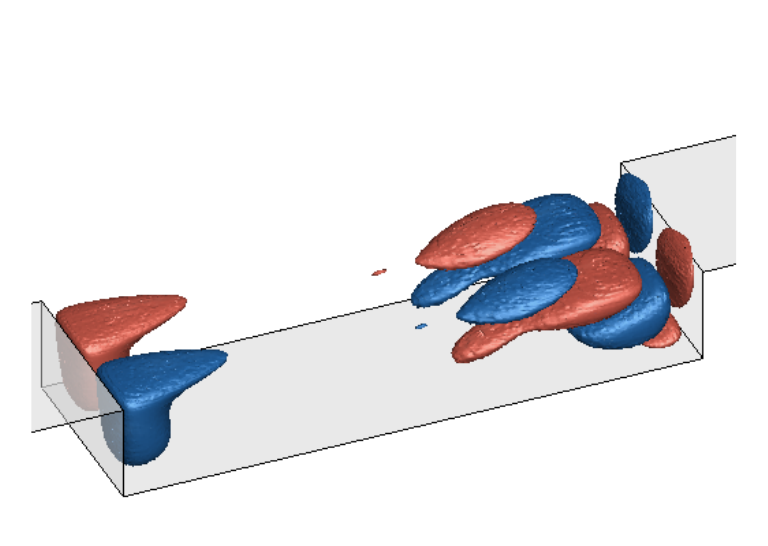}  \\ 
\begin{center}$(a)$ 3D nonlinear simulation \end{center} & \begin{center} $(b)$ 3D stationary mode ($\lambda/D=2.0$) \end{center}
\end{tabular}
\caption{Comparison of the iso-surface of the instantaneous spanwise velocity $w/u_\infty=\pm0.083$ from the 3D nonlinear simulation and $\hat w/u_\infty=\pm0.083\times 10^{-3}$ from the dominant 3D eigenmode (stationary mode) with $\lambda/D=2.0$ for the cavity with $L/D=6$ and $M_\infty=0.6$ and $Re_\theta=19$. Red and blue colors represent positive and negative values, respectively.}
\label{fig:3Dnon_LD6}
\end{figure}

For the cavity with $L/D=6$ ($Re_\theta=19$), the violent wake mode present in 2D DNS is not observed in 3D DNS. Instead, Rossiter mode II dominates the flow. When the flow has reached steady oscillations, there is no evident low-frequency peaks of the 3D structures in the nonlinear flows. In the comparison of the 3D spatial structures revealed by the 3D DNS and 3D eigenmodes as shown in figure \ref{fig:3Dnon_LD6}, we find that for the 3D nonlinear flows, the spatial structures located in the rear region of the cavity stay almost stationary in the spanwise direction with wavelength of $\lambda/D=2$, while those obtained from the linear stability analysis are distributed in the front and rear regions of the cavity. As shown in figure \ref{fig:dominant_LD6}, though all the 3D eigenmodes have negative values of growth rate, the spatial structures of 2D Rossiter modes (figure \ref{fig:beta0_eigenmode_LD6}) and 3D eigenmodes (figure \ref{fig:3Dviews}) overlap in the rear part of the cavity, which can lead to modal interactions and modify the instabilities predicted from biglobal stability analysis. 

In the 3D DNS, the large vortex roll-ups of wake mode present in the 2D DNS is no longer observable. Meanwhile, the spanwise motion becomes evident in the 3D flows. To uncover these spanwise effects on the long cavity, further analysis has been performed in our companion work \citep{Sun:TCFD16}, in which we found that the presence of spanwise motion (or three-dimensionalities) in the flow can preclude the formation of the wake mode, reducing the intense fluctuations of the flow. This observation suggests the potential to reduce cavity flow oscillations by introducing spanwise variations to the base flows. In order to weaken the strength of the impingement of shear layer on the trailing edge of the cavity, we can trigger the emergence of 3D global modes and achieve energy redistribution from spanwise vortices to streamwise vortical structures. To retain the energy inside the cavity, the spanwise wavelength of perturbation introduced to the flow should be in the range of $0.5-1.5$, since the spatial structures of 3D modes associated with these wavelengths mainly stay inside the cavity, which has been illustrated in figure \ref{fig:3Dviews}.  

%
%
%

To verify this concept of using 3D instability to reduce cavity flow oscillations, we note in passing that 3D slot-jets can be introduced along the leading edge of the cavity to alter the flow features. The performance of these control strategies via 3D steady blowing for cavity flows is assessed by \cite{Zhang:AIAA15} and \cite{George:AIAA15} through a large number of experiments and some representative large-eddy simulations. They find that the wavelengths for effective 3D flow control to reduce cavity flow oscillations are in good agreement with the findings of present global stability analysis.  
Although the idea of 3D flow control is not new \citep{ZDRAVKOVICH1981145}, the present work particularly uncovers the temporal instabilities affected by compressibility in the transonic regime. Moreover, the spanwise wavelength and frequency associated with 3D instabilities are identified, which can be useful in examining the influence of flow parameters for 3D flow control designs.
We believe that the insights from the global instability characteristics of compressible open-cavity flows can provide valuable information for designing physics-based flow control strategies.

\section{Conclusions}
\label{sec:summary}

Two- and three-dimensional biglobal instabilities of two-dimensional compressible open-cavity flows are examined for rectangular cavities with aspect ratios of $L/D=2$ and 6. 
Noteworthy in the present work is the focus on the 
global instability analysis for long cavity ($L/D=6$) flows in the transonic regime, which has not been examined in detail in the past, despite their importance in being 
representative of practical cavity applications on aircraft. Direct numerical simulations and biglobal stability analysis are used to uncover the influence of Mach number, Reynolds number, and spanwise wavelength on the 2D and 3D global instability modes with respect to the 2D steady states. In the 2D DNS, we find that an increase in Mach number in the subsonic regime  destabilizes the flow but stabilizes the flow in the transonic regime at low Reynolds number for both short ($L/D=2$) and long ($L/D=6$) cavities, which agrees with the findings from the work by \cite{Yamouni:JFM13} on flows over short cavities ($L/D=1$ and $2$).  
Moreover, the dominant Rossiter mode can shift due to the variation in the free stream Mach number. 

The 2D eigenmodes obtained from biglobal stability analysis exhibit excellent agreement with the properties of the 2D flow characteristics uncovered from 2D DNS, indicating that these 2D global modes are Rossiter modes driven by the shear layer. Through the 2D eigen-analysis with $\beta=0$, destabilization and subsequent stabilization effects via increasing Mach number is captured. Moreover, a shift in the dominant Rossiter mode is observed with Mach number variation, which explains the mode shifting behavior in 2D nonlinear simulation results.

In contrast to the 2D modes, the dominant 3D modes unraveled in the present work are 
instabilities that are largely unaffected by a change in Mach number. Although an increase in Mach number stabilizes all 3D eigenmodes, the overall behavior of 3D eigenmodes as a function of the spanwise wavelength $\lambda/D$ is almost independent of the Mach number. However, the type of dominant 3D mode is strongly dependent on $\lambda/D$. For the short cavity with $L/D=2$, the most-unstable/least-stable 3D modes have $\lambda/D\in[1.0,1.25]$ (mode $ii$) stemming from centrifugal instabilities for almost all Mach numbers considered, 
as reported by \cite{Bres:JFM08}. For the long cavity with $L/D=6$, the least-stable 3D modes have $\lambda/D=1.25$ and 2.0 for cases of $M_\infty \lesssim 0.3$ and $M_\infty \gtrsim 0.6$, respectively. Moreover, there are additional types of eigenmodes for this long cavity flow possessing various spatial structures compared to those of the shorter cavity. 

In 3D DNS, both frequencies and spatial structures of the leading 3D mode seen in the DNS correspond closely to the results of the linear biglobal stability analysis of the dominant 3D eigenmode for $L/D=2$. However, for $L/D=6$, 3D DNS reveals large roll up of the cavity shear layer, which indicates strong nonlinearities that depart from the linear analysis. Nevertheless, the insights from global stability analysis provide a potential pathway to reduce flow oscillations by introducing spanwise variations to the base flows, which has been demonstrated in the experiments by \cite{Lusk:EF12, Zhang:AIAA15} and \cite{George:AIAA15}. The analyses in this study provide valuable knowledge on global instabilities of 2D and 3D compressible open-cavity flows, which can be leveraged in designing physics-based flow control strategies in upcoming studies.

\section*{Acknowledgement}
This work was supported by the U.S. Air Force Office of Scientific Research (Grant number: FA9550-13-1-0091, Program Managers: Drs.~D.~Smith and I.~Leyva). The high performance computing resource was provided by the Research Computing Center at the Florida State University. The authors also acknowledge the insightful discussions with Dr.~Guillaume A.~Br\`es. 

\bibliographystyle{jfm}

\bibliography{ref}

\begin{thebibliography}{44}
\expandafter\ifx\csname natexlab\endcsname\relax\def\natexlab#1{#1}\fi
\def\au#1{#1} \def\ed#1{#1} \def\yr#1{#1}\def\at#1{#1}\def\jt#1{\textit{#1}}
  \def\bt#1{#1}\def\bvol#1{\textbf{#1}} \def\vol#1{#1} \def\pg#1{#1}
  \def\publ#1{#1}\def\arxiv#1{#1}\def\org#1{#1}\def\st#1{\textit{#1}}

\bibitem[Ahuja \& Mendoza(1995)]{Ahuja:NASA95}
{\sc \au{Ahuja, K.~K.} \& \au{Mendoza, J.}} \yr{1995}  \bt{Effects of cavity
  dimensions, boundary layer and temperature on cavity noise with emphasis on
  benchmark data to validate computational aeroacoustic codes}.  \org{{\em
  Tech. Rep.\/} 4653}.

\bibitem[{\AA}kervik {\em et~al.\/}(2006){\AA}kervik, Brandt, Henningson,
  H{\oe}pffner, Marxen \& Schlatter]{Akervik:PF06}
{\sc \au{{\AA}kervik, E.}, \au{Brandt, L.}, \au{Henningson, D.~S.},
  \au{H{\oe}pffner, J.}, \au{Marxen, O.} \& \au{Schlatter, P.}} \yr{2006}
  \at{Steady solutions of the {N}avier-{S}tokes equations by selective
  frequency damping}.  \jt{Phys. Fluids}  \bvol{18},  \pg{068--102}.

\bibitem[Arunajatesan {\em et~al.\/}(2014)Arunajatesan, Barone, Wagner, Casper
  \& Beresh]{Arunajatesan:AIAA14}
{\sc \au{Arunajatesan, S.}, \au{Barone, M.~F.}, \au{Wagner, J.~L.}, \au{Casper,
  K.~M.} \& \au{Beresh, S.~J.}} \yr{2014}  \bt{Joint experimental/computational
  inversigation into the effects of finite width on transonic cavity flow}.
  \publ{{AIAA} Paper 2014-3027}.

\bibitem[Beresh {\em et~al.\/}(2016)Beresh, Wagner \& Casper]{Beresh:JFM16}
{\sc \au{Beresh, S.~J.}, \au{Wagner, J.~L.} \& \au{Casper, K.~M.}} \yr{2016}
  \at{{Compressibility effects in the shear layer over a rectangular cavity}}.
  \jt{Journal of Fluid Mechanics}  \bvol{808},  \pg{116--152}.

\bibitem[Beresh {\em et~al.\/}(2015)Beresh, Wagner, Pruett, Henfling \&
  Spillers]{Beresh:AIAA15}
{\sc \au{Beresh, S.~J.}, \au{Wagner, J.~L.}, \au{Pruett, B. O.~M.},
  \au{Henfling, J.~F.} \& \au{Spillers, R.~W.}} \yr{2015}  \at{{Supersonic Flow
  over a Finite-Width Rectangular Cavity}}.  \jt{AIAA J.}  \bvol{53}~(2),
  \pg{296--310}.

\bibitem[Bergamo {\em et~al.\/}(2015)Bergamo, Gennaro, Theofilis \&
  Medeiros]{Bergamo:AST15}
{\sc \au{Bergamo, L.~F.}, \au{Gennaro, E.~M.}, \au{Theofilis, V.} \&
  \au{Medeiros, M. A.~F.}} \yr{2015}  \at{{Compressible modes in a square
  lid-driven cavity}}.  \jt{Aerospace Science and Technology}  \bvol{44}~(C),
  \pg{125--134}.

\bibitem[Br\`es(2007)]{Bres:2007}
{\sc \au{Br\`es, G.~A.}} \yr{2007}  \at{Numerical simulations of
  three-dimensional instabilities in cavity flows}. PhD thesis, California
  Institute of Technology.

\bibitem[Br\`es \& Colonius(2008)]{Bres:JFM08}
{\sc \au{Br\`es, G.~A.} \& \au{Colonius, T.}} \yr{2008}  \at{Three-dimensional
  instabilities in compressible flow over open cavities}.  \jt{J. Fluid Mech.}
  \bvol{599},  \pg{309--339}.

\bibitem[Br\`es {\em et~al.\/}(2017)Br\`es, Ham, Nichols \& Lele]{Bres:AIAAJ17}
{\sc \au{Br\`es, G.~A.}, \au{Ham, F.~E.}, \au{Nichols, J.~W.} \& \au{Lele,
  S.~K.}} \yr{2017}  \at{Unstructured large-eddy simulations of supersonic
  jets}.  \jt{AIAA J.}  \bvol{55}~(4).

\bibitem[Cattafesta {\em et~al.\/}(1997)Cattafesta, Garg, Choudhari \&
  Li]{Cattafesta:AIAA97}
{\sc \au{Cattafesta, L.~N.}, \au{Garg, S.}, \au{Choudhari, M.} \& \au{Li, F.}}
  \yr{1997}  \bt{Active control of flow-induced cavity resonance}.
  \publ{{AIAA} Paper 1997-1804}.

\bibitem[Cattafesta \& Sheplak(2011)]{Cattafesta:ARFM11}
{\sc \au{Cattafesta, L.~N.} \& \au{Sheplak, M.}} \yr{2011}  \at{Actuators for
  active flow control}.  \jt{Annu. Rev. Fluid Mech.}  \bvol{43},
  \pg{247--272}.

\bibitem[Cattafesta {\em et~al.\/}(2008)Cattafesta, Song, Williams, Rowley \&
  Alvi]{Cattafesta:PAS08}
{\sc \au{Cattafesta, L.~N.}, \au{Song, Q.}, \au{Williams, D.~R.}, \au{Rowley,
  C.~W.} \& \au{Alvi, F.~S.}} \yr{2008}  \at{Active control of flow-induced
  cavity oscillations}.  \jt{Prog. Aero. Sci.}  \bvol{44},  \pg{479--502}.

\bibitem[Colonius {\em et~al.\/}(1999)Colonius, Basu \& Rowley]{Colonius:99}
{\sc \au{Colonius, T.}, \au{Basu, A.~J.} \& \au{Rowley, C.~W.}} \yr{1999}
  \bt{Numerical investigation of the flow past a cavity}.  \publ{{AIAA} Paper
  1999-1912}.

\bibitem[Freund(1997)]{Freund:AIAAJ97}
{\sc \au{Freund, J.~B.}} \yr{1997}  \at{Proposed inflow/outflow boundary
  condition for direct computation of aerodynamic sound}.  \jt{AIAA J.}
  \bvol{35}~(4),  \pg{740--742}.

\bibitem[George {\em et~al.\/}(2015)George, Ukeiley, Cattafesta \&
  Taira]{George:AIAA15}
{\sc \au{George, B.}, \au{Ukeiley, L.}, \au{Cattafesta, L.~N.} \& \au{Taira,
  K.}} \yr{2015}  \bt{Control of three-dimensional cavity flow using leading
  edge slot blowing}.  \publ{{AIAA} Paper 2015-1059}.

\bibitem[Gharib \& Roshko(1987)]{Gharib:JFM87}
{\sc \au{Gharib, M.} \& \au{Roshko, A.}} \yr{1987}  \at{The effect of flow
  oscillations on cavity drag}.  \jt{J. Fluid Mech.}  \bvol{177},
  \pg{501--530}.

\bibitem[Heller \& Bliss(1975)]{Heller:AIAA75}
{\sc \au{Heller, H.~H.} \& \au{Bliss, D.~B.}} \yr{1975}  \bt{The physical
  mechanism of flow-induced pressure fluctuations in cavities and concepts for
  their suppression}.  \publ{{AIAA} Paper 1975-491}.

\bibitem[Kegerise {\em et~al.\/}(2004)Kegerise, Spina, Garg \&
  Catttafesta]{Kegerise:04PF}
{\sc \au{Kegerise, M.~A.}, \au{Spina, E.~F.}, \au{Garg, S.} \& \au{Catttafesta,
  L.~N.}} \yr{2004}  \at{Mode-switching and nonlinear effects in compressible
  flow over a cavity}.  \jt{Physics of Fluids}  \bvol{16}~(3),  \pg{678--686}.

\bibitem[Khalighi {\em et~al.\/}(2011{\natexlab{{\em a\/}}})Khalighi, Ham,
  Moin, Lele, Schlinker, Reba \& J.]{Khalighi:ASME2011}
{\sc \au{Khalighi, Y.}, \au{Ham, F.}, \au{Moin, P.}, \au{Lele, S.},
  \au{Schlinker, R.}, \au{Reba, R.} \& \au{J., Simonich}}
  \yr{2011{\natexlab{{\em a\/}}}}  \bt{Noise prediction of pressure-mismatched
  jets using unstructured large eddy simulation}. {Proceedings of ASME Turbo
  Expo}, Vancouver.

\bibitem[Khalighi {\em et~al.\/}(2011{\natexlab{{\em b\/}}})Khalighi, Nichols,
  Ham, Lele \& Moin]{Khalighi:AIAA11}
{\sc \au{Khalighi, Y.}, \au{Nichols, J.~W.}, \au{Ham, F.}, \au{Lele, S.~K.} \&
  \au{Moin, P.}} \yr{2011{\natexlab{{\em b\/}}}}  \bt{Unstructured large eddy
  simulation for prediction of noise issued from turbulent jets in various
  configurations}. 17th {AIAA/CEAS} Aeroacoustics Conference.

\bibitem[Krishnamurty(1956)]{Krishnamurty:1956}
{\sc \au{Krishnamurty, K.}} \yr{1956}  \at{Sound radiation from surface cutouts
  in high speed flow}. PhD thesis, California Institute of Technology.

\bibitem[Lawson \& Barakos(2011)]{Lawson:PAS11}
{\sc \au{Lawson, S.~J.} \& \au{Barakos, G.~N.}} \yr{2011}  \at{Review of
  numerical simulations for high-speed, turbulent cavity flows}.  \jt{Prog.
  Aero. Sci.}  \bvol{47},  \pg{186--216}.

\bibitem[Lehoucq {\em et~al.\/}(1996-2007)Lehoucq, Maschhoff, Sorensen \&
  Yang]{Arpack:96}
{\sc \au{Lehoucq, R.}, \au{Maschhoff, K.}, \au{Sorensen, D.} \& \au{Yang, C.}}
  \yr{1996-2007} {ARPACK} software.

\bibitem[Liu {\em et~al.\/}(2016)Liu, G{\'o}mez \& Theofilis]{Liu:JFM16}
{\sc \au{Liu, Q.}, \au{G{\'o}mez, F.} \& \au{Theofilis, V.}} \yr{2016}
  \at{{Linear instability analysis of low-{$Re$} incompressible flow over a
  long rectangular finite-span open cavity}}.  \jt{J. Fluid Mech.}  \bvol{799},
   \pg{R2}.

\bibitem[Lusk {\em et~al.\/}(2012)Lusk, Cattafesta \& Ukeiley]{Lusk:EF12}
{\sc \au{Lusk, T.}, \au{Cattafesta, L.} \& \au{Ukeiley, L.~S.}} \yr{2012}
  \at{Leading-edge slot blowing on an open cavity in supersonic flow}.
  \jt{Experiments in Fluids}  \bvol{53},  \pg{187--199}.

\bibitem[Maull \& East(1963)]{Maull:JFM63}
{\sc \au{Maull, D.~J.} \& \au{East, L.~F.}} \yr{1963}  \at{Three-dimensional
  flow in cavities}.  \jt{J. Fluid Mech.}  \bvol{16},  \pg{620--632}.

\bibitem[Meseguer-Garrido {\em et~al.\/}(2014)Meseguer-Garrido, de~Vicente,
  Valero \& Theofilis]{Garrido:JFM14}
{\sc \au{Meseguer-Garrido, F.}, \au{de~Vicente, J.}, \au{Valero, E.} \&
  \au{Theofilis, V.}} \yr{2014}  \at{On linear instability mechanisms in
  incompressible open cavity flow}.  \jt{Journal of Fluid Mechanics}
  \bvol{752},  \pg{219--236}.

\bibitem[Rockwell \& Naudascher(1978)]{Rockwell:JFE78}
{\sc \au{Rockwell, D.} \& \au{Naudascher, E.}} \yr{1978}
  \at{Review-self-sustaining oscillations of flow past cavities}.  \jt{J.
  Fluids Eng.}  \bvol{100},  \pg{152--165}.

\bibitem[Rossiter(1964)]{Rossiter:ARCRM64}
{\sc \au{Rossiter, J.~E.}} \yr{1964}  \bt{Wind-tunnel experiments on the flow
  over rectangular cavities at subsonic and transonic speeds}. {\em Tech.
  Rep.\/} 3438.  \org{Aeronautical Research Council Reports and Memoranda}.

\bibitem[Rowley {\em et~al.\/}(2002)Rowley, Colonius \& Basu]{Rowley:JFM02}
{\sc \au{Rowley, C.~W.}, \au{Colonius, T.} \& \au{Basu, A.~J.}} \yr{2002}
  \at{On self-sustained oscillations in two-dimensional compressible flow over
  rectangular cavities}.  \jt{J. Fluid Mech.}  \bvol{455},  \pg{315--346}.

\bibitem[Samimy {\em et~al.\/}(2007)Samimy, Kim, Kastner, Adamovich \&
  Utkin]{Samimy:JFM07}
{\sc \au{Samimy, M.}, \au{Kim, J.~H.}, \au{Kastner, J.}, \au{Adamovich, I.} \&
  \au{Utkin, Y.}} \yr{2007}  \at{Active control of high-speed and
  high-{R}eynolds-number jets using plasma actuators}.  \jt{J. Fluid Mech.}
  \bvol{578},  \pg{305--330}.

\bibitem[Sarohia(1975)]{Sarohia:1975}
{\sc \au{Sarohia, V.}} \yr{1975}  \at{Experimental and analytical investigation
  of oscillations in flows over cavities}. PhD thesis, California Institute of
  Technology.

\bibitem[Sun {\em et~al.\/}(2014)Sun, Nair, Taira, Cattafesta, Br\`es \&
  Ukeiley]{Sun:AIAA14}
{\sc \au{Sun, Y.}, \au{Nair, A.~G.}, \au{Taira, K.}, \au{Cattafesta, L.~N.},
  \au{Br\`es, G.~A.} \& \au{Ukeiley, L.~S.}} \yr{2014}  \bt{Numerical
  simulation of subsonic and transonic open-cavity flows}.  \publ{{AIAA} Paper
  2014-3092}.

\bibitem[Sun {\em et~al.\/}(2016{\natexlab{{\em a\/}}})Sun, Taira, Cattafesta
  \& Ukeiley]{Sun:TCFD16}
{\sc \au{Sun, Y.}, \au{Taira, K.}, \au{Cattafesta, L.~N.} \& \au{Ukeiley,
  L.~S.}} \yr{2016{\natexlab{{\em a\/}}}}  \at{Spanwise effects on
  instabilities of compressible flow over a long rectangular cavity}.
  \jt{Theoretical and Computational Fluid Dynamics}  \pg{pp. 1--11}.

\bibitem[Sun {\em et~al.\/}(2016{\natexlab{{\em b\/}}})Sun, Zhang, Taira,
  Cattafesta, George \& Ukeiley]{Sun:AIAA16}
{\sc \au{Sun, Y.}, \au{Zhang, Y.}, \au{Taira, K.}, \au{Cattafesta, L.},
  \au{George, B.} \& \au{Ukeiley, L.}} \yr{2016{\natexlab{{\em b\/}}}}
  \bt{Width and sidewall effects on high speed cavity flows}.  \publ{{AIAA}
  Paper 2016-1343}.

\bibitem[Theofilis(2003)]{Theofilis:PAS03}
{\sc \au{Theofilis, V.}} \yr{2003}  \at{Advances in global linear instability
  analysis of nonparallel and three-dimensional flows}.  \jt{Prog. Aero. Sci.}
  \bvol{39},  \pg{249--315}.

\bibitem[Theofilis \& Colonius(2004)]{Theofilis:AIAA04}
{\sc \au{Theofilis, V.} \& \au{Colonius, T.}} \yr{2004}  \bt{Three-dimensional
  instabilities of compressible flow over open cavities: direct solution of the
  biglobal eigenvalue problem.}  \publ{{AIAA} Paper 2004-2544}.

\bibitem[Toro(2009)]{Toro:09}
{\sc \au{Toro, E.~F.}} \yr{2009} {\em Riemann solvers and numerical methods for
  fluid dynamics\/}.  \publ{Springer-Verlag}.

\bibitem[de~Vicente {\em et~al.\/}(2014)de~Vicente, Basley, Meseguer-Garrido,
  Soria \& Theofilis]{Vicente:JFM14}
{\sc \au{de~Vicente, J.}, \au{Basley, J.}, \au{Meseguer-Garrido, F.},
  \au{Soria, J.} \& \au{Theofilis, V.}} \yr{2014}  \at{Three-dimensional
  instabilities over a rectangular open cavity: from linear stability analysis
  to experimentation}.  \jt{J. Fluid Mech.}  \bvol{748},  \pg{189--220}.

\bibitem[White(1991)]{white91}
{\sc \au{White, F.}} \yr{1991} {\em Viscous Fluid Flow\/}.  \publ{McGraw-Hill}.

\bibitem[Yamouni {\em et~al.\/}(2013)Yamouni, Sipp \& Jacquin]{Yamouni:JFM13}
{\sc \au{Yamouni, S.}, \au{Sipp, D.} \& \au{Jacquin, L.}} \yr{2013}
  \at{Interaction between feedback aeroacoustic and acoustic resonance
  mechanisms in a cavity flow: a global stability analysis}.  \jt{J. Fluid
  Mech.}  \bvol{717},  \pg{134--165}.

\bibitem[Zdravkovich(1981)]{ZDRAVKOVICH1981145}
{\sc \au{Zdravkovich, M.~M.}} \yr{1981}  \at{Review and classification of
  various aerodynamic and hydrodynamic means for suppressing vortex shedding}.
  \jt{Journal of Wind Engineering and Industrial Aerodynamics}  \bvol{7}~(2),
  \pg{145 -- 189}.

\bibitem[Zhang \& Naguib(2011)]{Zhang:EF11}
{\sc \au{Zhang, K.} \& \au{Naguib, A.~M.}} \yr{2011}  \at{Effect of finite
  cavity width on flow oscillation in a low-{M}ach-number cavity flow}.
  \jt{Experiments in Fluids}  \bvol{51}~(5),  \pg{1209--1229}.

\bibitem[Zhang {\em et~al.\/}(2015)Zhang, Sun, Arora, Cattafesta, Taira \&
  Ukeiley]{Zhang:AIAA15}
{\sc \au{Zhang, Y.}, \au{Sun, Y.}, \au{Arora, N.}, \au{Cattafesta, L.},
  \au{Taira, K.} \& \au{Ukeiley, L.}} \yr{2015}  \bt{Suppression of cavity
  oscillations via three-dimensional steady blowing}.  \publ{{AIAA} Paper
  2015-3219}.

\end{thebibliography}

\end{document}